\documentclass[longauth,traditabstract]{aa}
\usepackage[english]{babel} 
\usepackage[nonamebreak]{natbib}
\usepackage[stable]{footmisc} 
\usepackage{anyfontsize}
\usepackage{amsmath}
\usepackage{txfonts} 
\usepackage{placeins} 
\bibpunct{(}{)}{;}{a}{}{,} 
\usepackage{epstopdf} 
\usepackage{ifthen} 
\usepackage[breaklinks,colorlinks, citecolor=blue]{hyperref} 
\usepackage{overpic}
\usepackage{fixltx2e} 
\usepackage[table,usenames,dvipsnames]{xcolor}

\def\setsymbol#1#2{\expandafter\def\csname #1\endcsname{#2}}
\def\getsymbol#1{\csname #1\endcsname}

\def\Planck{\textit{Planck}}

\def\all2013resultspapers{\nocite{planck2013-p01, planck2013-p02, planck2013-p02a, planck2013-p02d, planck2013-p02b, planck2013-p03, planck2013-p03c, planck2013-p03f, planck2013-p03d, planck2013-p03e, planck2013-p01a, planck2013-p06, planck2013-p03a, planck2013-pip88, planck2013-p08, planck2013-p11, planck2013-p12, planck2013-p13, planck2013-p14, planck2013-p15, planck2013-p05b, planck2013-p17, planck2013-p09, planck2013-p09a, planck2013-p20, planck2013-p19, planck2013-pipaberration, planck2013-p05, planck2013-p05a, planck2013-pip56, planck2013-p06b, planck2013-p01a}}

\newbox\tablebox    \newdimen\tablewidth
\def\leaderfil{\leaders\hbox to 5pt{\hss.\hss}\hfil}
\def\endPlancktable{\tablewidth=\columnwidth 
    $$\hss\copy\tablebox\hss$$
    \vskip-\lastskip\vskip -2pt}

\def\tablenote#1 #2\par{\begingroup \parindent=0.8em
    \abovedisplayshortskip=0pt\belowdisplayshortskip=0pt
    \noindent
    $$\hss\vbox{\hsize\tablewidth \hangindent=\parindent \hangafter=1 \noindent
    \hbox to \parindent{$^#1$\hss}\strut#2\strut\par}\hss$$
    \endgroup}
\def\doubleline{\vskip 3pt\hrule \vskip 1.5pt \hrule \vskip 5pt}

\def\L2{\ifmmode L_2\else $L_2$\fi}

\def\DeltaT{\ifmmode \Delta T\else $\Delta T$\fi}
\def\deltat{\ifmmode \Delta t\else $\Delta t$\fi}
\def\fknee{\ifmmode f_{\rm knee}\else $f_{\rm knee}$\fi}
\def\Fmax{\ifmmode F_{\rm max}\else $F_{\rm max}$\fi}
\def\solar{\ifmmode{\rm M}_{\mathord\odot}\else${\rm M}_{\mathord\odot}$\fi}
\def\Msolar{\ifmmode{\rm M}_{\mathord\odot}\else${\rm M}_{\mathord\odot}$\fi}
\def\Lsolar{\ifmmode{\rm L}_{\mathord\odot}\else${\rm L}_{\mathord\odot}$\fi}
\def\inv{\ifmmode^{-1}\else$^{-1}$\fi}
\def\mo{\ifmmode^{-1}\else$^{-1}$\fi}
\def\sup#1{\ifmmode ^{\rm #1}\else $^{\rm #1}$\fi}
\def\expo#1{\ifmmode \times 10^{#1}\else $\times 10^{#1}$\fi}
\def\,{\thinspace}
\def\lsim{\mathrel{\raise .4ex\hbox{\rlap{$<$}\lower 1.2ex\hbox{$\sim$}}}}
\def\gsim{\mathrel{\raise .4ex\hbox{\rlap{$>$}\lower 1.2ex\hbox{$\sim$}}}}

\def\simprop{\mathrel{\raise .4ex\hbox{\rlap{$\propto$}\lower 1.2ex\hbox{$\sim$}}}}
\def\deg{\ifmmode^\circ\else$^\circ$\fi}
\def\pdeg{\ifmmode $\setbox0=\hbox{$^{\circ}$}\rlap{\hskip.11\wd0 .}$^{\circ}
          \else \setbox0=\hbox{$^{\circ}$}\rlap{\hskip.11\wd0 .}$^{\circ}$\fi}
\def\arcs{\ifmmode {^{\scriptstyle\prime\prime}}
          \else $^{\scriptstyle\prime\prime}$\fi}
\def\arcm{\ifmmode {^{\scriptstyle\prime}}
          \else $^{\scriptstyle\prime}$\fi}
\newdimen\sa  \newdimen\sb
\def\parcs{\sa=.07em \sb=.03em
     \ifmmode \hbox{\rlap{.}}^{\scriptstyle\prime\kern -\sb\prime}\hbox{\kern -\sa}
     \else \rlap{.}$^{\scriptstyle\prime\kern -\sb\prime}$\kern -\sa\fi}
\def\parcm{\sa=.08em \sb=.03em
     \ifmmode \hbox{\rlap{.}\kern\sa}^{\scriptstyle\prime}\hbox{\kern-\sb}
     \else \rlap{.}\kern\sa$^{\scriptstyle\prime}$\kern-\sb\fi}
\def\ra[#1 #2 #3.#4]{#1\sup{h}#2\sup{m}#3\sup{s}\llap.#4}
\def\dec[#1 #2 #3.#4]{#1\deg#2\arcm#3\arcs\llap.#4}
\def\deco[#1 #2 #3]{#1\deg#2\arcm#3\arcs}
\def\rra[#1 #2]{#1\sup{h}#2\sup{m}}

\def\dots{\relax\ifmmode \ldots\else $\ldots$\fi}
\def\WHzsr{\ifmmode $W\,Hz\mo\,sr\mo$\else W\,Hz\mo\,sr\mo\fi}
\def\mHz{\ifmmode $\,mHz$\else \,mHz\fi}
\def\GHz{\ifmmode $\,GHz$\else \,GHz\fi}
\def\mKs{\ifmmode $\,mK\,s$^{1/2}\else \,mK\,s$^{1/2}$\fi}
\def\muKs{\ifmmode \,\mu$K\,s$^{1/2}\else \,$\mu$K\,s$^{1/2}$\fi}
\def\muKRJs{\ifmmode \,\mu$K$_{\rm RJ}$\,s$^{1/2}\else \,$\mu$K$_{\rm RJ}$\,s$^{1/2}$\fi}
\def\muKHz{\ifmmode \,\mu$K\,Hz$^{-1/2}\else \,$\mu$K\,Hz$^{-1/2}$\fi}
\def\MJysr{\ifmmode \,$MJy\,sr\mo$\else \,MJy\,sr\mo\fi}
\def\MJysrmK{\ifmmode \,$MJy\,sr\mo$\,mK$_{\rm CMB}\mo\else \,MJy\,sr\mo\,mK$_{\rm CMB}\mo$\fi}
\def\microns{\ifmmode \,\mu$m$\else \,$\mu$m\fi}

\def\muK{\ifmmode \,\mu$K$\else \,$\mu$\hbox{K}\fi}
\def\microK{\ifmmode \,\mu$K$\else \,$\mu$\hbox{K}\fi}
\def\muW{\ifmmode \,\mu$W$\else \,$\mu$\hbox{W}\fi}
\def\kms{\ifmmode $\,km\,s$^{-1}\else \,km\,s$^{-1}$\fi}
\def\kmsMpc{\ifmmode $\,\kms\,Mpc\mo$\else \,\kms\,Mpc\mo\fi}
\providecommand{\sorthelp}[1]{}

\newcommand{\resub}[1]{#1}
\newcommand{\newresub}[1]{#1}

\newcommand{\Nside}{\ensuremath{N_{\mathrm{side}}}} 
\newcommand{\fsky}{{f_{\rm sky}}}
\newcommand{\be}{\begin{equation}} 
\newcommand{\ee}{\end{equation}}
\newcommand{\hatn}{\vec{\hat{n\,}}}

\newcommand{\nilc}{{\tt NILC}} 
\newcommand{\sevem}{{\tt SEVEM}}
\newcommand{\smica}{{\tt SMICA}} 
\newcommand{\commander}{{\tt Commander}}

\begin{document}

\title{\Planck\ intermediate results. XLI.\\ 
A map of lensing-induced $B$-modes} 
\author{\small
Planck Collaboration: P.~A.~R.~Ade\inst{82}
\and
N.~Aghanim\inst{54}
\and
M.~Ashdown\inst{64, 6}
\and
J.~Aumont\inst{54}
\and
C.~Baccigalupi\inst{81}
\and
A.~J.~Banday\inst{91, 10}
\and
R.~B.~Barreiro\inst{59}
\and
N.~Bartolo\inst{30, 60}
\and
S.~Basak\inst{81}
\and
E.~Battaner\inst{92, 93}
% ...
}
\institute{\small
APC, AstroParticule et Cosmologie, Universit\'{e} Paris Diderot, CNRS/IN2P3, CEA/lrfu, Observatoire de Paris, Sorbonne Paris Cit\'{e}, 10, rue Alice Domon et L\'{e}onie Duquet, 75205 Paris Cedex 13, France\goodbreak
\and
Aalto University Mets\"{a}hovi Radio Observatory and Dept of Radio Science and Engineering, P.O. Box 13000, FI-00076 AALTO, Finland\goodbreak
\and
African Institute for Mathematical Sciences, 6-8 Melrose Road, Muizenberg, Cape Town, South Africa\goodbreak
\and
Agenzia Spaziale Italiana Science Data Center, Via del Politecnico snc, 00133, Roma, Italy\goodbreak
\and
Aix Marseille Universit\'{e}, CNRS, LAM (Laboratoire d'Astrophysique de Marseille) UMR 7326, 13388, Marseille, France\goodbreak
\and
Astrophysics Group, Cavendish Laboratory, University of Cambridge, J J Thomson Avenue, Cambridge CB3 0HE, U.K.\goodbreak
\and
Astrophysics \& Cosmology Research Unit, School of Mathematics, Statistics \& Computer Science, University of KwaZulu-Natal, Westville Campus, Private Bag X54001, Durban 4000, South Africa\goodbreak
\and
CGEE, SCS Qd 9, Lote C, Torre C, 4$^{\circ}$ andar, Ed. Parque Cidade Corporate, CEP 70308-200, Bras\'{i}lia, DF,Ê Brazil\goodbreak
\and
CITA, University of Toronto, 60 St. George St., Toronto, ON M5S 3H8, Canada\goodbreak
\and
CNRS, IRAP, 9 Av. colonel Roche, BP 44346, F-31028 Toulouse cedex 4, France\goodbreak
\and
CRANN, Trinity College, Dublin, Ireland\goodbreak
\and
California Institute of Technology, Pasadena, California, U.S.A.\goodbreak
\and
Centro de Estudios de F\'{i}sica del Cosmos de Arag\'{o}n (CEFCA), Plaza San Juan, 1, planta 2, E-44001, Teruel, Spain\goodbreak
\and
Computational Cosmology Center, Lawrence Berkeley National Laboratory, Berkeley, California, U.S.A.\goodbreak
\and
DSM/Irfu/SPP, CEA-Saclay, F-91191 Gif-sur-Yvette Cedex, France\goodbreak
\and
DTU Space, National Space Institute, Technical University of Denmark, Elektrovej 327, DK-2800 Kgs. Lyngby, Denmark\goodbreak
\and
D\'{e}partement de Physique Th\'{e}orique, Universit\'{e} de Gen\`{e}ve, 24, Quai E. Ansermet,1211 Gen\`{e}ve 4, Switzerland\goodbreak
\and
Departamento de Astrof\'{i}sica, Universidad de La Laguna (ULL), E-38206 La Laguna, Tenerife, Spain\goodbreak
\and
Departamento de F\'{\i}sica, Universidad de Oviedo, Avda. Calvo Sotelo s/n, Oviedo, Spain\goodbreak
\and
Department of Astronomy and Astrophysics, University of Toronto, 50 Saint George Street, Toronto, Ontario, Canada\goodbreak
\and
Department of Astrophysics/IMAPP, Radboud University Nijmegen, P.O. Box 9010, 6500 GL Nijmegen, The Netherlands\goodbreak
\and
Department of Physics \& Astronomy, University of British Columbia, 6224 Agricultural Road, Vancouver, British Columbia, Canada\goodbreak
\and
Department of Physics and Astronomy, Dana and David Dornsife College of Letter, Arts and Sciences, University of Southern California, Los Angeles, CA 90089, U.S.A.\goodbreak
\and
Department of Physics and Astronomy, University College London, London WC1E 6BT, U.K.\goodbreak
\and
Department of Physics, Gustaf H\"{a}llstr\"{o}min katu 2a, University of Helsinki, Helsinki, Finland\goodbreak
\and
Department of Physics, Princeton University, Princeton, New Jersey, U.S.A.\goodbreak
\and
Department of Physics, University of California, One Shields Avenue, Davis, California, U.S.A.\goodbreak
\and
Department of Physics, University of California, Santa Barbara, California, U.S.A.\goodbreak
\and
Department of Physics, University of Illinois at Urbana-Champaign, 1110 West Green Street, Urbana, Illinois, U.S.A.\goodbreak
\and
Dipartimento di Fisica e Astronomia G. Galilei, Universit\`{a} degli Studi di Padova, via Marzolo 8, 35131 Padova, Italy\goodbreak
\and
Dipartimento di Fisica e Scienze della Terra, Universit\`{a} di Ferrara, Via Saragat 1, 44122 Ferrara, Italy\goodbreak
\and
Dipartimento di Fisica, Universit\`{a} La Sapienza, P. le A. Moro 2, Roma, Italy\goodbreak
\and
Dipartimento di Fisica, Universit\`{a} degli Studi di Milano, Via Celoria, 16, Milano, Italy\goodbreak
\and
Dipartimento di Matematica, Universit\`{a} di Roma Tor Vergata, Via della Ricerca Scientifica, 1, Roma, Italy\goodbreak
\and
Discovery Center, Niels Bohr Institute, Blegdamsvej 17, Copenhagen, Denmark\goodbreak
\and
Discovery Center, Niels Bohr Institute, Copenhagen University, Blegdamsvej 17, Copenhagen, Denmark\goodbreak
\and
European Space Agency, ESAC, Planck Science Office, Camino bajo del Castillo, s/n, Urbanizaci\'{o}n Villafranca del Castillo, Villanueva de la Ca\~{n}ada, Madrid, Spain\goodbreak
\and
European Space Agency, ESTEC, Keplerlaan 1, 2201 AZ Noordwijk, The Netherlands\goodbreak
\and
Facolt\`{a} di Ingegneria, Universit\`{a} degli Studi e-Campus, Via Isimbardi 10, Novedrate (CO), 22060, Italy\goodbreak
\and
Gran Sasso Science Institute, INFN, viale F. Crispi 7, 67100 L'Aquila, Italy\goodbreak
\and
HGSFP and University of Heidelberg, Theoretical Physics Department, Philosophenweg 16, 69120, Heidelberg, Germany\goodbreak
\and
Helsinki Institute of Physics, Gustaf H\"{a}llstr\"{o}min katu 2, University of Helsinki, Helsinki, Finland\goodbreak
\and
INAF - Osservatorio Astronomico di Padova, Vicolo dell'Osservatorio 5, Padova, Italy\goodbreak
\and
INAF - Osservatorio Astronomico di Roma, via di Frascati 33, Monte Porzio Catone, Italy\goodbreak
\and
INAF - Osservatorio Astronomico di Trieste, Via G.B. Tiepolo 11, Trieste, Italy\goodbreak
\and
INAF/IASF Bologna, Via Gobetti 101, Bologna, Italy\goodbreak
\and
INAF/IASF Milano, Via E. Bassini 15, Milano, Italy\goodbreak
\and
INFN, Sezione di Bologna, Via Irnerio 46, I-40126, Bologna, Italy\goodbreak
\and
INFN, Sezione di Roma 1, Universit\`{a} di Roma Sapienza, Piazzale Aldo Moro 2, 00185, Roma, Italy\goodbreak
\and
INFN, Sezione di Roma 2, Universit\`{a} di Roma Tor Vergata, Via della Ricerca Scientifica, 1, Roma, Italy\goodbreak
\and
IUCAA, Post Bag 4, Ganeshkhind, Pune University Campus, Pune 411 007, India\goodbreak
\and
Imperial College London, Astrophysics group, Blackett Laboratory, Prince Consort Road, London, SW7 2AZ, U.K.\goodbreak
\and
Infrared Processing and Analysis Center, California Institute of Technology, Pasadena, CA 91125, U.S.A.\goodbreak
\and
Institut d'Astrophysique Spatiale, CNRS (UMR8617) Universit\'{e} Paris-Sud 11, B\^{a}timent 121, Orsay, France\goodbreak
\and
Institut d'Astrophysique de Paris, CNRS (UMR7095), 98 bis Boulevard Arago, F-75014, Paris, France\goodbreak
\and
Institute of Astronomy, University of Cambridge, Madingley Road, Cambridge CB3 0HA, U.K.\goodbreak
\and
Institute of Theoretical Astrophysics, University of Oslo, Blindern, Oslo, Norway\goodbreak
\and
Instituto de Astrof\'{\i}sica de Canarias, C/V\'{\i}a L\'{a}ctea s/n, La Laguna, Tenerife, Spain\goodbreak
\and
Instituto de F\'{\i}sica de Cantabria (CSIC-Universidad de Cantabria), Avda. de los Castros s/n, Santander, Spain\goodbreak
\and
Istituto Nazionale di Fisica Nucleare, Sezione di Padova, via Marzolo 8, I-35131 Padova, Italy\goodbreak
\and
Jet Propulsion Laboratory, California Institute of Technology, 4800 Oak Grove Drive, Pasadena, California, U.S.A.\goodbreak
\and
Jodrell Bank Centre for Astrophysics, Alan Turing Building, School of Physics and Astronomy, The University of Manchester, Oxford Road, Manchester, M13 9PL, U.K.\goodbreak
\and
Kavli Institute for Cosmological Physics, University of Chicago, Chicago, IL 60637, USA\goodbreak
\and
Kavli Institute for Cosmology Cambridge, Madingley Road, Cambridge, CB3 0HA, U.K.\goodbreak
\and
Kazan Federal University, 18 Kremlyovskaya St., Kazan, 420008, Russia\goodbreak
\and
LAL, Universit\'{e} Paris-Sud, CNRS/IN2P3, Orsay, France\goodbreak
\and
LERMA, CNRS, Observatoire de Paris, 61 Avenue de l'Observatoire, Paris, France\goodbreak
\and
Laboratoire AIM, IRFU/Service d'Astrophysique - CEA/DSM - CNRS - Universit\'{e} Paris Diderot, B\^{a}t. 709, CEA-Saclay, F-91191 Gif-sur-Yvette Cedex, France\goodbreak
\and
Laboratoire Traitement et Communication de l'Information, CNRS (UMR 5141) and T\'{e}l\'{e}com ParisTech, 46 rue Barrault F-75634 Paris Cedex 13, France\goodbreak
\and
Laboratoire de Physique Subatomique et Cosmologie, Universit\'{e} Grenoble-Alpes, CNRS/IN2P3, 53, rue des Martyrs, 38026 Grenoble Cedex, France\goodbreak
\and
Laboratoire de Physique Th\'{e}orique, Universit\'{e} Paris-Sud 11 \& CNRS, B\^{a}timent 210, 91405 Orsay, France\goodbreak
\and
Lawrence Berkeley National Laboratory, Berkeley, California, U.S.A.\goodbreak
\and
Lebedev Physical Institute of the Russian Academy of Sciences, Astro Space Centre, 84/32 Profsoyuznaya st., Moscow, GSP-7, 117997, Russia\goodbreak
\and
Max-Planck-Institut f\"{u}r Astrophysik, Karl-Schwarzschild-Str. 1, 85741 Garching, Germany\goodbreak
\and
National University of Ireland, Department of Experimental Physics, Maynooth, Co. Kildare, Ireland\goodbreak
\and
Nicolaus Copernicus Astronomical Center, Bartycka 18, 00-716 Warsaw, Poland\goodbreak
\and
Niels Bohr Institute, Blegdamsvej 17, Copenhagen, Denmark\goodbreak
\and
Niels Bohr Institute, Copenhagen University, Blegdamsvej 17, Copenhagen, Denmark\goodbreak
\and
Nordita (Nordic Institute for Theoretical Physics), Roslagstullsbacken 23, SE-106 91 Stockholm, Sweden\goodbreak
\and
Optical Science Laboratory, University College London, Gower Street, London, U.K.\goodbreak
\and
SISSA, Astrophysics Sector, via Bonomea 265, 34136, Trieste, Italy\goodbreak
\and
School of Physics and Astronomy, Cardiff University, Queens Buildings, The Parade, Cardiff, CF24 3AA, U.K.\goodbreak
\and
School of Physics and Astronomy, University of Nottingham, Nottingham NG7 2RD, U.K.\goodbreak
\and
Sorbonne Universit\'{e}-UPMC, UMR7095, Institut d'Astrophysique de Paris, 98 bis Boulevard Arago, F-75014, Paris, France\goodbreak
\and
Space Research Institute (IKI), Russian Academy of Sciences, Profsoyuznaya Str, 84/32, Moscow, 117997, Russia\goodbreak
\and
Space Sciences Laboratory, University of California, Berkeley, California, U.S.A.\goodbreak
\and
Special Astrophysical Observatory, Russian Academy of Sciences, Nizhnij Arkhyz, Zelenchukskiy region, Karachai-Cherkessian Republic, 369167, Russia\goodbreak
\and
Sub-Department of Astrophysics, University of Oxford, Keble Road, Oxford OX1 3RH, U.K.\goodbreak
\and
The Oskar Klein Centre for Cosmoparticle Physics, Department of Physics,Stockholm University, AlbaNova, SE-106 91 Stockholm, Sweden\goodbreak
\and
UPMC Univ Paris 06, UMR7095, 98 bis Boulevard Arago, F-75014, Paris, France\goodbreak
\and
Universit\'{e} de Toulouse, UPS-OMP, IRAP, F-31028 Toulouse cedex 4, France\goodbreak
\and
University of Granada, Departamento de F\'{\i}sica Te\'{o}rica y del Cosmos, Facultad de Ciencias, Granada, Spain\goodbreak
\and
University of Granada, Instituto Carlos I de F\'{\i}sica Te\'{o}rica y Computacional, Granada, Spain\goodbreak
% ...
}

\abstract{
  The secondary cosmic microwave background (CMB) $B$-modes stem
  from the post-decoupling distortion of the polarization $E$-modes
  due to the gravitational lensing effect of large-scale
  structures. These lensing-induced $B$-modes constitute 
both a valuable probe of the dark matter distribution and an important 
contaminant for the extraction of the primary CMB $B$-modes from inflation. 
\Planck\ provides accurate nearly all-sky measurements of both the 
polarization $E$-modes and the integrated mass distribution via the 
reconstruction of the CMB lensing potential.
By combining these two data products, we have produced an all-sky
template map of the lensing-induced $B$-modes using a real-space
algorithm that minimizes the impact of sky masks. The
cross-correlation of this template with an observed (primordial and
secondary) $B$-mode map can be used to measure the lensing $B$-mode
power spectrum at multipoles up to $2000$. In particular, when
cross-correlating with the $B$-mode contribution directly derived from
the \Planck\ polarization maps, we obtain lensing-induced $B$-mode
power spectrum measurement at a significance level of $12\,\sigma$, which
agrees with the theoretical expectation derived from the
\Planck\ best-fit $\Lambda$CDM model. This unique nearly all-sky
secondary $B$-mode template, which includes the lensing-induced 
information from intermediate to small ($10\la\ell\la1000$) angular
scales, is delivered as part of the Planck 2015 public data
release. It will be particularly useful for experiments searching for
primordial $B$-modes, such as BICEP2/Keck Array or LiteBIRD, since it
will enable an estimate to be made of the lensing-induced
contribution to the measured total CMB $B$-modes.}

\date{\today}

\titlerunning{A map of lensing-induced $B$-modes}
\authorrunning{Planck Collaboration}
\maketitle

%
\section{Introduction}
\label{sec:introduction} 

Cosmic microwave background (CMB)
polarization anisotropies can be decomposed into curl-free $E$-modes
and gradient-free $B$-modes. In contrast to primordial $E$-modes,
primordial $B$-modes are sourced only by tensor perturbations
\citep{1985SvA....29..607P, 1997PhRvL..79.2180S, 1997PhRvL..78.2058K,
1997PhRvL..78.2054S} that can be formed in the pre-decoupling Universe
due to an early inflationary phase \citep{1975JETP...40..409G,
1979PZETF..30..719S, 1982PhLB..117..175S}.  Thus, primordial $B$-modes
of the CMB polarization are a direct probe of cosmological inflation
\citep[see][for details on inflationary
theory]{1981PhRvD..23..347G,1982PhLB..108..389L}.  The measurement of
the primordial $B$-mode power spectrum, which peaks at degree angular
scales, is the main target of a plethora of ground-based experiments
and satellite proposals.  There was great excitement in early 2014
when $B$-modes at the relevant angular scales detected by the BICEP2
experiment were interpreted as evidence of inflationary gravitational
waves \citep{2014PhRvL.112x1101A}.  Investigating the polarized dust
emission in the BICEP2 observation field using the 353-GHz data,
\Planck\footnote{\Planck\ (\url{http://www.esa.int/Planck}) is a
project of the European Space Agency (ESA) with instruments provided
by two scientific consortia funded by ESA member states and led by
Principal Investigators from France and Italy, telescope reflectors
provided through a collaboration between ESA and a scientific
consortium led and funded by Denmark, and additional contributions
from NASA (USA).}  revealed a higher dust contamination level than
expected from pre-\Planck\ foreground models~\citep{planck2014-XXX}.
In \citet{pb2015}, a joint analysis of the BICEP2/Keck Array data at
100 and 150\,GHz and the full-mission \Planck\ data (particularly the
353-GHz polarized data) has been conducted. This provides the
state-of-the-art constraints on the tensor-to-scalar ratio, $r$, which
is currently consistent with no detection of a primordial $B$-mode
signal.  When combined with the limit derived from the temperature
data (as discussed in \citealt{planck2013-p11} and
\citealt{planck2014-a15}), the current $95\,\%$ upper limit is
$r<0.08$, which already rules out some of the simplest inflationary
potentials~\citep{planck2014-a24}. We stand at the threshold of a
particularly exciting epoch that is marked by several ongoing or
near-future ground-based experimental efforts, based on technology
that is sensitive and mature enough to probe the primordial $B$-modes
to theoretically interesting levels.

In addition to the primordial contribution, a secondary
contribution is expected from the post-decoupling distortion of the
CMB polarization due to the effect of gravitational lensing
\citep[see][for a review of CMB lensing]{2006PhR...429....1L}.  In
particular, the lensing of the primordial CMB polarization $E$-modes
leads to an additional $B$-mode contribution.  The secondary $B$-mode
contribution to the $C_\ell^{BB}$ power spectrum dominates over the
primary one at $\ell\ga150$, even for large values of the
tensor-to-scalar ratio, ($r\sim1$).  Thus, it must be corrected for, in
order to measure the imprint of primordial tensor modes. This
correction is generally referred to as `delensing'.

The secondary $B$-mode power spectrum can be estimated 
by cross-correlating the total observed $B$-mode map
with a template constructed by combining a tracer of the gravitational
potential and an estimate of the primordial $E$-modes. Using such a
cross-correlation approach, the SPTpol team
\citep{2013PhRvL.111n1301H} reported the first estimate of the lensing
$B$-mode power spectrum, consisting of a roughly $<8\,\sigma$
measurement in the multipole range $300<\ell<2750$ using
\emph{Herschel}-SPIRE data as the mass tracer. 
\resub{
The POLARBEAR collaboration detected the lensing $B$-modes using 
CMB polarization data by fitting an amplitude relative 
to the theoretical expectations to their CMB polarization 
trispectrum measurements, and reported a $4.2\,\sigma$ rejection 
of the null-hypothesis~\citep{2014PhRvL.113b1301A}. 
Similarly, the ACTPol team reported $3.2\,\sigma$ evidence
of the lensing $B$-mode signal 
within its first season data, using the correlation of 
the lensing potential estimate and the cosmic infrared
background fluctuations measured by \Planck\ \citep{ACT2015}.
} 
Finally, using the full mission temperature and polarization data, 
\Planck\ obtained a template-based cross-correlation measurement of the
lensing $B$-mode power spectrum that covers the multipole range
$100<\ell<2000$, at a significance level of approximately
$10\,\sigma$, as described in \cite{planck2014-a17}.

\newresub{
Secondary $B$-modes dominate any potential primordial $B$-modes at high
multipoles, thus the high-$\ell$ $BB$ power spectrum of the observed
polarization maps can also be used to make a lensing-induced $B$-mode
measurement.
}
The POLARBEAR collaboration reported the first $BB$ 
measurement (at around $2\,\sigma$) of the $B$-mode power spectrum 
in the multipole range $500 < \ell < 2100$ \citep{2014ApJ...794..171T}.  
The SPTpol experiment also made  
a $BB$ estimate of the lensing $B$-modes in the
range $300<\ell<2300$, representing a $>4\,\sigma$ detection
\cite{2015ApJ...807..151K}.  Moreover, a non-zero lensing $B$-mode
signal has been found in the BICEP2/Keck Array data with around
$7\,\sigma$ significance, by fitting a freely-floating CMB lensing
amplitude in the joint analysis with \Planck\ data~\citep{pb2015}.

\newresub{For current or future experiments targeting the detection of
primordial $B$-modes, a precise estimation of the secondary CMB
$B$-modes at large and intermediate angular scales is required in
order to separate the secondary contributions from potential
primordial $B$-modes.}
\resub{
On the one hand, large angular scale experiments lack the high-resolution 
$E$-mode measurements that are required to measure the lensing-induced  
$B$-mode signal. On the other hand, for high-resolution experiments, 
partial sky coverage limits
their ability to extract the $B$-mode power spectrum and to
reconstruct the lensing potential at large angular scales. 
}
Thus, such experiments would benefit from a
pre-estimated secondary $B$-mode template, covering angular scales
from a few degrees down to sub-degree scales and matching their sky
coverage.

We present an all-sky secondary
$B$-mode template spanning from intermediate to large angular scales,
synthesized using the full mission \Planck\ data.  In
\cite{planck2014-a17}, the lensing $B$-mode estimate was band-limited
to $\ell>100$, in order to conservatively alleviate any low-$\ell$
systematic effects.  In contrast, here the focus is on improving the
reliability at intermediate angular scales ($10<\ell<200$).  We also
extend the lensing $B$-mode results of \cite{planck2014-a17} by
producing a lensing $B$-mode map, by performing extensive
characterization and robustness tests of this template map and by
discussing its utility for $B$-mode oriented experiments. This
$B$-mode map is delivered as part of the \Planck\ 2015 data
release.

The outline of this paper is as follows.
Section~\ref{Sect:datasims} describes the data and simulations that we
use.  We detail the methodology for the template synthesis in
Sect.~\ref{Sect:methodo}, \resub{and describe 
the construction of the mask in Sect.~\ref{Sect:masks}}. 
The lensing $B$-mode template reconstruction method is validated 
using simulations in Sect.~\ref{Sect:simu}. We
present the template we have obtained from \Planck\ foreground-cleaned data
in Sect.~\ref{Sect:results}, and assess its robustness against
foreground contamination and the choice of the data to cross-correlate
with in Sect.~\ref{Sect:tests}.  Section~\ref{Sect:keytool} addresses
the implications of the template for external experiments targeting
primordial $B$-mode searches.  We summarize and conclude in
Sect.~\ref{conclusions}.

%
\section{Data and simulations}
\label{Sect:datasims}

\noindent {\fontsize{10.5}{10.5}\selectfont {\sf \Planck\ sky maps: }} 
We have used foreground-cleaned CMB temperature and polarization maps derived from the \Planck\ satellite
full mission frequency channel maps from 30 to 857\,GHz in temperature
and 30 to 353\,GHz in polarization
\citep{planck2014-a01,planck2014-a03,planck2014-a04,planck2014-a05,
planck2014-a06,planck2014-a07,planck2014-a08,planck2014-a09}.  Our
main results are based on Stokes $I$, $Q$, and $U$ maps constructed
using the \smica\ component-separation algorithm
\citep{2003MNRAS.346.1089D} in temperature and polarization
simultaneously \citep{planck2014-a11}. The maps are at $5\arcm$
resolution in $N_{\mathrm{side}}=2048$ {\tt HEALPix} pixelization
\citep{gorski2005}.\footnote{\url{http://healpix.jpl.nasa.gov}}. For
the sake of assessing the robustness of our results, we have also utilized
foreground-cleaned maps that are produced using the other \Planck\
component-separation methods, namely \commander, \nilc, and \sevem\
\citep{planck2013-p06,planck2014-a11,planck2014-a12}. 
\resub{
The current publicly available \Planck\ HFI polarization maps, which 
are part of the \Planck\ 2015 data release,
are high-pass filtered at $\ell \la 30$ because of residual 
systematic effects on angular scales greater 
than $10\deg$~\citep{planck2014-a08, planck2014-a09}. 
However, we have used polarization maps covering all angular scale for our analysis,
since the results have proved not to be sensitive to CMB $E$-mode polarization
at large angular scales.\footnote{The $E$-mode polarization at $\ell<30$
contributes only at the sub-percent level to the template-based 
lensing $B$-mode power spectrum at $\ell=10$, and has even lower contribution 
at higher multipoles.}}

\vspace{0.5cm}
\noindent {\fontsize{10.5}{10.5}\selectfont {\sf Full Focal Plane simulations: }} 
For methodological validation and for the bias correction of the lensing
potential at the map level (known as the `mean-field' correction),
we have relied on the eighth \Planck\ Full Focal Plane (FFP8
hereafter) suite of \newresub{simulations}, as described in
\cite{planck2014-a14}.
\resub{Specifically, we have used FFP8 Monte-Carlo realizations of the
  Stokes $I$, $Q$, and $U$ outputs of the \smica\ component-separation
  method. These have been obtained by processing through the
  \smica\ algorithm the simulated
beam-convolved CMB and noise realizations of the nine \Planck\ frequency channels, as 
described in \cite{planck2014-a11}. As a result, both the noise
realizations and the beam transfer
function are representative of those of the \smica\ foreground-cleaned maps. 
Finally, the calibration of the \Planck\ 2015 data has been taken into
account by rescaling the CMB realizations as in \citet{planck2014-a17}.}

\vspace{0.5cm}
\resub{
\noindent {\fontsize{10.5}{10.5}\selectfont {\sf Fiducial cosmology: }} 
For normalizing the lensing potential map estimate and computing the filter 
and transfer functions of the $B$-mode template, we have used fiducial power spectra 
derived from the 2015 \Planck\ base $\Lambda$CDM cosmological parameters that 
have been determined from the combination of the 2015 temperature and
`lowP' likelihoods, as described in \citet{planck2014-a15}.  
}

%
\section{$B$-mode map reconstruction}
\label{Sect:methodo}

\vspace{1mm}
\subsection{Formalism} 
The secondary $B$-modes of CMB polarization arise from a leakage of a fraction of the $E$-modes into
$B$-modes due to the polarization remapping induced by the CMB lensing
effect. In terms of the polarization spin-two components $_{\pm 2} P
\equiv Q \pm iU$, and at first order in the lensing potential $\phi$,
the lensing-induced contribution reads 
\be 
_{\pm 2}P^{\rm{lens}}(\hatn) = \nabla _{\pm 2} P^{\rm{prim}}(\hatn) \cdot \nabla \phi(\hatn), 
\ee
where $_{\pm 2} P^{\rm{prim}}(\hatn)$ are the
polarization fields that would be observed in the absence of the
lensing effect \citep{1998PhRvD..58b3003Z, 2006PhR...429....1L}.
Rewriting the secondary polarization fields in terms of the
rotationally invariant $E$- and $B$-mode fields, so that
\be _{\pm 2} P^{\rm{lens}}(\hatn) = \nabla \left[ \sum_{\ell' m'}
\left( E^{\rm{prim}}_{\ell' m'} \pm i B^{\rm{prim}}_{\ell' m'} \right)
{}_{\pm 2}Y_{\ell 'm'} \right] \cdot \nabla \phi(\hatn),
\label{Eq:Pntheo} 
\ee
and considering their spin-two spherical-harmonic coefficients
\be _{\pm 2} P^{\rm{lens}}_{\ell m} = \int d\hatn {\, }_{\pm 2}
P^{\rm{lens}}(\hatn) {\, }_{\pm 2}Y_{\ell m}^*(\hatn),
\label{Eq:Ptheo} 
\ee
one finds that the gradient-free $B$-mode polarization receives a
secondary contribution that depends on the primordial $B$-modes and
the unlensed curl-free $E$-modes, of the form
\be B^{\rm lens}_{\ell m} = \frac{1}{2i} \left( _{+2}
P^{\rm{lens}}_{\ell m} - _{-2} P^{\rm{lens}}_{\ell m} \right).
\label{Eq:Btheo} 
\ee

\vspace{1mm}
\subsection{Algorithm}
\label{methodo:algorithm} 
First, we state the assumptions on which our
algorithm is based.  Since the $E$-mode amplitude is at least an order
of magnitude greater than the primordial $B$-mode amplitude, the
$B^{\rm lens}_{\ell m}$ contribution that comes from the $E$-mode
remapping largely dominates over the one from the lensing perturbation
of the primordial $B$-modes. From now on, the latter can safely be
neglected, consistent with the assumptions in \cite{planck2014-a17}.
We replace the primordial $E$-modes that appear in Eq.~(\ref{Eq:Pntheo})
with the total $E$-modes, $E=E^{\rm prim}+\delta E$. This amounts to
neglecting the second-order contribution to $B^{\rm{lens}}$ due to
$\delta E$, that is to say the lensing perturbation of the $E$-modes
themselves.

We then consider pure $E$ polarization fields:
\be _{\pm 2}P^{E}(\hatn) \equiv Q^{E}(\hatn) + i U^{E}(\hatn) \equiv
\sum_{\ell m} E_{\ell m} {\, }_{\pm 2}Y_{\ell m}(\hatn),
\label{Eq:PE} 
\ee
which define pure-$E$ Stokes parameters $Q^E$, $U^E$.

Implementing the above assumptions into Eq.~(\ref{Eq:Ptheo}) and using
the definition given in Eq.~(\ref{Eq:PE}), we build a secondary
polarization estimator that has the generic form
\be _{\pm 2} \hat P^{\rm{lens}}_{\ell m} = \mathcal{B}_{\ell}^{-1}
\int d\hatn \, \, \nabla _{\pm 2} \widetilde{P^{E}}(\hatn) \cdot
\nabla \widetilde{\phi}(\hatn) \, {\, }_{\pm 2}Y_{\ell m}^*(\hatn) ,
\label{Eq:Pesti} 
\ee
where $_{\pm 2} \widetilde{P^{E}}$ and $\widetilde{\phi}$ are the
filtered versions of the pure-$E$ polarization and lensing potential
fields, respectively, whereas $\mathcal{B}_{\ell}$ is a transfer
function ensuring that the estimator is unbiased. These quantities are
defined below in Sect.~\ref{methodo:filter}. We finally define secondary
CMB $B$-mode template Stokes maps $(Q^{\rm{lens}}(\hatn),
U^{\rm{lens}}(\hatn))$ by preserving the $B$-mode contribution in
Eq.~(\ref{Eq:Pesti}) and transforming back to real space.

In summary, we reconstruct the all-sky $B$-mode template using a
dedicated pipeline that consists of:
\begin{itemize}
\item[(i)] estimation of the deflection field using the filtered
reconstructed gravitational potential, $\nabla
\widetilde{\phi}(\hatn)$;
\item[(ii)] computation of the gradient of the filtered pure $E$-mode
input maps, $\nabla _{\pm 2} \widetilde{P^{E}}(\hatn)$;
\item[(iii)] calculation of the analytical transfer function;
\item[(iv)] construction of the polarization template using Eq.~(\ref{Eq:Pesti});
\item[(v)] formation of a secondary $B$-mode template using
Eq.~(\ref{Eq:Btheo}).
\end{itemize} These steps are further detailed in the rest of
this section.

\subsection{All-sky lensing potential reconstruction}
\label{methodo:lensing}
\resub{The paths of CMB photons are weakly deflected by the matter encountered
along the way from the last-scattering surface. As a result, the primary CMB observables 
are remapped according to the gradient of the gravitational potential $\phi$
integrated along the line-of-sight \citep{1987A&A...184....1B}.
This induces higher-order correlations within the CMB observables;
namely, a non-vanishing connected part of the four-point
correlation function, or equivalently in the spherical harmonic domain, a 
trispectrum, in the CMB maps. These can be used, 
in turn, to reconstruct the intervening mass 
distribution~\citep{1997A&A...324...15B, 1999PhRvD..59l3507Z}.} 
Specifically, to first order in $\phi$, the lensing induces a correlation between
the observed (lensed) CMB maps and the gradient of the primary 
(unlensed) maps. Building upon this property, quadratic estimators 
have been proposed to extract a lensing potential estimate from the observed
map~\citep{2001ApJ...557L..79H, 2001PhRvD..64h3005H, 2002ApJ...574..566H, 2003PhRvD..67h3002O}. 
We have reconstructed the lensing potential over a large portion of the sky
using the all-sky quadratic estimator described in \cite{2003PhRvD..67h3002O}, which has been modified to deal
with cut skies.

\subsubsection{Inpainting of the temperature map}
\label{methodo:lensing:inpaint} 

\begin{figure}[!h]
  \begin{center}
    \includegraphics[width=\columnwidth]{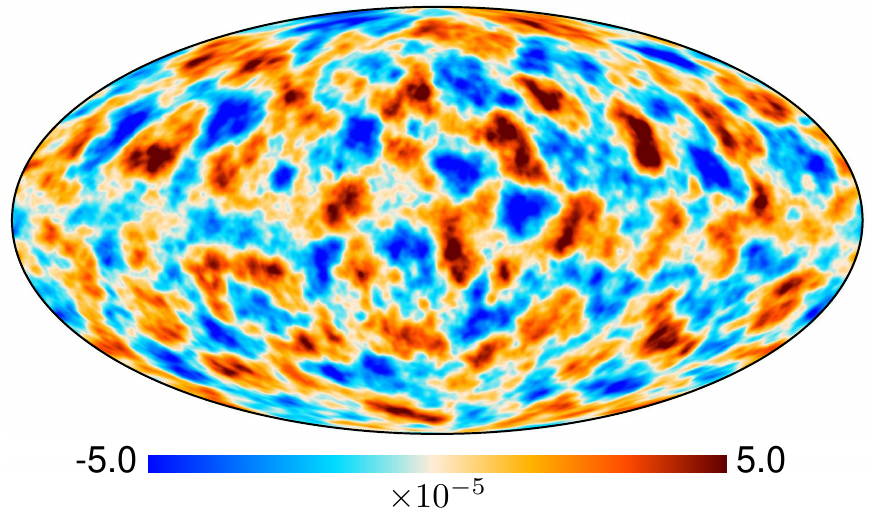}
    \caption{Wiener-filtered lensing potential estimated from the
\smica\ foreground-cleaned temperature map using the $f_{\rm{sky}}
\simeq 80\,\%$ lensing mask. The lensing potential estimate, which is shown 
in Galactic coordinates, is effectively inpainted using a 
lensing extraction method that relies on an inpainting of 
the input temperature map, as discussed in Sect.~\ref{methodo:lensing:inpaint}.}
    \label{Fig:phi}
  \end{center}
\end{figure}

In \cite{planck2014-a17}, foreground-contaminated
regions have been masked out at the stage of the inverse-variance filtering
of the input CMB maps, by allocating infinite variance to masked
pixels. The reconstructed $\phi$ is thus null-valued in the pixels
inside the analysis mask. For the sake of synthesizing Stokes maps
$(Q^{\rm{lens}}(\hatn), U^{\rm{lens}}(\hatn))$, as described in
Sect.~\ref{methodo:algorithm}, using such a masked $\phi$ estimate
would induce prohibitive amounts of mode mixing. Thus, we have used the {\tt
METIS} method, in which the masked CMB maps are restored to a complete
sky coverage, before their ingestion into the quadratic estimator, by
means of an inpainting procedure based on the `sparsity'
concept~\citep{inpainting:abrial06}.  More details are given in
\citet{2010AA...519A...4P}, where this method has been first described and
in \citet{planck2013-p12}, where it has been used to perform consistency
tests.  As a result of this procedure, the {\tt
METIS} method provides a $\phi$ estimate that is effectively inpainted to
cover the full sky. To illustrate this property, Fig.~\ref{Fig:phi}
shows a Wiener-filtered version of the $\phi$ estimate reconstructed
from the \smica\ temperature map using the baseline `L80'
lensing mask that we describe below in Sect.~\ref{Sect:masks}. 

As well as offering the advantage of mitigating the bias
that the mask induces in the $\phi$ map, which is discussed in
Sect.~\ref{methodo:lensing:mf}, this allows us to construct the secondary polarization template map 
using Eq.~(\ref{Eq:Pesti}), alleviating the
need for further processing steps to deal with partial sky
coverage. 

\subsubsection{Mean-field debiasing}
\label{methodo:lensing:mf} 

\newresub{
The quadratic estimator is based on the fact that, 
for a fixed lens distribution, the lensing breaks the statistical
isotropy of the CMB maps; specifically, it introduces off-diagonal terms of
the CMB covariance~\citep{2002ApJ...574..566H}. Therefore this is also 
sensitive to any other source of statistical anisotropies in the maps.} 
For \Planck\ data, the bias induced at the 
$\phi$ map level by any known sources of statistical anisotropies, which is referred 
to as the `mean-field bias', is dominated by the effects of masking
\citep{planck2013-p12}. The mean-field bias can be estimated from
Monte-Carlo (MC) simulations that include all the instrumental and
observational effects that can lead to a sizeable mean-field (e.g. the mask,
the spatial inhomogeneity in the noise, which yields 
the first sub-dominant mean field, and the beam asymmetry), 
by averaging the $\phi$ estimates obtained on MC realizations.  
We have used a set of 100 FFP8 realizations to obtain an estimate 
$\langle\bar{\phi}_{LM}\rangle_{\rm MC}$ of the mean-field 
modes\footnote{We use multipole indices $LM$ for the lensing potential 
in order to differentiate them from the $\ell m$ indices used for the CMB fields} 
for the non-normalized $\phi$ modes, labelled $\bar{\phi}_{LM}$. 
This mean-field estimate has then been subtracted 
from $\bar{\phi}$ to obtain the unbiased estimate $\hat \phi_{LM}$ of $\phi$ 
in the spherical harmonic domain, given by 
\be 
\hat \phi_{LM} = \mathcal{A}_L \left(
\bar{\phi}_{LM} - \langle\bar{\phi}_{LM}\rangle_{\rm MC} \right), 
\ee
\resub{where $\mathcal{A}_L$ is the normalization function, which ensures
that the estimator is unbiased. This is related to the normalization
$A_L^{\alpha}$ given in \citet{2003PhRvD..67h3002O},\footnote{It is also
related to the response function of the quadratic estimator for off-diagonal
terms of the CMB covariance $\mathcal{R}_L$, as defined in equation~A.16 of
\citet{planck2014-a17}, via $\mathcal{A}_L=\mathcal{R}_L^{-1}$.}
 via $\mathcal{A}_L=[L(L+1)]^{-1}A_L^{\alpha}$, 
and has been analytically calculated using the fiducial cosmology described 
in Sect.~\ref{Sect:datasims}.} 
\resub{To handle the slight difference between FFP8 input 
cosmology and the fiducial cosmology considered here, the mean-field 
 $\langle\bar{\phi}_{LM}\rangle_{\rm MC}$ has been multiplied by the ratio
 of the normalization functions derived from the input FFP8 and
fiducial cosmological models.}

\subsubsection{Mass tracer choice}

In addition to the lensing potential quadratic estimate, other tracers
of the underlying lensing potential could be
used to form a lensing-induced $B$-mode template. \newresub{The cosmic
infrared background (CIB) emission is of particular interest for
this purpose since, in contrast to most of
the large-scale structure tracers, which probe only a limited range of
redshifts; its redshift distribution has a broad overlap with the
lensing potential kernel~\citep{Song2003}.}
Using the best-fit halo-based model of \citet{planck2011-6.6},
\citet{planck2013-p13} have reported a roughly $80\,\%$ correlation
of the CIB fluctuations measured by \Planck\ with the lensing
potential, in agreement with the model expectations.
\resub{For the sake of the lensing $B$-mode
measurement, however, the uncertainties in the CIB modelling will have
a large impact on the signal estimate, unless the CIB model parameters
are marginalized within a joint analysis including lensing cross- and
auto-power spectrum measurements of the CIB~\citep{Sherwin2015}.}  
Moreover, the foreground residuals are another concern for any lensing
$B$-mode template synthesized from the CIB, as also discussed in
\citet{Sherwin2015}. The CIB signal is the most precisely measured at high
frequencies, where the Galactic dust emission is also important. Any
Galactic dust residuals in the CIB template map could be correlated
either with the polarized dust residuals or with the
intensity-to-polarization leakage of the dust emission in the CMB $E$-
and $B$-mode maps.

\resub{Similarly, lensing potential estimates can also be extracted by
means of a quadratic estimator on the \Planck\ polarization maps. This,
combined with the temperature-based $\phi$ estimate, reduces the power 
spectrum of the $\phi$ reconstruction noise by roughly $25\,\%$. 
However, for measuring the lensing $B$-mode power spectrum using a 
template-based approach,
resorting to a polarization-based $\phi$ estimate would require us to 
correct for a non-negligible Gaussian 
bias~\citep{planck2014-a17, Namikawa2014}.}

\resub{Here we have chosen to employ only the lensing potential
estimate to produce a lensing $B$-mode template that is
model-independent and more robust to foreground residuals; and to use
the temperature data only for estimating the lensing potential, which
ensures desirable properties for our
template, as discussed in Sect.~\ref{Sect:simu}. Relying on the independent
analysis presented in \citet{planck2014-a17}, we show in
Sect.~\ref{Sect:tests:masstrac} that these choices induce at worst a marginal
increase in the statistical uncertainties of the lensing $B$-mode
measurements.}

\subsection{Lensing-induced polarization fields}
\label{methodo:filter} 
Using the lensing potential estimate discussed
above and the observed $E$-mode map, we have reconstructed the template of
the secondary polarization field as in Eq.~(\ref{Eq:Pesti}). To reduce
uncertainties, we have used filtered versions of those maps that are defined
as 
\be 
\widetilde{\phi}(\hatn) = \int d\hatn \, \, f_L^{\phi} \,
\hat{\phi}_{LM} \, Y_{LM}^*(\hatn), 
\ee 
and 
\be
 _{\pm 2}
\widetilde{P^{E}}(\hatn) = \int d\hatn \,\, f_\ell^{E}\, E_{\ell m}\,
{\, }_{\pm 2}Y_{\ell m}^*(\hatn).  
\ee

The filter functions for $\phi$ and $E$, $f_L^{\phi}$ and
$f_\ell^{E}$, are Wiener filters, based on the optimality arguments
developed in \citet{cmbpol}. \resub{In \citet{planck2014-a17}, 
the $\phi$ estimates at $L<8$ have been discarded because of instability to the 
choice of the method used to correct the mean-field. Consistently, 
we have filtered the $\phi$ estimate to $L \leq 10$ in order to 
conservatively avoid any mean-field related issues.}
\newresub{We have used a tapering filter, labelled $f_{10}$, that smoothly goes
  from zero at $L=5$ to unity at $L=15$ by means of a power-of-cosine function centred
  in $L=10$.}
\resub{Using this filter, we still preserve $99\,\%$ of the available 
information for measuring the cross-correlation $B$-mode power spectrum.}
The filtered pure-$E$ polarization fields $_{\pm 2} \widetilde{P^{E}}$
have been directly obtained from the full-sky
observed $Q$ and $U$ maps by transforming into the
spin-weighted spherical harmonic basis, filtering the $E$-modes, and
reforming Stokes parameter fields with a null $B$-mode component. 
\resub{No deconvolution of the beam has been performed at this stage, which 
yields further filtering of the $E$-modes at high multipoles. 
Our filtering functions are defined by
\begin{align} 
f_L^{\phi} &=  \frac{f_{10}C_L^{\phi, \rm{fid}}}{C_L^{\phi, \rm{fid}} + N_L^{\phi}}, \nonumber \\ 
f_\ell^{E} &=  \frac{C_\ell^{E, \rm{fid}}}{C_\ell^{E, \rm{fid}} + N_L^{E}},
\label{Eq:filters}
\end{align}
where $C_L^{\phi}$ and $C_\ell^{E}$ are the fiducial $\phi$ and $E$-mode 
power spectra; $N_L^{\phi}$ is the $\phi$ reconstruction noise 
power spectrum calculated following \citet{2003PhRvD..67h3002O}; 
and $N_L^{E}$ is the pixel- and beam-deconvolved power spectrum 
of the $E$-mode noise.}

These steps have been performed using the fast spin-weighted spherical
harmonic transform capability of the {\tt HEALPix} library to generate
$Q^E$ and $U^E$ and compute their derivatives. Thanks to this simple
implementation, a $B$-mode template map at a resolution of $5\arcm$
can be reconstructed from the $\phi$ estimate and the observed $Q$ and
$U$ maps in a reasonable amount of computing time, which enables the
use of Monte-Carlo simulations. For example, the computing time is
about two minutes using eight cores of the Linux AMD64 machines of the
Institut National de Physique Nucl{\'e}aire et de Physique des
Particules (IN2P3) Computing Center\footnote{\url{http://cc.in2p3.fr/?lang=en}}.

Using a harmonic approach, as in \citet{2000PhRvD..62d3007H}, we
compute the transfer function that appears in Eq.~(\ref{Eq:Pesti}) by
imposing the condition that the secondary $B$-mode template
$\hat{B}^{\rm{lens}}~\equiv~({\,}_{+2}\hat{P}^{\rm{lens}}
-{\,}_{-2}\hat{P}^{\rm{lens}})/(2i)$ satisfies 
\be
 \langle B_{\ell
m}^* \hat B_{\ell' m'}^{\rm{lens}}\rangle = \delta_{\ell \ell'}
\delta_{mm'} \, C_\ell^{B, \rm{fid}}, 
\ee 
where $C_\ell^{B, \rm{fid}}$ is the fiducial lensing-induced $B$-mode power spectrum ($r=0$).

In terms of the fiducial power spectra $C_\ell^{X, \rm{fid}}$, $X =
\{E, B, \phi \}$, we obtain a transfer function 
\be 
\mathcal{B}_\ell =
\frac{1}{C_\ell^{B, \rm{fid}}} \sum_{L\ell'} f_L^{\phi} C_L^{\phi,
\rm{fid}} \, f_{\ell'}^{E} B_{\ell'} C_{\ell'}^{E, \rm{fid}} {\,
}_2F_{\ell L \ell'},
\label{Eq:beff} 
\ee 
where $B_{\ell'}$ is the beam function of the
polarization maps, and ${\, }_2F_{\ell L \ell'}$ is a geometrical term
defined in \citet{2000PhRvD..62d3007H}.

\subsection{$B$-mode template synthesis}
\label{methodo:results} 
Our secondary $B$-mode template is obtained as
in Eq.~(\ref{Eq:Btheo}) that is
\be \hat{B}_{\ell m}^{\rm{lens}} = \frac{1}{2i} \left( _{+2}
\hat{P}^{\rm{lens}}_{\ell m} - _{-2} \hat{P}^{\rm{lens}}_{\ell m}
\right).
\label{Eq:Besti} \ee
This is computed from $_{\pm 2}\hat{P}_{\ell m}^{\rm{lens}}$, the
spin-weighted spherical harmonic transforms of our real-space
secondary polarization estimates $\left( Q^{\rm{lens}} \pm i
U^{\rm{lens}}\right)$ that are corrected for the transfer function
given in Eq.~(\ref{Eq:beff}).

\subsection{Cross-correlation power spectrum of the template}

\resub{We form the cross-correlation power spectrum between 
the template $B$-modes given in Eq.~(\ref{Eq:Besti}) and 
the observed $B$-modes, $B^{\rm obs}_{\ell m}$, using}
\be 
\hat{C}_\ell^{BB^{\rm{lens}}} = \frac{1}{f_{\rm{sky}}^{\rm{eff}}(2
\ell +1)} \sum_{m=-\ell}^{\ell} B_{\ell m}^{\rm{obs}\, *} \tilde B_{\ell m}^{\rm{lens}}, 
\label{Eq:clb}
\ee
\resub{where the asterisk denotes complex conjugation; and 
$\tilde{B}_{\ell m}^{\rm{lens}}$ is a shorthand notation 
for the template $B$-mode harmonic coefficients obtained 
as in Eq.~(\ref{Eq:beff}), but using an apodized and masked version 
of the real-space secondary polarization estimates 
$\left( Q^{\rm{lens}} \pm i U^{\rm{lens}}\right)$. This apodized mask, the 
construction of which is described in Sect.~\ref{Sect:masks}, leaves an 
effective available sky fraction $f_{\rm{sky}}^{\rm{eff}}$ for analysis.
The $BB^{\rm{lens}}$ cross-correlation power spectrum represents an estimate of the 
lensing $B$-mode power spectrum. Moreover, this does
not require any noise term subtraction, as we verify in Sect.~\ref{Sect:simu}. }

\resub{The $BB^{\rm{lens}}$ power spectrum variance is estimated 
to a good approximation using}
\be
\sigma^2\left( \hat{C}_\ell^{BB^{\rm{lens}}}\right) \, = \,
\frac{1}{f_{\rm{sky}}^{\rm{eff}}(2\ell+1)} \left[
(C_\ell^{B})^2 + (\hat{C}_\ell^{B^{\rm{obs}}})(\hat{C}_\ell^{B^{\rm{lens}}}) \right],
\label{Eq:gaussvar} 
\ee 
\resub{where $\hat{C}_\ell^{B^{\rm{obs}}}$ is the 
auto-correlation power spectrum of the observed $B$-modes and
$\hat{C}_\ell^{B^{\rm{lens}}}$ the one of the template.
Equation~(\ref{Eq:gaussvar}) consists of a Gaussian variance prescription 
\citep[see e.g.][]{1995PhRvD..52.4307K}, which is not expected to 
rigorously apply to the lensing $B$-mode power spectrum estimate, 
because of its `sub-structure' (since it is based on the sum of the $TTEB$
trispectrum of the observed CMB signal). However, for a template map 
constructed on the \Planck\ polarization maps, given the polarization 
noise level, the higher-order terms that enter the variance are sub-dominant,
as is further discussed and tested in Sect.~\ref{Sect:simu}. }

%
%
\section{Construction of the $B$-mode mask }
\label{Sect:masks}
\resub{In this section, we detail the methodology for constructing 
the template mask. This has involved first preparing a series of 
foreground masks targeted at specific foreground emission in 
temperature. These foreground masks have been then combined to construct
an analysis mask for the lensing potential reconstruction. Finally,
the lensing analysis mask has been modified to define a mask for the 
template map. These three steps are detailed in 
Sects.~\ref{Sect:masks:foregrounds}--\ref{Sect:masks:template} 
below.}

\subsection{Foreground masks}
\label{Sect:masks:foregrounds}

\subsubsection*{Galactic mask} 
We have masked the regions of the sky that are strongly contaminated by the
diffuse Galactic emission, the carbon-monoxide (CO) transition line
emission and the extended nearby galaxy emission. 
\resub{This Galactic mask has been produced following the method described in
\cite{planck2013-p11}. It includes the diffuse Galactic mask described
in \citet{planck2013-p06}, which is produced by thresholding a
combination of CMB-corrected temperature maps at 30 and 353\,GHz
until a desired sky fraction is preserved. For our baseline
analysis, we have used a diffuse Galactic mask that retains about $80\,\%$ of
the sky. This diffuse Galactic mask discards
mainly low latitude regions. To mask the CO lines
contamination at intermediate latitudes, we have used the CO mask described
in Appendix A of \citet{planck2014-a13}, which has been obtained by
thresholding a smoothed version of the Type 3 CO map
\citep{planck2013-p03a} at $1\, \rm{K_{RJ}}\cdot\rm{km}\cdot s^{-1}$.
Furthermore, we have removed the emission from the most extended nearby
galaxies, including the two Magellanic clouds (LMC, SMC) and M31, by
cutting a radius that covers each galaxy in the 857-GHz map,
as described in \citet{planck2014-a13}. Our Galactic mask, which is
the merge of the diffuse Galactic, CO-line, and extended nearby galaxy
masks, consists of large cuts that extend over more than two degrees
on the sky, and preserves $79\,\%$ of the sky.}

\subsubsection*{Extragalactic object masks}
We have masked the infrared (IR) and radio point sources that have been detected 
in the temperature maps in frequency channels from 70 to 353\,GHz 
using the \Planck\ compact object catalogues, as well as the sky areas 
contaminated by the Sunyaev-Zeldovich (SZ) emission using both \Planck\ 
SZ cluster catalogues and Compton parameter maps. 
\resub{We started with the 
individual masks targeted at 100, 143, and 217\,GHz that have been used 
in \citet{planck2013-p12}. These masks have been produced using the \Planck\ Early Release 
Compact Source Catalogue (ERSC; \citet{planck2011-1.10}), the \Planck\ Catalogue 
of Compact Sources (PCCS; \citet{planck2013-p05}), and the \Planck\ catalogue 
of Sunyaev-Zeldovich sources (PSZ; \citet{planck2013-p05a}). 
These have been merged with the conservative point-source masks produced 
using the 2015 catalogue (PCCS2; \citet{planck2014-a35}), which are presented 
in \citet{planck2014-a13}. 
The SZ emission hav been further removed using the template 
Compton parameter y map for the detected galaxy clusters of the 2015 SZ catalogue 
(PSZ2; \citet{planck2014-a36}) that is described in \citet{planck2014-a28}. 
Individual SZ masks have been constructed for 100, 143, and 217\,GHz separately, 
by converting the template y-map into CMB temperature using the corresponding 
conversion factors listed in \citet{planck2014-a28}, and thresholding at $10\, \mu\rm{K}$.}
\resub{Finally, we have masked several small nearby galaxies including M33, M81, M82, M101, and CenA, 
as described in \citet{planck2014-a13}.} 

\resub{We have combined these individual masks to produce two extragalactic object masks that differ 
in the maximum cut radius. First, the `extended object mask' is made of holes
 the largest angular size of which ranges from two degrees to $30\arcmin$. 
This includes the extended SZ clusters and the small nearby galaxies. Second, the 
`compact object mask' is a collection of small holes of cut radius smaller than 
$30\arcmin$, and comprises the detected radio and IR point sources and the 
point-like SZ clusters. The latter represents a conservative point source mask 
that includes all the point sources detected above $S/N=5$ in the 100-, 
143-, and 217-GHz maps, as well as the point sources detected above 
$S/N=10$ in the adjacent frequencies (including 70 and 353\,GHz). 
This has been used in Sect.~\ref{Sect:tests:mask} to test the robustness of our 
results againts point source residuals in the \smica\ map.}

\subsection{Lensing analysis mask}
\label{Sect:masks:lensing}
For the lensing potential reconstruction, we have prepared a composite mask
targeted at the \smica\ temperature map. 
\resub{This map comes along with a `confidence' mask that defines the sky areas 
where the CMB solution is trusted. The \smica\ confidence mask, a thorough 
description of which is given in \citet{planck2014-a11},  
has been produced by thresholding the CMB power map that is obtained in squaring a 
band-pass filtered and smoothed version of the \smica\ map. 
Our lensing mask combines the Galactic, extended, and
compact object masks after some changes driven by the \smica\ confidence mask. 
Namely, the LMC and three molecular clouds at medium latitudes have been masked more 
conservatively in the \smica\ mask than in our initial Galactic mask. 
We have enlarged the latter accordingly. In the compact object mask, IR and radio 
source holes that are not also present in the \smica\ confidence mask
have been discarded. 
They mainly corresponds to point sources detected in the 353-GHz map, 
and are strongly weighted down in the \smica\ CMB map. 
The modified compact object mask removes $0.6\,\%$ of the sky in addition to the 
$21.4\,\%$ Galactic and the $0.4\,\%$ extended object cuts. The baseline lensing mask, 
which consists of the combination of these three masks, retains $77.6\,\%$ of
the sky and is labelled L80. }

\resub{In Sect.~\ref{Sect:tests:mask}, however, we test the 
stability of our results against source contamination by using a mask accounting for
more point-like and extended sources. This has been constructed as the union of 
the Galactic, extended, and compact objects and \smica\ confidence 
masks. This mask, in which the cut sky fraction due to point sources is 
as large as $1.5\,\%$, preserves $76\,\%$ of the sky for analysis and is named L80s. }

\subsection{Template analysis mask}
\label{Sect:masks:template}

\subsubsection{Methodology for constructing the mask}
\label{Sect:masks:template:method}
\resub{Given the lensing mask, the construction of the template mask is 
a trade-off between preserving a large sky fraction for analysis 
and alleviating the bias induced by the lensing reconstruction on 
an incomplete sky. Inpainting the masked temperature map from 
which the lensing potential is extracted, as described 
in Sect.~\ref{methodo:lensing}, strongly suppresses this effect. 
The point-like holes induce negligible bias in the lensing potential 
reconstruction provided they are treated by inpainting 
beforehand~\citep{aurelien}. Therefore, the compact object mask 
does not need to be propagated to the template map, 
as it is verified using Monte-Carlo \newresub{simulations} in Sect.~\ref{Sect:simu}. 
However, larger sky cuts performed by the lensing mask (e.~g., those 
arising from the Galactic and extended object masks) \newresub{have} 
to be applied to the $B$-mode template.
Furthermore, some artefacts may be seen near the mask boundaries, 
which we refer to as the mask leakage, and which arises from the convolution of 
the signal with the mask side-lobes when transforming in spherical 
harmonic space~\citep{2012AA...544A..27P}. From these considerations,
 we have tailored a method of constructing the mask template using 
Monte-Carlo \newresub{simulations}: starting with 
the combination of the Galactic and the extended object masks, we
progressively enlarged the mask beyond the boundaries, until we 
\newresub{observed} a negligible residual bias in the cross-correlation power
spectrum of the template. This has been obtained by extending the Galactic 
mask $3^\circ$ beyond the boundaries and the extended compact object 
mask $30\arcm$ beyond. In Sect.~\ref{Sect:simu}, we check that using 
a template mask that are constructed in this fashion allows no significant
impact in the $BB^{\rm{lens}}$ lensing $B$-mode power spectrum.}

\subsubsection{The baseline and test template masks}
\resub{
The baseline mask of the lensing $B$-mode template map has been constructed from 
the lensing analysis L80 mask using the method described above. 
Specifically, the Galactic mask has been enlarged 3\deg\
beyond the boundaries, removing $30\,\%$ of the sky, and 
the extended object mask has been also slightly widened by $30\arcmin$ beyond 
the boundaries, removing an additional $1.2\,\%$ of the sky. 
The baseline template mask preserves a sky fraction of $69\,\%$ and is named
B70.} 

\resub{
In Sect.~\ref{Sect:tests:mask}, however, we check that consistent
 results are obtained using either the Galactic mask as it is, without 
enlargement, or a more conservative Galactic mask that removes 40\,\% of
the sky. The most aggressive template mask, labelled B80, leaves 77\,\%
 of the sky for the analysis, 
and the most conservative one, named B60, leaves 58\,\%.
We also test against extragalactic contamination through using the lensing 
L80s mask for the lensing potential reconstruction and the corresponding 
template mask constructed using the method described 
in Sect.~\ref{Sect:masks:template:method} and labelled B70s.
In addition to these masks that are targeted at the temperature map, in 
Sect.~\ref{Sect:tests:mask}, we also use a mask tailored for polarization 
in order to test against polarized foreground contamination. For this purpose, 
the B70 mask has been combined with the \smica\ confidence mask for
polarization, a detailed description of which is given in
\citet{planck2014-a11}. 
The latter removes regions of the sky where the polarization power (which is 
computed by squaring a low-pass filter-smoothed version of the \smica\  
polarization map) exceeds $5\,\mu\rm{K}^2$. The combined B70 and \smica\ 
confidence mask for polarization is labelled B70p and
retains $68\,\%$ of the sky. }     

\resub{
When computing the lensing $B$-mode power spectrum, we have employed apodized 
versions of the masks, using a cosine taper of 2\deg\ width for the
Galactic mask and $30\arcmin$ width for the extended object mask. 
As a consequence, effective sky fractions available for the power spectrum
are 5--10\,\% lower than the sky fraction of the template 
analysis masks. A list of the template masks discussed here 
is given in Table~\ref{Tab:mask}, together with 
the corresponding sky fraction, both with and without apodization.   
}
\begin{table}[tb] 
\begingroup 
\newdimen\tblskip \tblskip=5pt
\caption{Template analysis masks. The columns labelled $f_{\rm sky}$ and
$f_{\rm sky}^{\, \rm eff}$ give the sky fractions that are
preserved by the masks and their apodized versions, respectively.}  
\label{Tab:mask} 
\nointerlineskip 
\vskip -3mm
\footnotesize 
\setbox\tablebox=\vbox{ 
  \newdimen\digitwidth
  \setbox0=\hbox{\rm 0} 
  \digitwidth=\wd0 
  \catcode`*=\active 
  \def*{\kern\digitwidth}
  \newdimen\signwidth
  \setbox0=\hbox{+}
  \signwidth=\wd0
  \catcode`!=\active
  \def!{\kern\signwidth}
  \halign{ \hbox to 0.9in{#\leaderfil}\tabskip=2em& 
    \hfil#\hfil& 
    \hfil#\hfil\tabskip=0pt\cr 
    \noalign{\doubleline} 
    \omit Mask label\hfil& $f_{\rm sky}$& $f_{\rm sky}^{\, \rm eff}$\cr 
%
\noalign{\vskip 4pt\hrule\vskip 5pt} 
B60& 0.58&0.54\cr 
B70& 0.69& 0.65\cr
B70p& 0.68& 0.64\cr 
B70s& 0.68& 0.61\cr 
B80& 0.77& 0.69\cr 
\noalign{\vskip 5pt\hrule\vskip 4pt} 
}}
\endPlancktable 
\endgroup
\end{table}
%

%
%

\section{Validation on simulations}
\label{Sect:simu}

We now validate the pipeline described in Sect.~\ref{Sect:methodo}
\resub{ and test the impact of masking as described in Sect.~\ref{Sect:masks:template}} 
using a Monte-Carlo (MC) approach. We analyse our template \resub{by forming 
the $BB^{\rm{lens}}$ power spectrum given in Eq.~(\ref{Eq:clb})}, which has two
objectives:
\begin{itemize}
\item[(i)]{to assess that the template encloses the expected lensing
$B$-mode information, thus validating the assumptions on which the
synthesis method relies;}
\item[(ii)]{to demonstrate its utility for measuring the secondary
$B$-mode power spectrum.}
\end{itemize}

We use 
\resub{two independent sets of 100 \smica\ $I$, $Q$, and $U$ simulations, part of} 
the FFP8 MC simulation set described in Sect.~\ref{Sect:datasims}.

\subsection{Full-range power spectrum estimate}
\label{Sect:simu:cl} 

We have applied the pipeline described above to the \smica\ 
temperature and Stokes parameter simulations to obtain a set of 100
$\hat{B}^{\rm{lens}}$ estimates. \resub{For the lensing reconstruction 
on the temperature simulations, we have used the baseline L80 lensing
mask, and employed the second simulation set for the 
mean-field bias correction.}  
Then, lensing $B$-mode power spectrum estimates have been formed by cross-correlating
 $\hat{B}^{\rm{lens}}$ and the input $B$-modes \resub{as in 
Eq.~(\ref{Eq:clb}), and using the apodized 
version of the B70 mask.}
We have estimated the lensing $B$-mode $BB^{\rm{lens}}$ band-powers by multiplying 
Eq.~(\ref{Eq:clb}) by $\ell(\ell+1)/2\pi$ and averaging the multipoles 
over bins of width $\Delta \ell \geq 100$ to further reduce 
the pseudo-$C_\ell$ multipole mixing. 

\begin{figure}[!h]
  \begin{center}
   \includegraphics[trim = {0 0.3cm 0 0}, clip=true, width=\columnwidth]{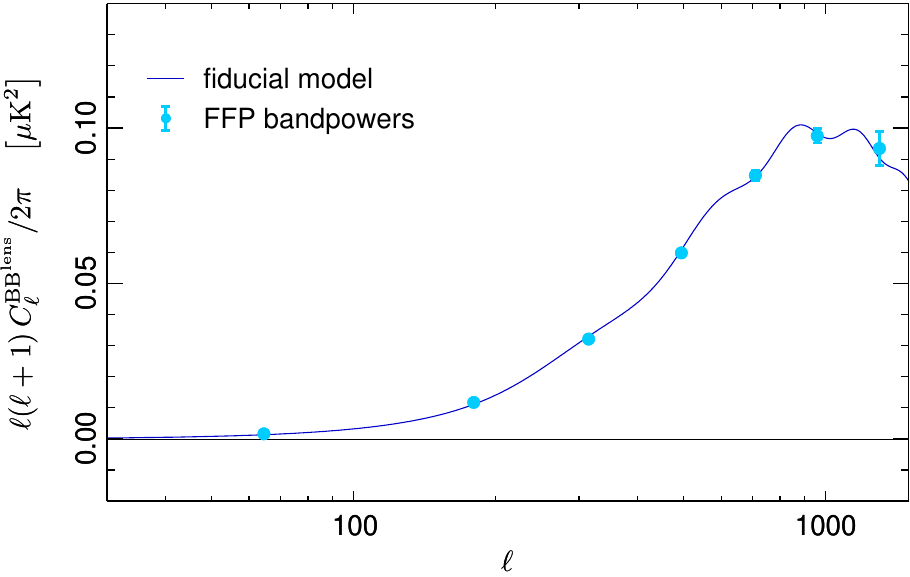}
   \includegraphics[width=\columnwidth]{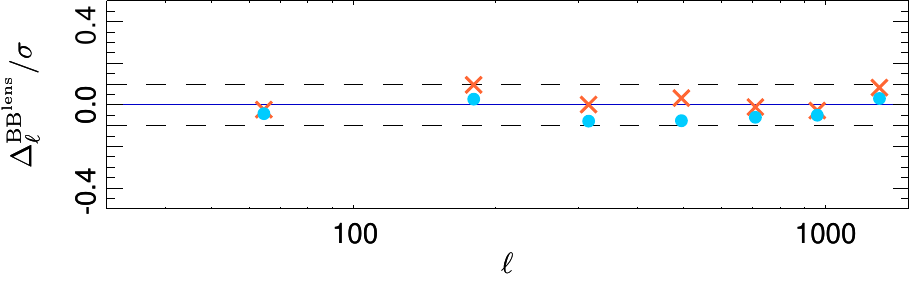}
   \caption{Lensing $B$-mode power spectrum obtained by cross-correlating 
the $B^{\rm{lens}}$ templates with the corresponding 
fiducial $B$-mode simulations (top), 
and residuals with respect to the model (bottom). 
\emph{Top}: averaged $BB{\rm{lens}}$ band-powers using the \resub{apodized} 
B70 template mask, with multipole bins of $\Delta \ell \geq 100$ 
(blue points). \resub{The error bars are the standard deviation on the mean 
of the band-power estimate set.} The dark blue curve is the fiducial 
$B$-mode power spectrum of our simulations, which assumes $r=0$. 
\emph{Bottom}: $BB{\rm{lens}}$ band-power residual
 with respect to the fiducial model, given in units of the 
$1\,\sigma$ error of a single realization. 
\resub{For comparison, we also show the band-power residuals obtained 
without masking, as discussed in Sect.~\ref{Sect:simu:cl}
 (red points).} 
Dashed lines show the $\pm 1\,\sigma$ range of 100 realizations, 
indicating the precision level to which we are able to test 
against bias.}
 \label{Fig:simuclbb}
 \end{center}
\end{figure}

Fig.~\ref{Fig:simuclbb} shows the averaged $BB^{\rm{lens}}$ 
band-powers obtained using the \resub{apodized} B70 mask. 
\resub{The error bars have been 
estimated using the standard deviation
of $100$ estimates of the $BB^{\rm{lens}}$ band-powers 
with the apodized B70 mask.} 
\resub{The averaged $BB^{\rm{lens}}$ band-power residuals,
  i.e. $\Delta C_b^{BB{\rm{lens}}} = \sum_{\ell \in b}
  C_\ell^{BB{\rm{lens}}} - C_\ell^{BB, \rm{fid}}$, are
plotted in the lower panel of Fig.~\ref{Fig:simuclbb}; this shows a
negligible residual bias of $\la 0.1\,\sigma$ up to multipoles of 
2000. Relative to the input lensing $B$-mode band-power, this 
corresponds to less than a percent.} 
To isolate the impact of the inpainting, we have also extracted  
the lensing $B$-modes using the lensing potential reconstructed from
the full-sky temperature map, that is to say without using any
mask. For comparison, the band-power residuals are also
plotted for full-sky case in Fig.~\ref{Fig:simuclbb}.

\resub{We have further tested that the $BB^{\rm{lens}}$ band-powers
constitute an unbiased estimate of the fiducial lensing $B$-modes 
by fitting an amplitude with respect to the fiducial model. The 
averaged amplitude $A_{B\rm{lens}}$ obtained on 100 estimates 
of the $BB^{\rm{lens}}$ band-powers using the B70 mask is $0.989 \pm 0.008$,
where the error has been
evaluated using the standard deviation of the $A_{B\rm{lens}}$ 
estimates normalized by the square root of the number of realizations. 
We therefore conclude that our pipeline provides us with an unbiased
 estimate of $A_{B\rm{lens}}$.} 

It is worth noting that the choice of extracting the lensing potential
from the temperature-only data ensures the absence of Gaussian
$N_L^{(0)}$-like bias, which must be corrected for in the case of a
lensing extraction using $B$-mode information. It also gives a strong
suppression of any higher-order bias terms.

\subsection{Statistical error budget}
\label{Sect:simu:error} 
First, we have quantified the statistical error associated with our 
lensing $B$-mode band-power measure via simulations by
computing the standard deviation of our MC band-power 
estimates. \resub{Then, these MC error bars have been compared to
semi-analytical errors evaluated in taking the square root of 
the Gaussian variance given in Eq.~(\ref{Eq:gaussvar}).} 
\resub{The auto-correlation $B$-mode power spectrum of the \smica\ simulation 
has been modelled using $C_\ell^{B^{\rm{obs}}} = C_\ell^{B} + B_\ell^{-2}N_\ell^{B}$,
where $C_\ell^{B}$ and $N_\ell^{B}$ are 
the MC input signal and noise power spectra and $B_\ell$ is the beam
function. For the auto-correlation power spectrum of the template, 
we have used the averaged estimate on our MC set. Furthermore, we have
used the $\Delta \ell \geq 100$ binning function to average 
the variance over multipoles. We note that these large multipole bins 
contribute to drive our band-power estimates close to a normal 
distribution by virtue of the central-limit theorem, which in turn, 
brings the Gaussian variance of Eq.~(\ref{Eq:gaussvar}) closer 
to the true variance.}  

\begin{figure}[!h]
  \begin{center}
   \includegraphics[trim = {0 0.3cm 0 0}, clip=true, width=\columnwidth]{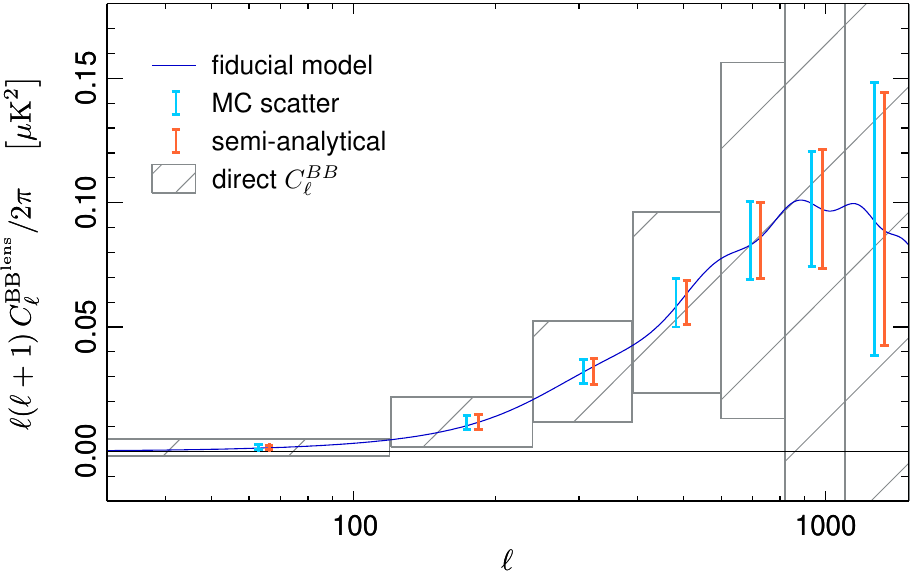}
   \includegraphics[width=\columnwidth]{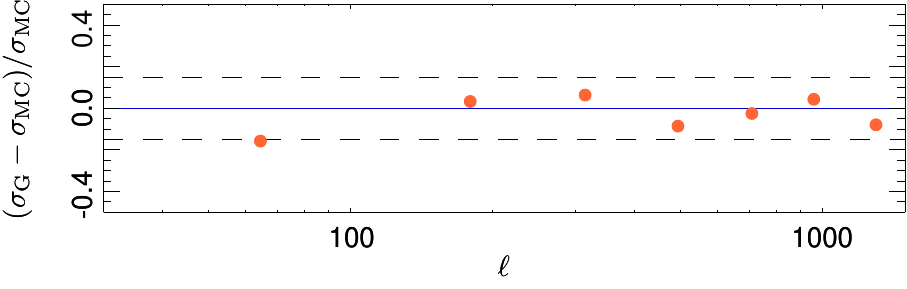}
   \caption{Error budget. We compare the MC derived
uncertainties $\sigma_{\rm{MC}}$ associated with the $BB^{\rm{lens}}$ 
measurements to \resub{the semi-analytical errors $\sigma_{\rm{G}}$ 
obtained using Eq.~(\ref{Eq:gaussvar})}.  \emph{Top:} averaged $BB^{\rm{lens}}$ 
band-powers using either $\sigma_{\rm{MC}}$ (blue) or $\sigma_{\rm{G}}$ (red) 
as error estimates. For illustration, we also show
the expected sensitivity of the \Planck\ \smica\ CMB polarization map
to the $BB$ band-powers (grey boxes), 
whose uncertainties are a factor of four larger than those of 
the $BB^{\rm{lens}}$ band-powers using the lensing $B$-mode template. 
\resub{\emph{Bottom:} relative difference of $\sigma_{\rm{G}}$ with 
respect to $\sigma_{\rm{MC}}$. Dashed lines show a $15\,\%$ difference.}}
   \label{Fig:simuerror}
  \end{center}
\end{figure}

In Fig.~\ref{Fig:simuerror}, we plot the MC error bars and the
semi-analytical ones. For comparison purposes, we also show the errors
one would have obtained from a $BB$ power spectrum measurement by
computing the auto-power spectrum of the fiducial $B$-mode map.

The uncertainties in our $BB^{\rm{lens}}$ band-powers are well
approximated by the \resub{semi-analytical errors} at all multipoles. 
\resub{In the lower panel of Fig.~\ref{Fig:simuerror} we plot the relative 
difference between the semi-analytical errors using Eq.~(\ref{Eq:gaussvar})
and the error estimates obtained by \newresub{simulations}. 
Using Eq.~(\ref{Eq:gaussvar}) leads to a 16\,\% underestimate of the error
in the first multipole bin and less than 10\,\%
underestimation at higher multipoles. These results validate the use
of Eq.~(\ref{Eq:gaussvar}) to evaluate the template-based $B$-mode 
power spectrum uncertainties.} 
We also find that the uncertainties of the $BB^{\rm{lens}}$ power spectrum
 using the template map are approximately four times lower
than those of a total $B$-mode $BB$ measurement coming directly from the $B$-mode map.

Gathering the results of our MC analysis, we observe that, when used
in cross-correlation with the $B$-mode map, the lensing $B$-mode
template we compute provides a lensing $C_\ell^{B}$ measurement,
which: (i) does not rely on any bias subtraction; (ii) has nearly
optimal uncertainty; and (iii) has four times lower uncertainty than the 
auto-$C_\ell$ measurement on the fiducial $B$-mode map.

%
%

\section{\Planck-derived secondary $B$-mode template}
\label{Sect:results}

\subsection{Template synthesis}
\label{Sect:results:map} 
\begin{figure}[!h]
  \begin{center}
    \includegraphics[width=\columnwidth]{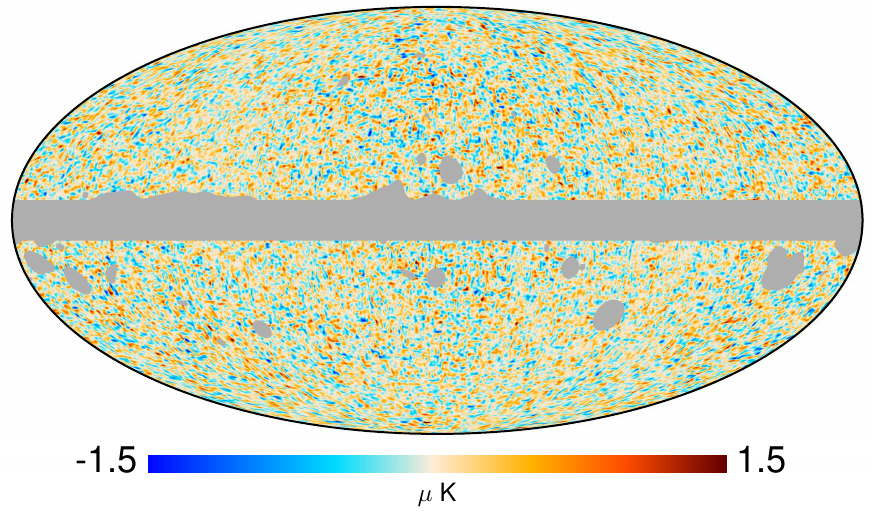}
    \includegraphics[width=\columnwidth]{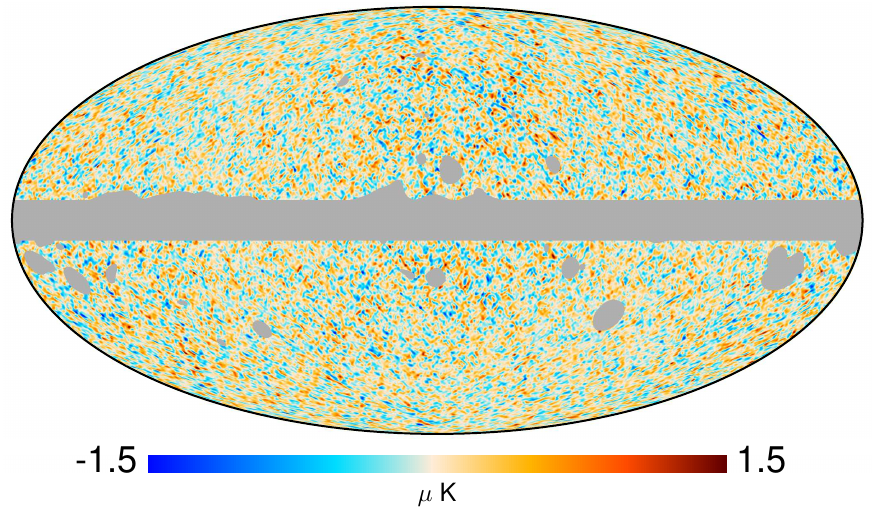}
    \caption{\smica\ lensing-induced $Q$ and $U$ templates that have
been convolved with a Gaussian beam of $60\arcm$~FWHM to
highlight the large angular scales (corresponding to multipoles below 200). 
No spurious patterns are observed at large angular scales.  
}
    \label{Fig:QUlens}
  \end{center}
\end{figure}

We have produced the template map of the lensing
$B$-modes by applying the pipeline described in
Sect.~\ref{Sect:methodo} to the foreground-cleaned temperature and
polarization maps obtained using the \smica\ component-separation
method, as described in Sect.~\ref{Sect:datasims}. We have first obtained the
$\hat{Q}^{\rm{lens}}$ and $\hat{U}^{\rm{lens}}$ templates defined in
Sect.~\ref{methodo:algorithm}, filtered versions of which are plotted
in Fig.~\ref{Fig:QUlens}. Specifically, the maps have been smoothed
using a Gaussian beam of $1\deg$ FWHM to highlight any
low-$\ell$ systematic effects, such as those due to
intensity-to-polarization leakage. This first inspection indicates
that these templates are not affected by any obvious low-$\ell$
systematic effects. \resub{More rigorous tests, however, are performed in Sect.~\ref{Sect:tests:freq}  using
intensity-to-polarization leakage corrected maps.}

\begin{figure}[!h]
  \begin{center}
    \includegraphics[width=\columnwidth]{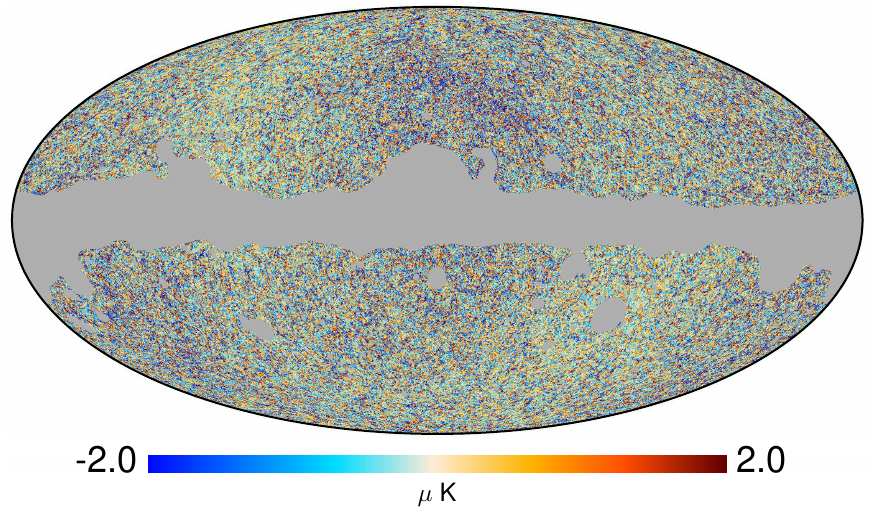}
    \includegraphics[width=\columnwidth]{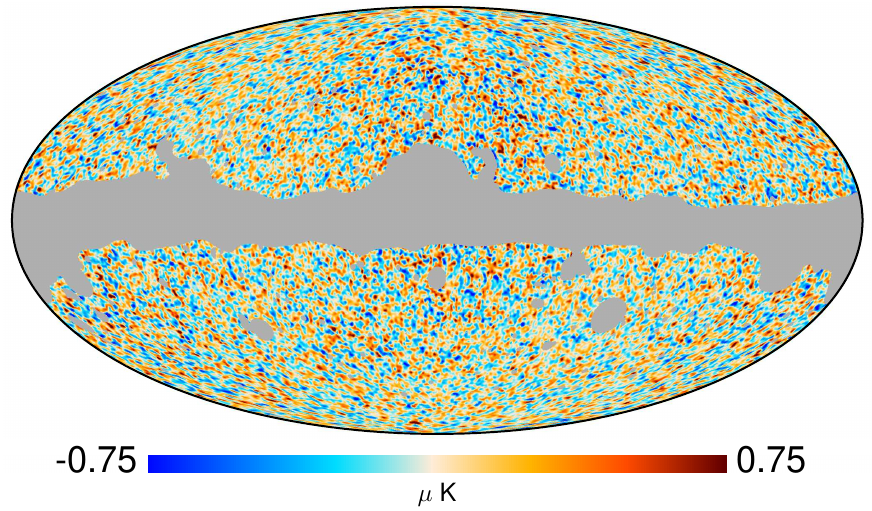}
    \caption{\Planck-derived $B$-mode template map computed using the
\smica\ foreground-cleaned CMB maps. For illustration, the map has
been convolved with a Gaussian beam of $10\arcm$ (upper panel) and
$60\arcm$ (lower panel) FHWM. The grey area represents the L80 mask,
which was used at the lensing potential reconstruction stage. \resub{No obvious foreground residuals 
are seen in the high-resolution map, nor any obvious systematic effects in the 
low-resolution one.} 
}
    \label{Fig:results}
  \end{center}
\end{figure}

Then we have used the \smica\ $\hat{Q}^{\rm{lens}}$ and
$\hat{U}^{\rm{lens}}$ templates to make our secondary $B$-mode
template, as described in Sect.~\ref{methodo:results}. For
illustration purposes, we have produced a $B$-mode template map by
inverting $\hat{B}_{\ell m}^{\rm{lens}}$ back to pixel-space through
an inverse spherical harmonic transform.  Although our $B$-mode
template contains information in the multipole range $10 < \ell <
2000$, Fig.~\ref{Fig:results} shows two filtered versions of the
$B$-mode map to highlight different ranges of angular
scale. The high-resolution map (which is simply slightly smoothed
using a Gaussian beam of $10\,\arcm$ FWHM) should show any important
foreground contamination at small angular scales, whereas the
low-resolution one (which is smoothed using a $1\deg$ Gaussian beam
and downgraded to $\Nside=256$ {\tt HEALPix} resolution) should reveal
any large angular scale systematic effects. \resub{No evident systematic 
effects are observed in the maps plotted in Fig.~\ref{Fig:results}. In 
Sect.~\ref{Sect:tests}, we further assess the template robustness 
against various systematic effects by means of a series of tests at 
the power spectrum level.}

\subsection{Variance contributions}
\label{Sect:results:varmap}

\resub{In this section, we describe and quantify the various contributions
that enter the template map variance. In particular, ranking the sources of
variance helps us in quantifying the template variance dependence
in the choice of the lensing potential estimates discussed in
Sect.~\ref{methodo:lensing}. This also anticipates the discussion that we
develop in Sect.~\ref{Sect:keytool} on the utility of the \Planck\ template
for other experiments that are attempting to measure the lensing $B$-modes,
compared to the use of a template that combines the \Planck\ lens
reconstruction and the experiment's $E$-mode measurement.}

\begin{figure}[!h]
  \begin{center}
    \includegraphics[width=\columnwidth]{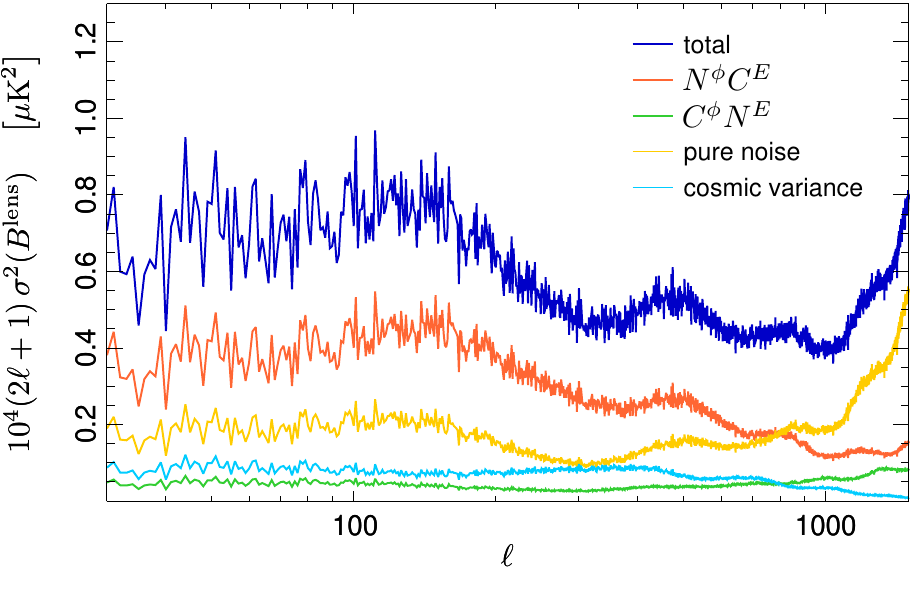}
    \caption{Template map variance budget. The total variance of the
template, plotted in dark blue, is split into the contribution of
the lens reconstruction noise (orange) and $E$-mode noise (green)
cross-terms; and the pure noise (yellow) and cosmic variance
(light blue) terms defined in Sect.~{\ref{Sect:results:varmap}}.}
    \label{Fig:mapvar}
  \end{center}
\end{figure}

\resub{Using the auto-power spectrum estimate
$\hat{C}_\ell^{B^{\rm{lens}}}$ computed on the apodized masked
template, the template variance is} 
\be 
\sigma^2\left(
B^{\rm{lens}}\right) \, = \,
\frac{1}{f_{\rm{sky}}^{\rm{eff}}(2\ell+1)}
\hat{C}_\ell^{B^{\rm{lens}}},
\label{Eq:gaussvarB} 
\ee 
\resub{where $f_{\rm{sky}}^{\rm{eff}}$ is the
effective sky fraction that is preserved by the apodized B70
mask.  This receives three main contributions: cosmic variance of both
the CMB $E$-modes and the lenses; instrumental noise; and lens
reconstruction noise. For the sake of completeness, other sub-dominant
contributions could include foreground residuals, higher-order terms and
secondary contractions of the CMB trispectrum (such as the so-called
$N^{(1)}$ bias of the lens reconstruction). Within this model,
$C_\ell^{B^{\rm{lens}}}$ can be analytically calculated using 
\be
C_\ell^{\hat{B}\rm{lens}} = \mathcal{B}_\ell^{-2} \, \sum_{L\ell'}
(f_L^{\phi})^2 (C_L^{\phi,{\rm fid}}+N_L^{\phi})\, (f_{\ell'}^{E}B_{\ell'})^2
(C_{\ell'}^{E,{\rm fid}}+N_{\ell'}^{E})\, _2F_{\ell L\ell'},
\label{Eq:autoclbblens}
\ee 
where the notation is the same as in Eqs.~(\ref{Eq:filters})
and (\ref{Eq:beff}).  
Although the variance is evaluated using the template power 
spectrum estimate, Eq.~(\ref{Eq:autoclbblens}) provides us with 
a useful tool for isolating
the relative contributions to the total variance of the template.} 
\resub{It can be decomposed into four terms: the `cosmic variance'
contribution, which arises from the product of the $\phi$ and $E$
power spectra; the `pure noise' contribution, which involves the
product of both noise spectra; and two cross-terms, namely the `$\phi$-noise
primed' contribution $N_L^{\phi}C_{\ell}^{E}$ and the `$E$-noise
primed' contribution $C_L^{\phi}N_{\ell}^{E}$. In
Fig.~\ref{Fig:mapvar}, we plot the different contributions to the total
variance of the template. }
\resub{The template variance is dominated by the $\phi$-noise primed
cross-term, which contributes an amount of about $60\,\%$ to the total
variance. To further quantify the template variance dependence on the
$\phi$ estimate, we have rescaled $N_{L}^{\phi}$ by a factor of $0.75$
in Eq.~(\ref{Eq:autoclbblens}), which represents a simple way to emulate the
precision gain that could be obtained by including the polarization-based
$\phi$ reconstruction and a more precise $\phi$ estimate at high multipoles
(e.g. using the CIB as a mass tracer). This has resulted in a $20\,\%$ reduction
of the template variance. This, in turn, has been propagated to the 
variance of the $BB^{\rm{lens}}$ measurement using
Eq.~(\ref{Eq:gaussvar}), and has induced a $12\,\%$ improvement of the signal detection significance,
in agreement with the independent analysis reported in
\citet{planck2014-a17} and discussed in Sect.~{\ref{Sect:tests:masstrac}}}. 

\resub{The $E$-mode noise power spectrum, which contributes mainly through
the pure noise term, provides less than $35\,\%$ of the variance up to
$\ell = 800$. As a consequence, very little leverage is
left for other experiments to improve on the template uncertainties
by producing a new template that combines each experiment's
$E$-mode measurement with the \Planck\ lensing potential
reconstruction. In Sect.~\ref{Sect:keytool}, we see that the
\Planck\ template offers other experiments a useful tool for
measuring the lensing $B$-modes.}

%
%
%
%
%

\section{Robustness tests of our template}
\label{Sect:tests}

We now perform a series of consistency tests to characterize
the $B$-mode template described in Sect.~\ref{Sect:results}, with a
view to using it for measuring the lensing $C_\ell^B$ in
cross-correlation with observed polarization maps. 
In Sect.~\ref{Sect:tests:mask}, we assess the template robustness 
against foreground residuals, first in varying the morphology and 
level of conservatism of the mask, and then in using 
foreground-cleaned CMB maps obtained with four independent 
component-separation algorithms. 
In Sect.~\ref{Sect:tests:freq}, we use the \Planck\ polarization maps in 
individual frequency channels to assess the stability 
of the lensing $C_\ell^B$ estimates with respect to the observed map 
to which our template is correlated.
\resub{In Sect.~\ref{Sect:tests:masstrac}, we discuss the consistency 
of the baseline lensing $B$-mode power spectrum presented here with 
the independent determination in \citet{planck2014-a17}, which used
various mass tracers.}
Finally, in Sect.~\ref{Sect:tests:ext}, the \Planck\ lensing $C_\ell^B$
measurements obtained using the $B$-mode template is compared 
to external measurements.

\subsection{Template tests using  $B$-mode band-power estimates}
\label{Sect:tests:valid}

\resub{With the aim of testing the $B$-mode template, we have followed the 
template-based cross-correlation power spectrum approach that we
developed to validate our pipeline via simulations,
as described in Sect.~\ref{Sect:simu:cl}. The tests that have been performed 
there also allow for an end-to-end assessment of the entire pipeline, 
whose main specifications are recalled below. Using FPP8 MC simulations for
the \smica\ method, together with the B70 template mask, we found that:
(i) unbiased lensing $B$-mode band-power estimates are obtained by 
cross-correlating the $B$-mode template estimate with the input $B$-mode map;
and (ii) the semi-analytical Gaussian variance given in Eq.~(\ref{Eq:gaussvar})
provides a good approximation for the uncertainties. To test the 
consistency of the band-power estimates with theoretical expectations, we
fitted an amplitude with respect to the fiducial $C_\ell^B$ band-powers, and
found an average value of $0.989 \pm 0.008$, which indicates that the lensing 
$B$-mode signal is accurately retrieved.
}

\resub{Here, we have used the $B$-mode template presented in
Sect.~\ref{Sect:results} in cross-correlation with 
the \smica\ foreground-cleaned polarization maps, and have applied the 
B70 template mask to obtain a baseline lensing $B$-mode 
band-power estimate, which is compared below to various other estimates. 
We further check the consistency between the baseline 
and alternative estimates by comparing the measured 
amplitude difference to the expected variance of 
the difference in Monte-Carlo simulations.}

\subsection{Robustness against foreground residuals}
\label{Sect:tests:mask} 
\resub{The robustness of the baseline measurements against the impact
of foregrounds is assessed by comparison with alternative estimates that have different 
residuals. A first sequence of alternative estimates have been obtained by employing
masks of various levels of conservatism with respect to the Galactic 
and extragalactic foregrounds in temperature and polarization. 
These are detailed in Sects.~\ref{Sect:tests:mask:gal}
to \ref{Sect:tests:mask:pol}. The 
corresponding fitted $A_{B\rm{lens}}$ amplitudes and differences 
from the B70 estimate are listed in Table~\ref{Tab:signi}.
Secondly, we consider measurements obtained with
foreground-cleaned maps using different component-separation methods and 
give the resulting amplitudes in Sect.~\ref{Sect:tests:mask:pol}.}

\subsubsection{Galactic contamination tests} 
\label{Sect:tests:mask:gal}
\resub{
We test for residual foreground contamination around the Galactic 
plane by comparing the B70 lensing $B$-mode estimate to estimates 
using the more conservative B60 mask and the more aggressive B80 mask,
discussed in Sect.~\ref{Sect:masks}. 
We recall that B80 includes a Galactic mask that preserves $80\,\%$
of the sky; this Galactic mask has been extended by $3\deg$ beyond its
boundaries for constructing B70, while B60 includes a larger 
diffuse Galactic mask that retains about $60\,\%$ of the sky. 
We have employed the apodized version of the masks using cosine tapers, as described
in Sect.~\ref{Sect:masks}. For B80, however, we have used slightly larger apodization widths, namely $3\deg$ for the Galactic mask (instead of $2\deg$) 
and $1\deg$ for the extended compact object mask (instead of $30\arcm$).  
}

The lensing $C_\ell^B$ band-power estimates using the B60, B70, and B80
masks are shown in Fig.~\ref{Fig:masking}, and the fitted amplitudes 
are listed in Table~\ref{Tab:signi}. Both the B60 and B80 estimates 
are consistent with the B70 estimate. In particular, the agreement 
between the B60 and B70 lensing $B$-mode estimates indicates 
that any impact of
Galactic foreground residuals lies well below the uncertainties. 
The consistency between the B70 and B80 estimates further
indicates that: (i) an 80\,\% Galactic cut suffices to avoid any
Galactic foreground residuals in the \smica\ map; and (ii) any leakage
near the border of the mask (treated by inpainting) has negligible
impact.

\subsubsection{Extragalactic object contamination tests} 
\label{Sect:tests:mask:ps}
\resub{
For extracting the lensing potential $\phi$ map estimate that is 
used for the $B$-mode template synthesis, we have employed the L80 lensing mask
described in Sect.~\ref{Sect:masks:lensing}. 
This cuts out any point sources included in the intersection of 
the conservative compact object mask described in
 Sect.~\ref{Sect:masks:foregrounds} and the \smica\ confidence mask. 
Here we test the stability of the $B$-mode estimate through the use 
of a more conservative point source mask that involves the union of 
these two masks. Namelly, we have produced an alternative estimate that relies on a $\phi$ 
reconstruction using the L80s lensing mask, 
which includes the union of the compact object and the \smica\ masks 
(see Sect.~\ref{Sect:masks:lensing}). In L80s, the sky area that is 
removed due to the expected point source contamination is increased by a factor of $2.3$
compared to the L80 case. The extended object mask is also slightly enlarged by
$25\,\%$ due to areas of radius between $30\arcmin$ and $120\arcmin$ located 
at intermediate Galactic latitudes that are present in the \smica\ confidence
mask.  The $B$-mode band-powers have then been estimated using the corresponding B70s 
mask for the template (which removes $1\,\%$ more sky fraction
than the baseline B70 mask, including about a hundred additional 
extragalactic objects). 
The B70s band-power estimate, which is plotted in Fig.~\ref{Fig:masking}, 
is consistent with the baseline B70 estimate within uncertainties. 
The fitted amplitude difference from the B70 estimate, which is given in 
Table~\ref{Tab:signi}, is well within the expected difference estimated from
simulations. This validates the choices we made to include the expected point 
sources to our baseline mask, and indicates that our analysis is free from 
any sizeable bias due to extragalactic objects.    
}

\subsubsection{Polarized foreground emission tests} 
\label{Sect:tests:mask:pol}

\begin{figure}[!h]
  \begin{center}
   \includegraphics[trim = {0 0.45cm 0 0}, clip=true, width=\columnwidth]{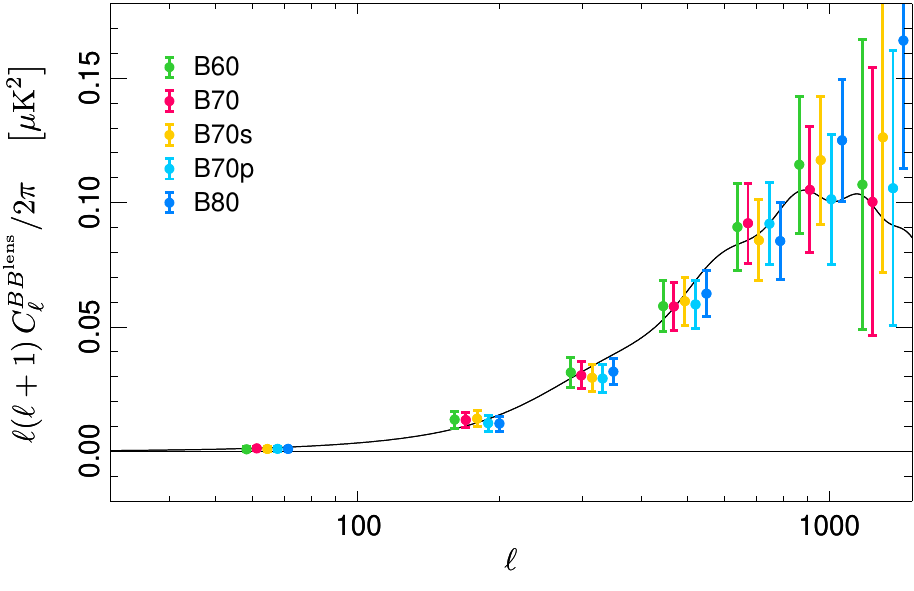}
   \includegraphics[width=\columnwidth]{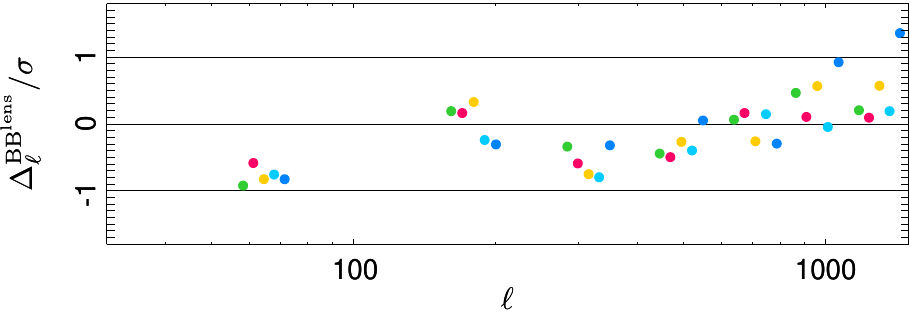}
   \caption{Cross-correlation of our $B$-mode template with the
\smica\ polarization map using masks of different levels of
conservatism. The data points show the lensing $C_\ell^B$ band-powers
estimated using masks that preserve a sky fraction of about 60\,\%
(labelled B60), 70\,\% (`B70'), and 80\,\% (`B80'), and that are
targeted to the foreground emission in temperature 
(see Sect.~\ref{Sect:tests:mask:gal}); as well as using the 68\,\% `B70s'
mask, which allows for a more conservative masking of the extragalactic
objects (see Sect.~\ref{Sect:tests:mask:ps}), and the 68\,\% 
polarization-targeted `B70p' mask (see Sect.~\ref{Sect:tests:mask:pol}). 
The residuals with respect to the model in units of the 
$1\sigma$ band-power uncertainties are shown in the lower panel. 
The good agreement of all cases within error-bars provides a robustness test against
foregrounds, as discussed in Sect.~\ref{Sect:tests:mask}.}
   \label{Fig:masking}
  \end{center}
\end{figure}

\resub{We test for the impact of polarized foreground residuals by comparing 
the B70 estimate to an estimate using the B70p mask targeted at the 
\smica\ polarization maps, as discussed in Sect.~\ref{Sect:masks:template}. 
The B70p mask discards a sky fraction $2.5\,\%$ larger than B70,
mainly because of extended areas at medium latitude that may be contaminated
by polarized foreground residuals. 
The lensing $B$-mode band-power estimates using the B70 and B70p masks 
are plotted in Fig.~\ref{Fig:masking} and the corresponding amplitude 
fits are given in Table~\ref{Tab:signi}. The B70p estimate well agrees with the B70 estimate. We therefore expect 
no significant bias related to polarized foreground residuals.}

\begin{table}[tb] 
\begingroup 
\newdimen\tblskip \tblskip=5pt
\caption{Band-power amplitudes for various analysis masks, which are
listed in the first column. The column labelled $f_{\rm sky}^{\, \rm eff}$ 
give the effective sky fraction preserved by the apodized masks. The
measured band-power amplitudes are given in the column labelled
$A_{B\rm{lens}}$, and the corresponding lensing B-mode detection
significance levels in the column labelled S/N. The difference 
between the test amplitudes and the baseline B70 amplitude is given, 
where applicable, as $\Delta$, together with the expected 
difference $\sigma_\Delta$ estimated using Monte-Carlo \newresub{simulations}.}  
\label{Tab:signi} 
\nointerlineskip 
\vskip -3mm
\footnotesize 
\setbox\tablebox=\vbox{ 
  \newdimen\digitwidth
  \setbox0=\hbox{\rm 0} 
  \digitwidth=\wd0 
  \catcode`*=\active 
  \def*{\kern\digitwidth}
  \newdimen\signwidth
  \setbox0=\hbox{+}
  \signwidth=\wd0
  \catcode`!=\active
  \def!{\kern\signwidth}
  \halign{ \hbox to 0.9in{#\leaderfil}\tabskip=2em& 
    \hfil#\hfil&
    \hfil#\hfil& 
    \hfil#\hfil& 
    \hfil#\hfil\tabskip=0pt\cr 
    \noalign{\doubleline} 
    \omit Mask label\hfil& $f_{\rm sky}^{\, \rm eff}$& $A_{B\rm{lens}}$& S/N & $\Delta \pm \sigma_\Delta$\cr 
%
\noalign{\vskip 4pt\hrule\vskip 5pt} 
B70&  0.65& $0.96\pm0.08$& 11.9& n.~a.\cr
B60& 0.54& $0.98\pm0.09$& 11.1& $-0.02\pm0.04$\cr 
B70p&  0.64& $0.94\pm0.08$& 11.3& $*0.02\pm0.02$\cr 
B70s&  0.61& $0.97\pm0.08$& 11.8& $-0.01\pm0.04$\cr 
B80&  0.69& $1.00\pm0.08$& 12.8& $-0.04\pm0.03$\cr 
\noalign{\vskip 5pt\hrule\vskip 4pt} 
}}
\endPlancktable 
\endgroup
\end{table}

From the tests performed so far, the results of which are 
gathered in Table~\ref{Tab:signi}, we have found all the band-power
estimates, from the most conservative $f_{\rm sky}=0.58$
to the most optimistic $f_{\rm sky}=0.77$ analysis and
whatever the morphology of the mask, in good agreement with
the baseline estimate. This provides an indication that (i) 
the $B$-mode template is robust 
against any foreground contamination or any impact of the sky-cut, and   
(ii) it allows us to obtain reliable lensing $C_\ell^B$ measurements
for up to $70\,\%$ of the sky.

\begin{figure}[!h]
  \begin{center}
    \includegraphics[trim = {0 0.45cm 0 0}, clip=true, width=\columnwidth]{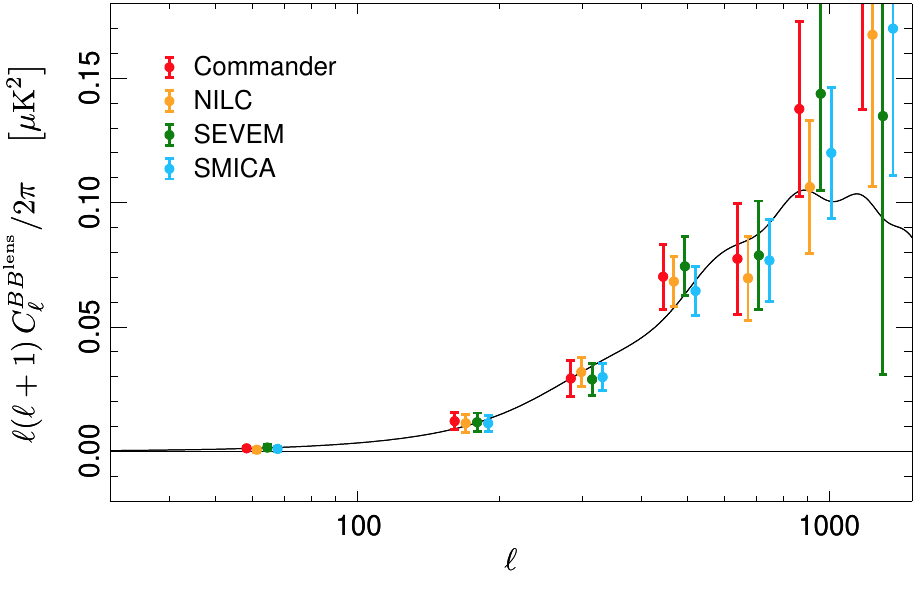}
    \includegraphics[width=\columnwidth]{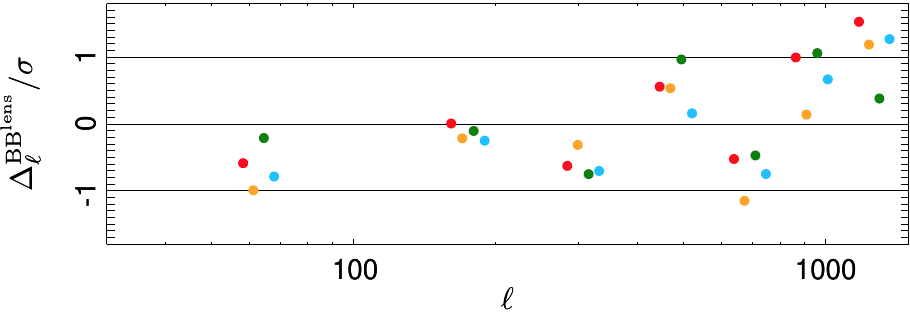}
   \caption{Consistency of our results using the four \Planck\
component-separation algorithms. The data points shown in the upper panel represent the lensing
$B$-mode band-power measurements obtained by cross-correlating the
lensing $B$-mode template estimates using the \commander\ (red), 
\nilc\ (orange), \sevem\ (forest green), and \smica\ (blue) cleaned
polarization map with the corresponding $B$-mode map. The residuals 
with respect to the model in units of the $1\sigma$ band-power uncertainties
 are shown in the lower panel. The consistency of the four
estimates strongly indicates the robustness of our
baseline template against polarized foreground residuals, 
as discussed in Sect.~\ref{Sect:tests:mask:pol}.}
    \label{Fig:4methods}
  \end{center}
\end{figure}

We further test our template robustness against foreground residuals
by comparing results using the four \Planck\ component-separation
methods, as described in \citet{planck2014-a11}. We have focused on the
polarized foreground residuals; we have used different
foreground-cleaned Stokes parameter maps, but the same temperature
map: specifically, the lensing $B$-mode templates have been synthesized from
the foreground-cleaned Stokes $Q$ and $U$ maps using different codes
and a fixed lensing potential extraction obtained from the
\smica\ temperature map.
\resub{For this specific test, we have produced a common mask that combines 
the B80 mask and the union of the confidence masks provided by each of 
the component-separation methods described in \citet{planck2014-a11}.
The common mask retains a sky fraction of $75\,\%$ for analysis, and 
its apodized version preserves an effective sky fraction of $62\,\%$.}
Fig.~\ref{Fig:4methods} shows the resulting lensing $B$-mode power
spectra, obtained through cross-correlation of the lensing $B$-mode
templates with the input Stokes $Q$ and $U$ maps from the
corresponding component-separation method.

The four Stokes $Q$ and $U$ foreground-cleaned solutions lead to
consistent lensing $B$-mode power spectrum measurements (within
1$\,\sigma$) over the entire multipole range probed.  We measure fits
to the amplitude with respect to the fiducial model of:
%
\begin{align} 
A_{B{\rm lens}} &= 1.01 \pm 0.11 \quad (\rm{\tt Commander}); \nonumber \\ 
A_{B{\rm lens}} &= 0.97 \pm 0.09 \quad (\rm{\tt NILC}); \nonumber \\ 
A_{B{\rm lens}} &= 1.00 \pm 0.10 \quad (\rm{\tt SEVEM}), \nonumber \\ 
A_{B{\rm lens}} &= 0.97 \pm 0.08 \quad (\rm{\tt SMICA}). \nonumber
\end{align}
%
These correspond to $9.5\,\sigma$, $11.3\,\sigma$, $9.7\,\sigma$, and
$11.8\,\sigma$ detections of the lensing $B$-modes,
respectively. The consistency of the lensing $C_\ell^B$ measurements
based on four CMB solutions with different foreground residuals
indicates the immunity of our baseline lensing $B$-mode template to
polarized foreground emission.

\subsection{Stability with respect to the observed polarization }
\label{Sect:tests:freq} 
So far, we have considered the cross-correlation of the $B$-mode template synthesized from the Stokes
$I$, $Q$, and $U$ foreground-cleaned maps, with the same $Q$, $U$ maps. 
We now test the cross-correlation of our baseline template with
other CMB foreground-cleaned polarization maps. In particular, we
have used foreground and intensity-to-polarization leakage-corrected \Planck\
channel maps at 100, 143, and 217\,GHz, which also serve to test the
robustness of our template against low-$\ell$ systematic effects. 
\resub{These single channel maps rely on the pre-launch ground-based
measurements of the detector bandpasses to correct for the 
intensity-to-polarization leakage due to bandpass mismatch between the 
detectors within a frequency channel, as described in~\citet{planck2014-a09}. 
The polarized emission of the Galactic dust has been corrected for by using the 353-GHz map 
as a dust template.}  

\begin{figure}[!h]
  \begin{center}
    \includegraphics[trim = {0 0.45cm 0 0}, clip=true, width=\columnwidth]{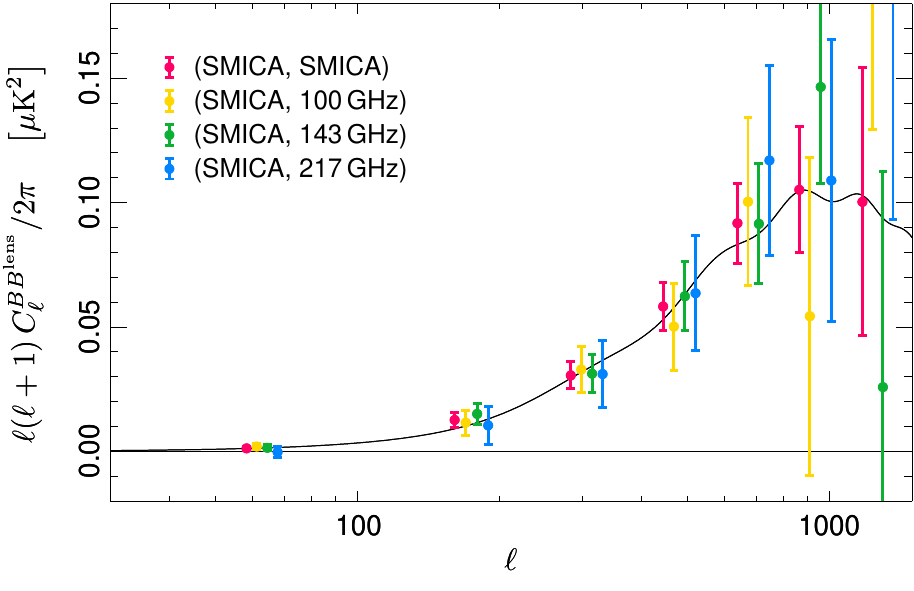}
    \includegraphics[width=\columnwidth]{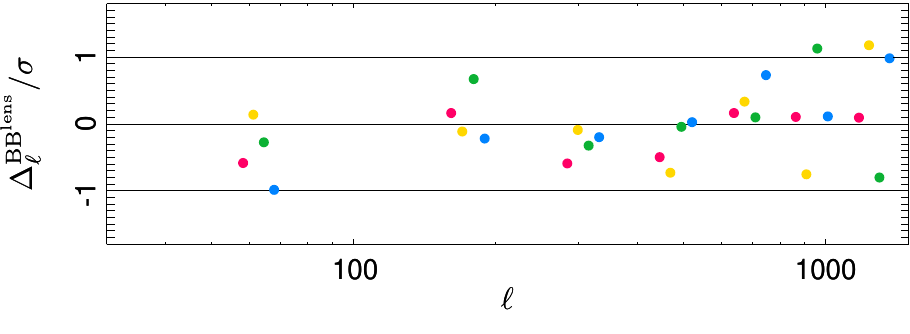}
   \caption{Stability to the change in the observed polarization
map. Data points shown in the upper panel are the $BB^{\rm lens}$ band-powers obtained
from the cross-correlation of our \smica\ $B$-mode template with the
\smica\ polarization map (red), and with single frequency
polarization maps at 100\,GHz (yellow), 143\,GHz (green), and 217\,GHz
(blue), which were corrected for foreground and
intensity-to-polarization leakage. Residuals 
with respect to the model in units of the $1\sigma$ band-power uncertainties
 are shown in the lower panel.}
 \label{Fig:singlefreq}
  \end{center}
\end{figure}

In Fig.~\ref{Fig:singlefreq}, we compare the baseline $BB^{\rm
lens}$ band-powers from the B70 \smica\ analysis to the band-powers
obtained by correlating our \smica\ $B$-mode template with 
\resub{the $B$-mode maps at $100$, $143$, and $217~\rm{GHz}$ built 
out of the corresponding single channel foreground-cleaned polarization maps and}
using the same B70 mask. The band-power estimates are in good agreement with each other
within uncertainties, indicating the robustness of our template
against polarization systematic effects (that mainly affect the low
multipoles).
We find that our lensing $B$-mode template provides stable
measurements of the CMB lensing $B$-modes independent of the choice of
the polarization maps with which it is correlated.

\subsection{Consistency with previous \Planck\ results}
\label{Sect:tests:masstrac} 
 
\resub{First of all, we have checked that the
baseline lensing $B$-mode power spectrum presented here is consistent
with the independent determination of \citet{planck2014-a17} over the
$100<\ell<2000$ multipole range, which provides us with further
validation of our methodology choices.}\footnote{In 
\citet{planck2014-a17}, one of the null tests conducted for checking
the robustness of the lensing potential power spectrum has showed mild
evidence of a difference from zero, specifically a non-zero signal has been
seen in the $TT$ curl-mode power spectrum (which contains the $TTTT$
curl-mode trispectrum) with a significance above $2\,\sigma$. This
feature, however, has no real impact in the lensing-induced power
spectrum being tested here. First, it contains the $TTEB$ trispectrum,
which differs from the trispectrum affected by the curl-mode null-test
failure. Additionally, any impact on the variance of the
lensing-induced $B$-mode power spectrum, via some $TTTT$ trispectrum
dependent terms, would be totally overwhelmed by the dominant Gaussian
noise term.}  

\resub{In \citet{planck2014-a17}, lensing-induced $B$-mode power spectrum
measurements have been obtained using various mass tracers: lensing
potential reconstructions using either both temperature and
polarization data or temperature data only; and CIB fluctuations
measured by \Planck\ in the 545-GHz channel. The latter have relied on a
model of the CIB emission for calculating its cross-correlation
with the lensing potential. All cases have been consistent with theoretical 
expectations and in good agreement with each other, which validates 
the stability of the measurement with respect to the mass tracer choice.
The additional information brought by the polarization-based $\phi$ estimates 
yields a $10\,\%$ improvement of the lensing B-mode detection significance compared to the 
case using the temperature-based $\phi$ estimate, whereas the use of the 
CIB fluctuations as a mass tracer lowers the detection significance 
by roughly the same amount.}
\resub{Thus, including the polarization information for reconstructing $\phi$
or using the CIB as a mass tracer would not have substantially improved the 
uncertainties of the $B$-mode template that we have produced.}

\subsection{Consistency with external results}
\label{Sect:tests:ext} 
Our $BB^{\rm lens}$ band-power
estimates are a measurement of the lensing $B$-mode power spectrum
that we now compare to measurements reported by other experiments.

\resub{As stated in Sect.~\ref{sec:introduction}, the available 
$B$-mode measurements come in two flavours: 
the $BB$ power spectrum of the observed polarization maps 
measures the total $B$-mode signal; whereas the $BB^{\rm{lens}}$
power spectrum between a template and the observed polarization maps 
probes the secondary contribution only.} 

\begin{figure}[!h]
  \begin{center}
    \begin{overpic}[width=\columnwidth]{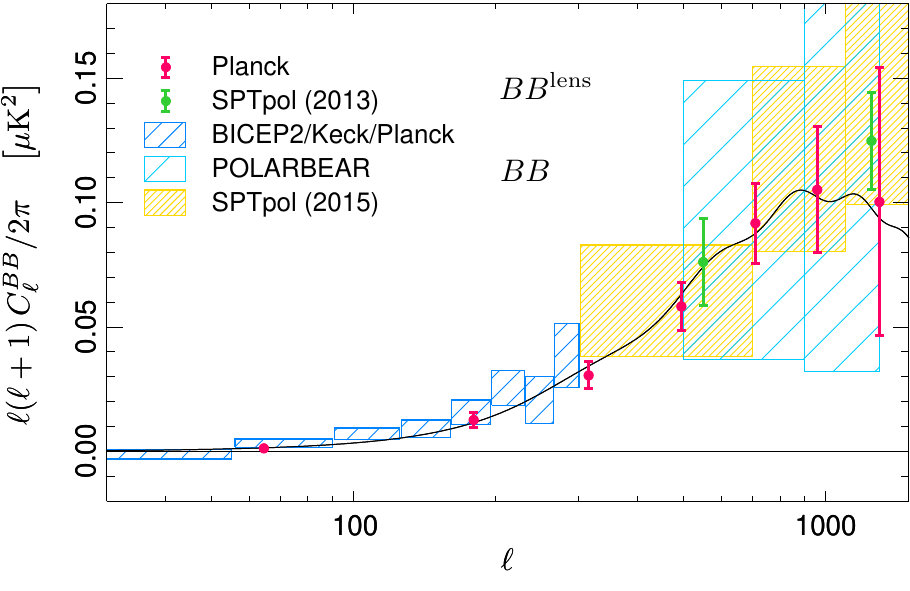}
\put(50.5,54.5){\large{$\left.
            \begin{array}{@{}ll@{}} \\
               
            \end{array}\right\}$}} \put(50.5,45){\small{$\left.
            \begin{array}{@{}ll@{}} \\ \\ \\
 
            \end{array}\right\}$}}
    \end{overpic}
    \vspace{0.2cm}
    \includegraphics[trim = {-1.3mm 0 0 0}, clip=true, width=0.998\columnwidth]{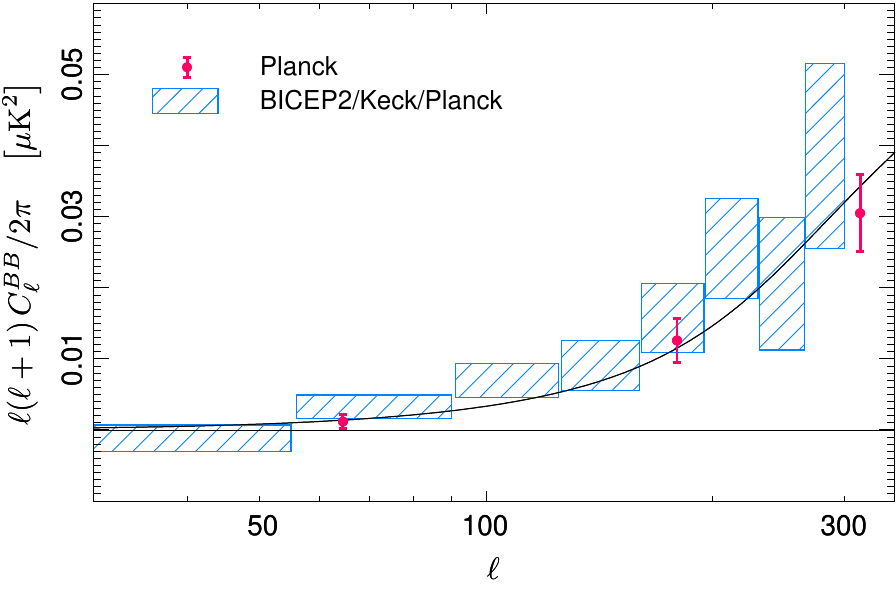}
   \caption{Consistency with external $B$-mode power
spectrum measurements on the full multipole range (top) and at $\ell<350$ (bottom). 
We compare our baseline $BB^{\rm{lens}}$ estimate (red points)
to the SPTpol template-based results (green points)
\citep{2013PhRvL.111n1301H}; and to the $BB$ power spectrum
 measurements from BICEP2/Keck Array (blue boxes)
\citep{BK5}, POLARBEAR (light blue boxes) \citep{2014ApJ...794..171T},
and SPTpol (yellow boxes) \citep{2015ApJ...807..151K}, as discussed in
Sect.~\ref{Sect:tests:ext}.  The black line shows the theoretical
lensing $B$-mode power spectrum for the base $\Lambda$CDM best-fit
\Planck\ model (with $r=0$).}
    \label{Fig:external}
  \end{center}
\end{figure}

\resub{In Fig.~\ref{Fig:external}, we gather the composite $C_\ell^{B}$  
measurements obtained by BICEP2/Keck Array, POLARBEAR, SPTpol,
and the \Planck\ mission, which represents the landscape of current 
CMB $B$-mode band-power estimates.}
\resub{The $BB$ power spectrum measurements of BICEP2/Keck
Array,\footnote{\url{http://bicepkeck.org/keck_2015_release.html}}
POLARBEAR~\citep{2014ApJ...794..171T} and SPTpol~\citep{2015ApJ...807..151K} 
are collected by the brace labelled $BB$ in Fig.~\ref{Fig:external}.
For BICEP2/Keck Array, we plot the $B$-mode band-powers in the multipole
range $20< \ell <335$, obtained from the combination of
the BICEP2 and Keck Array maps reported in \citet{BK5} and
corrected for polarized dust emission using \Planck\ data, as
described in \citet[][which is referred to as BKP hereafter]{pb2015}.}
\resub{The $BB^{\rm{lens}}$ measurements of SPTpol 
\citep{2013PhRvL.111n1301H} and \Planck\ (this analysis) are gathered by the 
brace labelled $BB^{\rm{lens}}$. }
For \Planck\ ${BB^{\rm lens}}$ band-powers, 
we select the B70 analysis whose robustness 
has been assessed in Sect.~\ref{Sect:tests:mask}.
The $4.2\,\sigma$ lensing $B$-mode detection 
reported by the POLARBEAR team~\citep{2014PhRvL.113b1301A}
and the $3.2\sigma$ detection obtained by ACTpol~\citep{ACT2015} 
are not shown in Fig.~\ref{Fig:external} for clarity.

Our \Planck-derived $BB^{\rm lens}$ band-powers are in good
agreement with other $B$-mode measurements, using both the 
$BB$ or the $BB^{\rm lens}$ power spectrum methods. \Planck\ provides the
most precise measurement of the lensing-induced $B$-mode power spectrum to
date~\citep[as assessed in][]{planck2014-a17}. Covering a wide
multipole range $10<\ell<2000$, our band-powers consists of lensing
$C_\ell^{B}$ measurements at a significance level as high as
$12\,\sigma$. It is also worth focusing on the low-$\ell$ range. 
\resub{The $C_\ell^{B}$ measurements are expanded at $\ell>350$ 
in the lower panel of Fig.~\ref{Fig:external},} 
which illustrates the lensing $B$-mode template utility 
for the primordial-to-secondary $B$-mode
discrimination. Using \Planck\ data alone, which combines wide sky
coverage, necessary to probe low multipoles, and a good angular
resolution, needed for the lensing potential extraction, the lensing
$B$-mode template enables us to extend the measure of the
lensing-induced $C_\ell^{B}$ into the low multipole range.

%
%
\section{Implications for current and future $B$-mode experiments}
\label{Sect:keytool}

Here, we address the implications of the template for
experiments targeting primordial $B$-modes. A full joint analysis
using external data is beyond the scope of this paper. However, as
examples we discuss two different aspects. 
First, we address to what
extent the template can help experiments in measuring the lensing
$B$-mode power spectrum, in particular over the largest accessible
angular scales. 
Secondly, we forecast the improvement of the lensing amplitude 
measurement when using the \Planck\ lensing $B$-mode template, 
and we discuss whether this improvement can translate into a 
better sensitivity to the tensor-to-scalar ratio.

\subsection{Measurement of the lensing $B$-mode power spectrum}
\label{Sect:keytool:blensing} 
Since the \Planck\ lensing $B$-mode
template covers almost the entire sky, polarization measurements from 
any $B$-mode experiment can be cross-correlated with the template. 
This ensures a valuable lensing $B$-mode power spectrum measurement, 
including the intermediate angular scales, where the
lensing signal is of the same order of magnitude or sub-dominant with
respect to the polarized foreground emission \citep[see][hereafter
referred to as PIP-XXX]{planck2014-XXX}. 

For current ground-based experiments, the cross-correlation with the
\Planck\ template is a promising method for measuring the lensing
$B$-mode signal at the largest accessible angular scales. In general,
it leads to better results than a cross-correlation with a lensing
$B$-mode template built out of the \Planck\ lensing potential and each
experiment's $E$-mode map. This can be understood from two different
considerations. On the one hand, a study of the lensing $B$-mode
signal kernel \citep[as in][]{Fabbian2013, Simard2015} 
reveals that most of the signal at low and intermediate angular scales 
comes from products of the lensing potential power at low multipoles 
and the $E$-mode power at higher multipoles. 
For example, at $\ell=100$ (200), 80\,\%
(90\,\%) of the lensing $B$-mode power arises from $E$-mode power at
$\ell>335$. For degree-scale experiments, such as BICEP, it would be
of little value to use its own $E$-mode data to generate a
new lensing $B$-mode template. 
\resub{On the other hand, in Sect.~\ref{Sect:results:varmap}, 
we have found the $B$-mode template variance to be dominated 
by the contribution arising from the noise power spectrum 
of the reconstructed lensing potential $N_L^{\phi}$, even for 
the \Planck\ $E$-mode noise.} 
\resub{For high-resolution experiments covering a small 
fraction of the sky, we do not expect that 
a better sensitivity in the $E$-modes 
compensates for the uncertainties linked to the small sky coverage.
We have quantified this effect in using 
Eqs.~(\ref{Eq:gaussvarB}) and (\ref{Eq:autoclbblens})
to forecast the template uncertainties, specifically for an ideal 
experiment providing noiseless $E$-mode measurements for multipoles
from 30 to 2000 with a resolution of 1\arcmin\ and a sky coverage of 
1\,\%. We find the uncertainties of the ideal  
experiment's $E$-mode-based template to be about five times larger than
the \Planck\ template uncertainties up to multipoles of 700 
(and still $50\,\%$ larger than those of the \Planck\ template 
at $\ell=2000$).}
We conclude that, for experiments covering less than a percent of the sky, 
such as SPTpol or POLARBEAR, the synthesis of a lensing $B$-mode template 
that combines the experiment's $E$-mode data with \Planck\ $\phi$ estimate 
would degrade the signal-to-noise for measuring the lensing $B$-mode
power spectrum.

\subsubsection{Uncertainty forecasting method} 
We forecast the uncertainties of the lensing $B$-mode power spectrum 
measurement that current experiments can obtain from cross-correlating
their $B$-mode signal with the \Planck\ lensing $B$-mode template. As
in Sect.~\ref{Sect:tests:ext}, we consider the BICEP2/Keck Array,
POLARBEAR, and SPTpol examples. Error bars are evaluated using
Eq.~(\ref{Eq:gaussvar}). They include lensed cosmic variance from the
best-fit $\Lambda$CDM model and the statistical error from the
template (the template auto-power spectrum
$\hat{C}_\ell^{B{\rm{lens}}}$ factor) and from the experiment (the
experiment auto-spectrum $(C_\ell^B + N_\ell^B)$ factor). To
estimate the template variance within the experiment's field,
$\hat{C}_\ell^{B\rm{lens}}$ is analytically calculated using the
lensing potential and $E$-mode noise power spectra, rescaled to the
noise spatial inhomogeneity within the experiment's field. This
rescaling relies on the \smica\ hit count map.  We use a simplified
model for the $BB$ auto-spectrum of the experiment that includes the
lensed CMB $B$-modes, polarized dust, white noise, and Gaussian
beam. The specific model and analysis choices are detailed below. 

The polarized dust emission has been parametrized by a single free amplitude
in power, $A_{\rm{dust}}$, which is defined at the reference frequency
of 353\,GHz and at a multipole of $\ell=80$. Following PIP-XXX, the
dust power spectrum has been modelled as the power law
$C_\ell^{\rm{dust}}\propto \ell^{-2.42}$, with a spatially uniform
frequency scaling according to a modified black-body law, assuming a
fixed dust temperature $T_{\rm d}=19.6 \, \rm{K}$ and spectral index
$\beta_{\rm d} = 1.6$. For the BICEP2/Keck Array combination, we have fixed
the dust amplitude to the best-fit value obtained in the joint
BICEP2/Keck Array and \Planck\ analysis described in BKP, namely
$A_{\rm{dust}} = 3 \, \mu\rm{K}^2$. In PIP-XXX, individual dust
amplitudes have been fitted for a series of sky patches, and their
scaling with the mean 353-GHz dust intensity was described using an
empirical relation. However, there was a warning about the difficulty
in deriving precise dust amplitude estimates from this empirical
law. From figure~7 of PIP-XXX, we can see that, at the low dust
intensity values expected in the SPTpol and POLARBEAR fields, the
amplitudes range from $0.1$ to $10\, \mu\rm{K}^2$. For simplicity, we have
assumed the level of dust emission in the POLARBEAR and SPTpol fields
to be the same as in the BICEP2 field, and so have used the value $A_{\rm
dust} =3 \, \mu\rm{K}^2$. We checked that assuming the more
pessimistic $A_{\rm dust} = 10\, \mu\rm{K}^2$ leads to a minor
increase in the forecast-ed $BB^{\rm{lens}}$ band-power uncertainties.

The BICEP2/Keck Array combination has been modelled as reaching a noise
level of $3.4\,\mu\rm{K}.arcmin$ over $400\,{\rm deg}^2$, and as
having a Gaussian beam of $31\arcm$ FWHM for both BICEP2 and Keck
Array experiments \citep{BK4,BK5}. POLARBEAR has been modelled as reaching a
depth of $8\,\mu\rm{K.arcmin}$ over an effective sky area of $25\,{\rm
  deg}^2$ ($\fsky = 0.06\,\%$) with $3\parcm5$ resolution at 150\,GHz
\citep{2014ApJ...794..171T}. For SPTpol, we have modelled the noise spectrum
from the characteristics reported in \citet{2013PhRvL.111n1301H}, by
considering the combination of the 95-GHz observation ($1\parcm83$
resolution and $25\,\mu\rm{K.arcmin}$ noise level) and the 150-GHz
observation ($1\parcm06$ resolution and $10\,\mu\rm{K.arcmin}$ noise
level) over a sky area of $100\,{\rm deg}^2$ ($\fsky = 0.24\,\%$). We have
used the same multipole binning as chosen by each experiment in
existing publications, and have added a low-$\ell$ bin
including multipoles up to the largest accessible angular scales
defined by the experiment's sky coverage. For POLARBEAR and SPTpol, we
have considered the $100 < \ell < 2000$ multipole range.

\subsubsection{Lensing $B$-mode band-power forecasts}
\begin{figure}[!h]
  \begin{center}
    \includegraphics[width=\columnwidth]{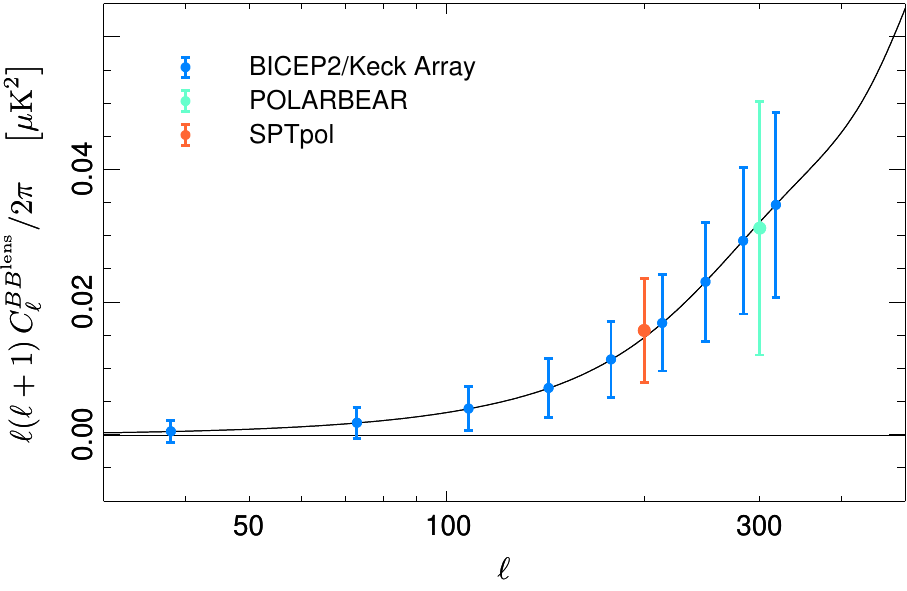}
   \caption{Forecasts of lensing $B$-mode band-powers from
cross-correlating the \Planck\ lensing $B$-mode template with external
data from $B$-mode targeted experiments. Blue circles show the
BICEP2/Keck Array band-powers forecasts using the same multipole
binning as that of BICEP2/Keck Array data points shown in
Fig.~\ref{Fig:external}. $BB^{\rm{lens}}$ band-power
measurements can be extended to an additional $\ell=100$--500 bin
using POLARBEAR (turquoise) and to an additional
$\ell=100$--300 bin using the SPTpol (dark orange) in cross-correlation
with the template.}
   \label{Fig:ext-xcorr}
  \end{center}
\end{figure}

Figure~\ref{Fig:ext-xcorr} shows the forecast-ed $BB^{\rm{lens}}$ 
band-powers over multipoles up to $500$, 
for which the \Planck\ template is the most useful in helping other
experiments to measure the lensing $B$-mode power spectrum. 
We find that the BICEP2/Keck Array could be used in combination with the
\Planck\ template to obtain $BB^{\rm{lens}}$ band-powers in the multipole range 
$\ell=20$--335 measured at a significance level of about $6\,\sigma$ 
(sensitivity to $A_{B\rm{lens}}$ of 0.17). \resub{
Therefore, our forecast-ed sensitivity to the lensing $B$-mode signal 
is comparable to that obtained in the BKP analysis, 
where a lensing amplitude of $A_{\rm lens} = 1.13 \pm 0.18$ has been measured. 
However, we show in Sect.~\ref{Sect:keytool:fisher} that we could improve 
the sensitivity to the lensing $B$-mode signal by adding the $B$-mode template 
to a BICEP2/Keck Array and \Planck\ joint analysis.} 
The \Planck\ template could enables sub-degree angular scale experiments, 
such as BICEP2/Keck Array, to probe the $\ell$-dependence of 
the lensing $B$-mode signal over their full multipole range, 
including multipoles at which the
lensing-induced signal is sub-dominant compared to other sources of
$B$-mode signal (such as polarized dust). 
Moreover, the cross-correlation with the template should allow experiments 
targeting higher multipoles (such as POLARBEAR or SPTpol) to measure the lensing
$B$-mode signal at intermediate angular scales ($100<\ell<300$),
extending their $BB^{\rm{lens}}$ estimates down to as low
multipoles as their sky coverage 
would permit.

\subsection{Direct delensing capabilities}
\label{Sect:keytool:direct} 
`Delensing' consists of subtracting the
lensing $B$-mode template from the $B$-mode map of an experiment in
order to try to highlight the primordial $B$-modes.  Because of the noise
level of the template, poor efficiency is expected from a direct
delensing approach \citep[e.g.][]{Marian2007}.  We have quantified
the expected impact on the tensor-to-scalar ratio uncertainty in 
calculating the improvement factor as defined in \citet{Smith2009} as
\begin{equation} \alpha = \frac{C_\ell^{B,{\rm lens}} +
N_\ell^{B}}{C_\ell^{B,{\rm res}} + N_\ell^{B}}.
\end{equation}
Here $C_\ell^{B,{\rm lens}}$ and $C_\ell^{B,{\rm res}}$
are the lensing $B$-mode power spectrum, and its residual after
subtraction of the template from the external data, while $N_\ell^{B}$
is the noise power spectrum of the experiment. We find a maximum
improvement factor (corresponding to $N_\ell^{B}=0$) of 5\,\% at $\ell
< 200$, reaching a maximum of 10\,\% around $\ell=300$, in agreement
with the expectations derived in \citet{Smith2009}.

\subsection{Improvement of the parameter accuracies}
\label{Sect:keytool:fisher} 
We now go beyond the lensing $B$-mode power spectrum measurement and
quantify whether the use of the
template as an additional data set (together with other $B$-mode data
and the \Planck\ dust template) can tighten cosmological parameter
constraints, and in particular, the amplitude of the lensing potential
power spectrum that scales the lensing-induced $B$-modes,
$A_{\rm{lens}}$.  We discuss whether more precise measurements of the
lensing scaling translate into tightened constraints on the
tensor-to-scalar ratio.

We derive forecasts for particular experiments by performing a Fisher
analysis\footnote{We note that in this approach, the
parameter likelihood is assumed to be Gaussian close to its maximum,
so that the parameter constraints can be derived by calculating the
distribution Hessian taken at the fiducial values of the
parameters. Owing to this assumption, parameter confidence contours are
ellipses.} \citep[for a description of this method, see
e.g.][]{1997Tegmark}. We consider a three-parameter model, $\{r,\,
A_{\rm{lens}}, \, A_{\rm{dust}}\}$, consisting of the tensor-to-scalar
ratio and the amplitudes of the lensing potential and polarized dust
power spectra. We compare the parameter constraints obtained in
two cases: (1) the data sets consist of $B^{\rm{exp}}$, the external
data from the $B$-mode targeted experiment, and $\hat{B}^{\rm{dust}}$,
the \Planck\ $B$-mode data at 353\,GHz, considered as a polarized dust
template; and (2) the \Planck\ lensing $B$-mode template,
$\hat{B}^{\rm{lens}}$, is used in combination with the other two data
sets. As examples of external experiments, we consider the
BICEP2/Keck Array combination, hereafter referred to as BK, described
in \citet{BK5}; and a wide sky coverage experiment, such as
LiteBIRD \citep{2014LiteBIRD}. The rationale driving these choices is
twofold.  Because of the lensing $B$-mode template noise level in a 1\,\%
sky area and from the results obtained above, we do not expect the
inclusion of the template to bring a large improvement of the
sensitivity to $r$ for BK. However, we will be able to validate our
simple fiducial analysis by comparing `Case 1' to the floating lensing
amplitude analysis described in BKP.
By contrast, a more substantial
improvement is expected for a large sky coverage $B$-mode experiment
such as LiteBIRD. 
For definiteness, we consider two different fiducial cosmologies in
what follows,
which consist of a $r=0.05$ model and $r=10^{-4}$ model.

\subsubsection{Fisher analysis} 
We consider the Fisher information
matrix of the form \citep[see e.g.][]{1997Tegmark}
\be
F_{ij} =
\sum_\ell \frac{1}{2}{(2\ell +1)}f_{\rm{sky}} \, {\rm
  Tr}[\tens{C}\tens{C}_{,i}\tens{C}\tens{C}_{,j}]
\ee
for a fiducial
data covariance matrix $\tens{C}$ and its derivatives with respect to
the parameters labelled by $i$ and $j$.  Given the data set $\{B_{\ell
m}^{\rm exp}, \hat{B}_{\ell m}^{\rm dust}, \hat{B}_{\ell m}^{\rm
lens}\}$, we need to model the $B$-mode auto-power spectra for the
external data, $C_\ell^{\rm exp}$, for the \Planck\ 353-GHz map,
$C_\ell^{353}$, and for the \Planck\ lensing $B$-mode template within
the experiment's sky coverage, $C_\ell^{\hat{B}{\rm lens}}$, as well
as the corresponding cross-power spectra.
 
The fiducial model consists of an extension of the six-parameter
$\Lambda$CDM model considered so far, including primordial
gravitational waves (GW) of amplitude $r$, a freely floating amplitude of the
lensing potential power spectrum, $A_{\rm lens}$, and a free polarized
dust amplitude in power $A_{\rm dust}$ (defined at the reference
frequency of 353\,GHz and at a multipole of $\ell=80$). In this
fiducial framework, the $B$-mode auto-power spectrum of the experiment
under consideration is 
\be 
C_\ell^{\rm exp} = \frac{r}{r_{\rm{fid}}}
C_\ell^{{\rm GW},r=r_{\rm{fid}}} + A_{\rm lens} C_\ell^{\rm lens} + \alpha
A_{\rm dust} R(\nu, 353)^2 C_\ell^{\rm dust} + N_\ell^{\rm exp}, 
\ee
where $C_\ell^{{\rm GW}, r=r_{\rm{fid}}}$ and $C_\ell^{\rm{lens}}$ are the gravitational
wave and lensing power spectra at the fiducial $r=r_{\rm{fid}}$ and $A_{\rm
lens}=1$ values. \resub{We have considered two different fiducial values for $r$, 
either $r_{\rm{fid}}=0.05$ or $r_{\rm{fid}}=10^{-4}$.} 
The dust power spectrum, $C_\ell^{\rm dust}$, and
its frequency scaling with the reference 353-GHz frequency, $R(\nu,
353)$ have been modelled as in Sect.~\ref{Sect:keytool:blensing}, following
PIP-XXX. In addition, for multi-band experiments with
foreground-cleaning capabilities, the dust power spectrum has been assumed
to be cleaned up to a residual level defined by the $\alpha$ factor
($\alpha = 1$ for a single frequency experiment). Finally,
$N_\ell^{\rm exp}$ is the $B$-mode noise power spectrum of each
experiment.

The \Planck\ 353-GHz $B$-mode auto-power spectrum in each
experiment's sky coverage has been modelled as \be C_\ell^{353} = A_{\rm
dust} C_\ell^{\rm dust} + N_\ell^{353}, \ee where the full-sky noise
power spectrum at 353\,GHz has been scaled to take the spatial
inhomogeneity within the experimental field into account. We have neglected the
sub-dominant CMB $B$-mode polarization signal.

Finally, the \Planck\ lensing $B$-mode template auto-power spectrum
within the experimental field has been analytically calculated using
Eq.~(\ref{Eq:autoclbblens}), in which 
the noise power spectra of the lensing potential $N_{L}^{\phi}$
and of the $E$-mode $N_{\ell}^{E}$ have been scaled to deal with the
spatial inhomogeneity within the experimental field. The
cross-correlation of the experiment's data with the 353-GHz dust
template and lensing $B$-mode template are $(\alpha)^{0.5} A_{\rm
dust} R(\nu, 353) C_\ell^{\rm dust}$ and
$C_\ell^{\rm{lens}}$. Following the assumptions of the model, we have 
neglected the sub-dominant cross-correlation of the lensing $B$-mode and
353-GHz dust templates.

\subsubsection{Examples of BK and LiteBIRD}
For existing data, we have used a data-driven model. Both the BK and the \Planck\ 353-GHz noise power
spectra in the BK field have been extrapolated from the noise band-powers
released along with the BKP
likelihood.\footnote{\url{http://bicepkeck.org/bkp2_2015_release.html}}.
For the future project LiteBIRD, we have modelled the noise power spectrum
from the foreseen instrumental characteristics defined in
\citet{2014LiteBIRD}. Only the 100- and 140-GHz bands have been considered,
the two lowest and two highest bands being discarded, assuming that
they are used for foreground cleaning.  Following
\citet{2014LiteBIRD}, the 100-GHz band reaches a depth of
$3.7\,\mu\rm{K.arcmin}$ and a resolution defined by a Gaussian beam of
$45\arcm$ FWHM, while the 140-GHz band has a noise level of
$4.7\,\mu\rm{K.arcmin}$ and $32\arcm$ FWHM. The polarized dust in each
frequency band has been assumed to be cleaned over an effective sky area of
63\,\% up to the same residual level as in the BK field. This
corresponds to a mildly conservative \citep{2009Dunkley} 17\,\%
residual level in the map domain (corresponding to $\alpha = 2.9\,\%$
in power) for the dust amplitude $A_{\rm dust} = 104.5\,\mu\rm{K}^2$,
obtained in PIP-XXX via a power-law fit in the large retained science
region `LR63'.

With the fiducial model established and using the numerical analysis
method described, we can infer the sensitivity to the parameters as
$\Delta \theta_i = \sqrt{(F^{-1})_{ii}}$ for $\theta_i \in \{ r,
A_{\rm lens}, A_{\rm dust}\}$. The results for BK and LiteBIRD \resub{for the 
two fiducial values of $r$ and} in the
two cases considered, depending on whether the lensing $B$-mode
template is used or not, are presented in Table~\ref{Tab:fisher}.
%
\begin{table}[tb]
\begingroup 
\newdimen\tblskip \tblskip=5pt
\caption{Fisher analysis inferred parameter uncertainties for the
BICEP2/Keck Array experiment and for the LiteBIRD project 
\resub{using the $r=0.05$ and $r=10^{-4}$ fiducial models.} 
Forecasts are given for two different data sets: `Case 1' consists of the
experiment's data and the \Planck\ 353-GHz map; and `Case 2' consists
of the same two ingredients plus the \Planck\ lensing $B$-mode
template.}  
\label{Tab:fisher} 
\nointerlineskip 
\vskip -3mm
\footnotesize 
\setbox\tablebox=\vbox{ 
  \newdimen\digitwidth
  \setbox0=\hbox{\rm 0} 
  \digitwidth=\wd0 
  \catcode`*=\active
  \def*{\kern\digitwidth}
  \newdimen\signwidth 
  \setbox0=\hbox{+} 
  \signwidth=\wd0
  \catcode`!=\active 
  \def!{\kern\signwidth}
  \halign{ 
    \hbox to 0.6in{#\leaderfil}\tabskip=2em&
    \hfil#\hfil&
    \hfil#\hfil& 
    \hfil#\hfil& 
    \hfil#\hfil\tabskip=0pt\cr 
\noalign{\doubleline} 
\omit&\multispan2 \hfil BICEP2/Keck$^{\rm a}$ \hfil& \multispan2 \hfil LiteBIRD$^{\rm b}$\hfil\cr
\noalign{\vskip -6pt}
\omit&\multispan2\hrulefill&\multispan2\hrulefill\cr
Parameter\hfil& Case 1& Case 2& Case 1& Case 2\cr 
\noalign{\vskip 4pt\hrule\vskip 5pt} 
\omit&\multispan4\hfil$r=0.05$\hfil\cr
\noalign{\vskip 3pt}
$\Delta r$& 0.031& 0.030& 0.0021*& 0.0018*\cr
$\Delta A_{\rm lens}$& 0.19*& 0.12*& 0.042**& 0.033**\cr 
$\Delta A_{\rm dust}$& 0.7**& 0.6**& 0.05***& 0.05***\cr 
\multispan5\hrulefill\hrulefill\cr
\omit&\multispan4\hfil$r=10^{-4}$\hfil\cr
\noalign{\vskip 3pt}
$\Delta r$& 0.025& 0.024& 0.00036& 0.00035\cr
$\Delta A_{\rm lens}$& 0.18*& 0.12*& 0.0205*& 0.0195*\cr 
$\Delta A_{\rm dust}$& 0.6**& 0.6**& 0.05***& 0.04***\cr 
\noalign{\vskip 5pt\hrule\vskip 4pt} 
}}
\endPlancktable 
\tablenote {{\rm a}} See \citet{BK5}. \par 
\tablenote {{\rm b}} See \citet{2014LiteBIRD}.\par
\endgroup
\end{table}

\begin{figure}[!h]
  \begin{center}
   \includegraphics[trim = {0 0 0.035cm 0}, clip=true]{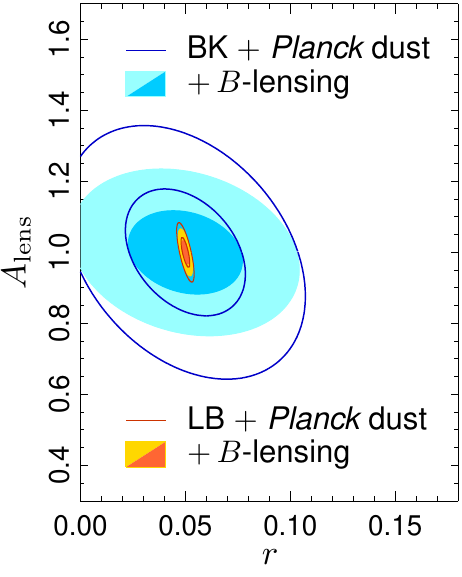}
   \includegraphics[trim = {0.16cm 0 0 0}]{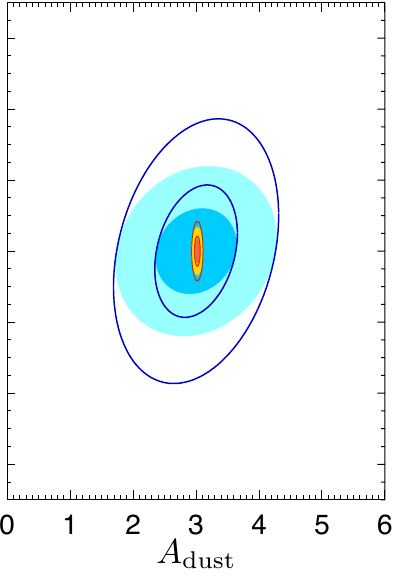}
   \includegraphics[trim = {0 0 0.035cm 0}, clip=true]{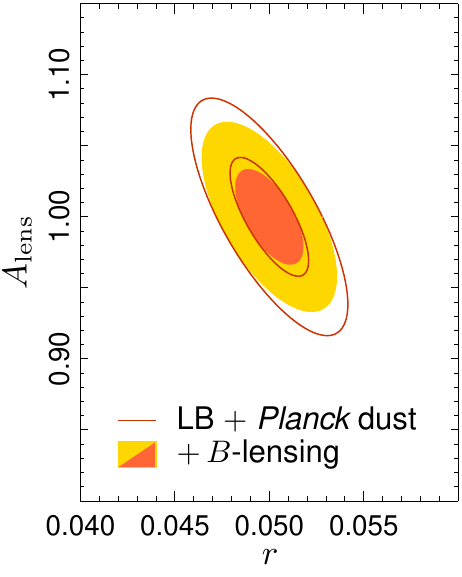}
   \includegraphics[trim = {0.3cm 0 0 0}]{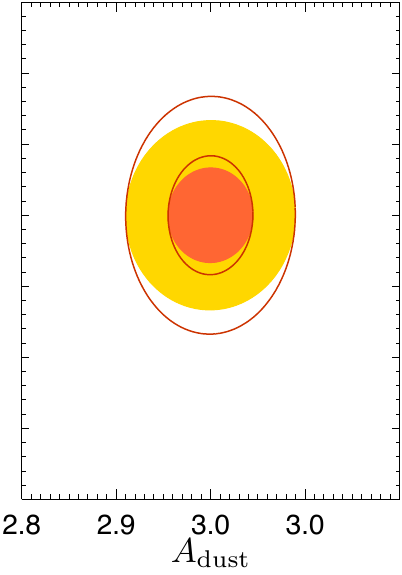}
   \caption{Constraints on the tensor-to-scalar ratio $r$ and the
amplitude of the polarized dust power $A_{\rm dust}$ within a model
with free lensing potential amplitude $A_{\rm{lens}}$. The
two-dimensional likelihood contours at 68\,\% and 95\,\% are forecast-ed for
BICEP2/Keck Array (BK, shades of blue) and for LiteBIRD (LB, shades of
red), in combination with the \Planck\ 353-GHz dust template only
(line contours) and the \Planck\ dust and lensing $B$-mode templates
(shaded contours). \resub{The lower panels show a zoom-in on the LiteBIRD
contours.}}
   \label{Fig:fisher}
  \end{center}
\end{figure}

For the BK analysis in Case~1 (without the lensing template) 
\resub{and using the $r=0.05$ fiducial model}, we find
$\Delta r = 0.031$, $\Delta A_{\rm lens} = 0.186$ and $\Delta A_{\rm
dust} = 0.7$, in agreement with results reported in BKP:
$r=0.048_{-0.032}^{+0.035}$, $A_{\rm lens} = 1.13 \pm 0.18$ (from the
free lensing amplitude extended analysis) and $A_{\rm dust} =
3.3_{-0.8}^{+0.9}$.  This indicates that our fiducial Fisher analysis
yields reliable parameter uncertainty estimates despite the underlying
simplifying assumptions. The use of the lensing template translates
into a 5\,\% improvement of the $r$ constraint, whereas the constraint
on $A_{\rm{lens}}$ is tightened by 36\,\%. 
\resub{Similar results are found using the more pessimistic
$r=10^{-4}$ fiducial model; Case~2 yields a $5\,\%$ improvement 
of the constraint on $r$ and a $35\,\%$ improvement of the 
constraint on $A_{\rm{lens}}$, compared to Case~1.} 
Over the multipole range covered by BK,
$A_{\rm{lens}}$ is weakly correlated with both $r$ and
$A_{\rm{dust}}$, as was noted in the BKP paper, so that the
improvement in the $A_{\rm lens}$ constraint translates into modest
improvement of the $r$ or $A_{\rm{dust}}$ constraints. This is
verified in Fig.~\ref{Fig:fisher}, where the two-dimensional contours
at 68\,\% and 95\,\% are shown in the $A_{\rm lens}$--$r$ and $A_{\rm
lens}$--$A_{\rm dust}$ planes.  However, Fig.~\ref{Fig:fisher} also
shows that the $A_{\rm lens}$--$r$ correlation is further reduced when
the lensing $B$-mode template is used, leading to more robust
constraints.

For large angular scale experiments whose multipole coverage is
limited to $\ell \la 200$, such as LiteBIRD, 
the implication of the template slightly differs.  
\resub{For the $r=10^{-4}$ fiducial model, the secondary 
$B$-modes, which fully dominate over the primary $B$-modes, 
are precisely measured even without the help of the template. 
In the $r=0.05$ fiducial model, however,} 
the lack of measurement over a lensing $B$-mode dominated multipole 
range increases the parameter degeneracy, 
in particular between $A_{\rm lens}$ and $r$,
 as seen in Fig.~{\ref{Fig:fisher}}. As a result, the 20\,\% improvement 
on the $A_{\rm lens}$ uncertainties, which arises from using the \Planck\ 
lensing $B$-mode template, translates into a 15\,\% improvement on the $r$
uncertainties.

\section{Conclusions}
\label{conclusions}

We have produced a nearly all-sky template of the CMB secondary
$B$-modes using the \Planck\ full-mission foreground-cleaned CMB
temperature and Stokes parameter maps.  For this purpose, we have
developed a dedicated pipeline that has been verified via specific
simulations. We show that the constructed template includes the
lensing $B$-mode contribution at all angular scales covered by
\Planck\ and shows no contamination from primordial $B$-modes. This
template has been used to compute the CMB lensing $B$-mode power
spectrum by cross-correlating it with the total foreground-cleaned
\Planck\ polarization $B$-mode maps (via the publicly available $Q$
and $U$ maps). We find that the resulting CMB lensing $B$-mode power
spectrum is insensitive to foreground contamination and independent of
the choice of the foreground-cleaned \Planck\ polarization $B$-mode
map used for the cross-correlation analysis. Furthermore, we find that
the results are in good agreement with the expected CMB lensing
$B$-mode power spectrum computed using the baseline \Planck\ 2015
best-fitting $\Lambda$-CDM model.  We obtain a 12$\,\sigma$ detection
of the lensing $B$-modes, in agreement with the results in the
companion \cite{planck2014-a17} paper.

\Planck\ provides a unique nearly all-sky lensing $B$-mode template,
containing all the lens-induced information from intermediate to large
angular scales.  This template, which is included as part of the
\Planck\ 2015 data release, will be a useful tool for current and
future ground-based experiments targeting the measurement of the
primordial CMB $B$-mode power spectrum. Indeed, this template can be
used to obtain a reliable measurement of the lensing
$B$-mode power spectrum with future experiments or to improve the
precision with which they can detect the lensing $B$-modes
in their own data, by tightening the constraints on the lensing
amplitude. This, in turn, can help in the much more challenging
endeavour of constraining the tensor-to-scalar ratio.

\begin{acknowledgements} The Planck Collaboration acknowledges the
support of: ESA; CNES, and CNRS/INSU-IN2P3-INP (France); ASI, CNR, and
INAF (Italy); NASA and DoE (USA); STFC and UKSA (UK); CSIC, MINECO, JA
and RES (Spain); Tekes, AoF, and CSC (Finland); DLR and MPG (Germany);
CSA (Canada); DTU Space (Denmark); SER/SSO (Switzerland); RCN
(Norway); SFI (Ireland); FCT/MCTES (Portugal); ERC and PRACE (EU). A
description of the Planck Collaboration and a list of its members,
indicating which technical or scientific activities they have been
involved in, can be found at
\url{http://www.cosmos.esa.int/web/planck}. This paper made use of
the {\tt HEALPix} software package. We thank the anonymous referee for their
helpful comments and thoughtful suggestions that contributed to improve this paper.
\end{acknowledgements}

\bibliographystyle{aat}

\bibliography{B_lensing_bib,Planck_bib}

\def\eprinttmppp@#1arXiv:@{#1}
\providecommand{\arxivlink[1]}{\href{http://arxiv.org/abs/#1}{arXiv:#1}}
\def\eprinttmp@#1arXiv:#2 [#3]#4@{\ifthenelse{\equal{#3}{x}}{\ifthenelse{
\equal{#1}{}}{\arxivlink{\eprinttmppp@#2@}}{\arxivlink{#1}}}{\arxivlink{#2}
  [#3]}}
\providecommand{\eprintlink}[1]{\eprinttmp@#1arXiv: [x]@}
\providecommand{\eprint}[1]{\eprintlink{#1}}
\providecommand{\adsurl}[1]{\href{#1}{ADS}}
\begin{thebibliography}{74}
\expandafter\ifx\csname natexlab\endcsname\relax\def\natexlab#1{#1}\fi

\bibitem[{{Abrial} {et~al.}(2007){Abrial}, {Moudden}, {Starck}, {Bobin},
  {Fadili}, {Afeyan}, \& {Nguyen}}]{inpainting:abrial06}
{Abrial}, P., {Moudden}, Y., {Starck}, J., {et~al.}, {Morphological Component
  Analysis and Inpainting on the sphere: Application in Physics and
  Astrophysics}. 2007, J. Fourier Anal. and Applic., 13, 729

\bibitem[{{Ade} {et~al.}(2014{\natexlab{a}}){Ade}, {Aikin}, {Barkats},
  {Benton}, {Bischoff}, {Bock}, {Brevik}, {Buder}, {Bullock}, {Dowell},
  {Duband}, {Filippini}, {Fliescher}, {Golwala}, {Halpern}, {Hasselfield},
  {Hildebrandt}, {Hilton}, {Hristov}, {Irwin}, {Karkare}, {Kaufman}, {Keating},
  {Kernasovskiy}, {Kovac}, {Kuo}, {Leitch}, {Lueker}, {Mason}, {Netterfield},
  {Nguyen}, {O'Brient}, {Ogburn}, {Orlando}, {Pryke}, {Reintsema}, {Richter},
  {Schwarz}, {Sheehy}, {Staniszewski}, {Sudiwala}, {Teply}, {Tolan}, {Turner},
  {Vieregg}, {Wong}, {Yoon}, \& {Bicep2 Collaboration}}]{2014PhRvL.112x1101A}
{Ade}, P.~A.~R., {Aikin}, R.~W., {Barkats}, D., {et~al.}, {Detection of B-Mode
  Polarization at Degree Angular Scales by BICEP2}. 2014{\natexlab{a}}, \prl,
  112, 241101, \eprint{1403.3985}

\bibitem[{{Ade} {et~al.}(2014{\natexlab{b}}){Ade}, {Akiba}, {Anthony},
  {Arnold}, {Atlas}, {Barron}, {Boettger}, {Borrill}, {Chapman}, {Chinone},
  {Dobbs}, {Elleflot}, {Errard}, {Fabbian}, {Feng}, {Flanigan}, {Gilbert},
  {Grainger}, {Halverson}, {Hasegawa}, {Hattori}, {Hazumi}, {Holzapfel},
  {Hori}, {Howard}, {Hyland}, {Inoue}, {Jaehnig}, {Jaffe}, {Keating},
  {Kermish}, {Keskitalo}, {Kisner}, {Le Jeune}, {Lee}, {Linder}, {Leitch},
  {Lungu}, {Matsuda}, {Matsumura}, {Meng}, {Miller}, {Morii}, {Moyerman},
  {Myers}, {Navaroli}, {Nishino}, {Paar}, {Peloton}, {Quealy}, {Rebeiz},
  {Reichardt}, {Richards}, {Ross}, {Schanning}, {Schenck}, {Sherwin},
  {Shimizu}, {Shimmin}, {Shimon}, {Siritanasak}, {Smecher}, {Spieler},
  {Stebor}, {Steinbach}, {Stompor}, {Suzuki}, {Takakura}, {Tomaru}, {Wilson},
  {Yadav}, {Zahn}, \& {Polarbear Collaboration}}]{2014PhRvL.113b1301A}
{Ade}, P.~A.~R., {Akiba}, Y., {Anthony}, A.~E., {et~al.}, {Measurement of the
  Cosmic Microwave Background Polarization Lensing Power Spectrum with the
  POLARBEAR Experiment}. 2014{\natexlab{b}}, \prl, 113, 021301

\bibitem[{{Benoit-L{\'e}vy} {et~al.}(2013){Benoit-L{\'e}vy}, {D{\'e}chelette},
  {Benabed}, {Cardoso}, {Hanson}, \& {Prunet}}]{aurelien}
{Benoit-L{\'e}vy}, A., {D{\'e}chelette}, T., {Benabed}, K., {et~al.}, {Full-sky
  CMB lensing reconstruction in presence of sky-cuts}. 2013, \aap, 555, A37,
  \eprint{1301.4145}

\bibitem[{{Bernardeau}(1997)}]{1997A&A...324...15B}
{Bernardeau}, F., {Weak lensing detection in CMB maps.} 1997, \aap, 324, 15,
  \eprint{astro-ph/9611012}

\bibitem[{{BICEP2 and Keck Array Collaborations}
  {et~al.}(2015{\natexlab{a}}){BICEP2 and Keck Array Collaborations}, {Ade},
  {Ahmed}, {Aikin}, {Alexander}, {Barkats}, {Benton}, {Bischoff}, {Bock},
  {Brevik}, {Buder}, {Bullock}, {Buza}, {Connors}, {Crill}, {Dowell},
  {Dvorkin}, {Duband}, {Filippini}, {Fliescher}, {Golwala}, {Halpern},
  {Harrison}, {Hasselfield}, {Hildebrandt}, {Hilton}, {Hristov}, {Hui},
  {Irwin}, {Karkare}, {Kaufman}, {Keating}, {Kefeli}, {Kernasovskiy}, {Kovac},
  {Kuo}, {Leitch}, {Lueker}, {Mason}, {Megerian}, {Netterfield}, {Nguyen},
  {O'Brient}, {Ogburn}, {Orlando}, {Pryke}, {Reintsema}, {Richter}, {Schwarz},
  {Sheehy}, {Staniszewski}, {Sudiwala}, {Teply}, {Thompson}, {Tolan}, {Turner},
  {Vieregg}, {Weber}, {Willmert}, {Wong}, \& {Yoon}}]{BK5}
{BICEP2 and Keck Array Collaborations}, {Ade}, P.~A.~R., {Ahmed}, Z., {et~al.},
  {BICEP2/Keck Array V: Measurements of B-mode Polarization at Degree Angular
  Scales and 150 GHz by the Keck Array}. 2015{\natexlab{a}}, \apj, 811, 126,
  \eprint{1502.00643}

\bibitem[{{BICEP2 and Keck Array Collaborations}
  {et~al.}(2015{\natexlab{b}}){BICEP2 and Keck Array Collaborations}, {Ade},
  {Aikin}, {Barkats}, {Benton}, {Bischoff}, {Bock}, {Bradford}, {Brevik},
  {Buder}, {Bullock}, {Dowell}, {Duband}, {Filippini}, {Fliescher}, {Golwala},
  {Halpern}, {Hasselfield}, {Hildebrandt}, {Hilton}, {Hui}, {Irwin}, {Kang},
  {Karkare}, {Kaufman}, {Keating}, {Kefeli}, {Kernasovskiy}, {Kovac}, {Kuo},
  {Leitch}, {Lueker}, {Megerian}, {Netterfield}, {Nguyen}, {O'Brient},
  {Ogburn}, {Orlando}, {Pryke}, {Richter}, {Schwarz}, {Sheehy}, {Staniszewski},
  {Sudiwala}, {Teply}, {Thompson}, {Tolan}, {Turner}, {Vieregg}, {Weber},
  {Wong}, \& {Yoon}}]{BK4}
{BICEP2 and Keck Array Collaborations}, {Ade}, P.~A.~R., {Aikin}, R.~W.,
  {et~al.}, {BICEP2/Keck Array. IV. Optical Characterization and Performance of
  the BICEP2 and Keck Array Experiments}. 2015{\natexlab{b}}, \apj, 806, 206,
  \eprint{1502.00596}

\bibitem[{{BICEP2/Keck Array and Planck Collaborations}(2015)}]{pb2015}
{BICEP2/Keck Array and Planck Collaborations}, {Joint Analysis of BICEP2/Keck
  Array and Planck Data}. 2015, \prl, 114, 101301, \eprint{1502.00612}

\bibitem[{{Blanchard} \& {Schneider}(1987)}]{1987A&A...184....1B}
{Blanchard}, A. \& {Schneider}, J., {Gravitational lensing effect on the
  fluctuations of the cosmic background radiation}. 1987, \aap, 184, 1

\bibitem[{{Delabrouille} {et~al.}(2003){Delabrouille}, {Cardoso}, \&
  {Patanchon}}]{2003MNRAS.346.1089D}
{Delabrouille}, J., {Cardoso}, J.-F., \& {Patanchon}, G., {Multidetector
  multicomponent spectral matching and applications for cosmic microwave
  background data analysis}. 2003, \mnras, 346, 1089, \eprint{astro-ph/0211504}

\bibitem[{{Dunkley} {et~al.}(2009){Dunkley}, {Amblard}, {Baccigalupi},
  {Betoule}, {Chuss}, {Cooray}, {Delabrouille}, {Dickinson}, {Dobler},
  {Dotson}, {Eriksen}, {Finkbeiner}, {Fixsen}, {Fosalba}, {Fraisse}, {Hirata},
  {Kogut}, {Kristiansen}, {Lawrence}, {Magalha\~{}Es}, {Miville-Deschenes},
  {Meyer}, {Miller}, {Naess}, {Page}, {Peiris}, {Phillips}, {Pierpaoli},
  {Rocha}, {Vaillancourt}, \& {Verde}}]{2009Dunkley}
{Dunkley}, J., {Amblard}, A., {Baccigalupi}, C., {et~al.} 2009, in American
  Institute of Physics Conference Series, Vol. 1141, American Institute of
  Physics Conference Series, ed. S.~{Dodelson}, D.~{Baumann}, A.~{Cooray},
  J.~{Dunkley}, A.~{Fraisse}, M.~G. {Jackson}, A.~{Kogut}, L.~{Krauss},
  M.~{Zaldarriaga}, \& K.~{Smith}, 222--264

\bibitem[{{Fabbian} \& {Stompor}(2013)}]{Fabbian2013}
{Fabbian}, G. \& {Stompor}, R., {High-precision simulations of the weak lensing
  effect on cosmic microwave background polarization}. 2013, \aap, 556, A109,
  \eprint{1303.6550}

\bibitem[{{G{\'o}rski} {et~al.}(2005){G{\'o}rski}, {Hivon}, {Banday},
  {Wandelt}, {Hansen}, {Reinecke}, \& {Bartelmann}}]{gorski2005}
{G{\'o}rski}, K.~M., {Hivon}, E., {Banday}, A.~J., {et~al.}, {HEALPix: A
  Framework for High-Resolution Discretization and Fast Analysis of Data
  Distributed on the Sphere}. 2005, \apj, 622, 759, \eprint{astro-ph/0409513}

\bibitem[{{Grishchuk}(1975)}]{1975JETP...40..409G}
{Grishchuk}, L.~P., {Amplification of gravitational waves in an isotropic
  universe}. 1975, Sov. J. Exp. Theor. Phys., 40, 409

\bibitem[{{Guth}(1981)}]{1981PhRvD..23..347G}
{Guth}, A.~H., {Inflationary universe: A possible solution to the horizon and
  flatness problems}. 1981, \prd, 23, 347

\bibitem[{{Hanson} {et~al.}(2013){Hanson}, {Hoover}, {Crites}, {Ade}, {Aird},
  {Austermann}, {Beall}, {Bender}, {Benson}, {Bleem}, {Bock}, {Carlstrom},
  {Chang}, {Chiang}, {Cho}, {Conley}, {Crawford}, {de Haan}, {Dobbs},
  {Everett}, {Gallicchio}, {Gao}, {George}, {Halverson}, {Harrington},
  {Henning}, {Hilton}, {Holder}, {Holzapfel}, {Hrubes}, {Huang}, {Hubmayr},
  {Irwin}, {Keisler}, {Knox}, {Lee}, {Leitch}, {Li}, {Liang}, {Luong-Van},
  {Marsden}, {McMahon}, {Mehl}, {Meyer}, {Mocanu}, {Montroy}, {Natoli},
  {Nibarger}, {Novosad}, {Padin}, {Pryke}, {Reichardt}, {Ruhl}, {Saliwanchik},
  {Sayre}, {Schaffer}, {Schulz}, {Smecher}, {Stark}, {Story}, {Tucker},
  {Vanderlinde}, {Vieira}, {Viero}, {Wang}, {Yefremenko}, {Zahn}, \&
  {Zemcov}}]{2013PhRvL.111n1301H}
{Hanson}, D., {Hoover}, S., {Crites}, A., {et~al.}, {Detection of B-Mode
  Polarization in the Cosmic Microwave Background with Data from the South Pole
  Telescope}. 2013, \prl, 111, 141301, \eprint{1307.5830}

\bibitem[{{Hu}(2000)}]{2000PhRvD..62d3007H}
{Hu}, W., {Weak lensing of the CMB: A harmonic approach}. 2000, \prd, 62,
  043007, \eprint{astro-ph/0001303}

\bibitem[{{Hu}(2001{\natexlab{a}})}]{2001PhRvD..64h3005H}
{Hu}, W., {Angular trispectrum of the cosmic microwave background}.
  2001{\natexlab{a}}, \prd, 64, 083005, \eprint{astro-ph/0105117}

\bibitem[{{Hu}(2001{\natexlab{b}})}]{2001ApJ...557L..79H}
{Hu}, W., {Mapping the Dark Matter through the Cosmic Microwave Background
  Damping Tail}. 2001{\natexlab{b}}, \apjl, 557, L79, \eprint{astro-ph/0105424}

\bibitem[{{Hu} \& {Okamoto}(2002)}]{2002ApJ...574..566H}
{Hu}, W. \& {Okamoto}, T., {Mass Reconstruction with Cosmic Microwave
  Background Polarization}. 2002, \apj, 574, 566, \eprint{astro-ph/0111606}

\bibitem[{{Kamionkowski} {et~al.}(1997){Kamionkowski}, {Kosowsky}, \&
  {Stebbins}}]{1997PhRvL..78.2058K}
{Kamionkowski}, M., {Kosowsky}, A., \& {Stebbins}, A., {A Probe of Primordial
  Gravity Waves and Vorticity}. 1997, \prl, 78, 2058, \eprint{astro-ph/9609132}

\bibitem[{{Keisler} {et~al.}(2015){Keisler}, {Hoover}, {Harrington}, {Henning},
  {Ade}, {Aird}, {Austermann}, {Beall}, {Bender}, {Benson}, {Bleem},
  {Carlstrom}, {Chang}, {Chiang}, {Cho}, {Citron}, {Crawford}, {Crites}, {de
  Haan}, {Dobbs}, {Everett}, {Gallicchio}, {Gao}, {George}, {Gilbert},
  {Halverson}, {Hanson}, {Hilton}, {Holder}, {Holzapfel}, {Hou}, {Hrubes},
  {Huang}, {Hubmayr}, {Irwin}, {Knox}, {Lee}, {Leitch}, {Li}, {Luong-Van},
  {Marrone}, {McMahon}, {Mehl}, {Meyer}, {Mocanu}, {Natoli}, {Nibarger},
  {Novosad}, {Padin}, {Pryke}, {Reichardt}, {Ruhl}, {Saliwanchik}, {Sayre},
  {Schaffer}, {Shirokoff}, {Smecher}, {Stark}, {Story}, {Tucker},
  {Vanderlinde}, {Vieira}, {Wang}, {Whitehorn}, {Yefremenko}, \&
  {Zahn}}]{2015ApJ...807..151K}
{Keisler}, R., {Hoover}, S., {Harrington}, N., {et~al.}, {Measurements of
  Sub-degree B-mode Polarization in the Cosmic Microwave Background from 100
  Square Degrees of SPTpol Data}. 2015, \apj, 807, 151, \eprint{1503.02315}

\bibitem[{{Knox}(1995)}]{1995PhRvD..52.4307K}
{Knox}, L., {Determination of inflationary observables by cosmic microwave
  background anisotropy experiments}. 1995, \prd, 52, 4307,
  \eprint{astro-ph/9504054}

\bibitem[{{Lewis} \& {Challinor}(2006)}]{2006PhR...429....1L}
{Lewis}, A. \& {Challinor}, A., {Weak gravitational lensing of the CMB}. 2006,
  \physrep, 429, 1, \eprint{astro-ph/0601594}

\bibitem[{{Linde}(1982)}]{1982PhLB..108..389L}
{Linde}, A.~D., {A new inflationary universe scenario: A possible solution of
  the horizon, flatness, homogeneity, isotropy and primordial monopole
  problems}. 1982, Phys. Lett. B, 108, 389

\bibitem[{{Marian} \& {Bernstein}(2007)}]{Marian2007}
{Marian}, L. \& {Bernstein}, G.~M., {Detectability of CMB tensor B modes via
  delensing with weak lensing galaxy surveys}. 2007, \prd, 76, 123009,
  \eprint{0710.2538}

\bibitem[{{Matsumura} {et~al.}(2014){Matsumura}, {Akiba}, {Borrill}, {Chinone},
  {Dobbs}, {Fuke}, {Ghribi}, {Hasegawa}, {Hattori}, {Hattori}, {Hazumi},
  {Holzapfel}, {Inoue}, {Ishidoshiro}, {Ishino}, {Ishitsuka}, {Karatsu},
  {Katayama}, {Kawano}, {Kibayashi}, {Kibe}, {Kimura}, {Kimura}, {Koga},
  {Kozu}, {Komatsu}, {Lee}, {Matsuhara}, {Mima}, {Mitsuda}, {Mizukami},
  {Morii}, {Morishima}, {Murayama}, {Nagai}, {Nagata}, {Nakamura}, {Naruse},
  {Natsume}, {Nishibori}, {Nishino}, {Noda}, {Noguchi}, {Ogawa}, {Oguri},
  {Ohta}, {Otani}, {Richards}, {Sakai}, {Sato}, {Sato}, {Sekimoto}, {Shimizu},
  {Shinozaki}, {Sugita}, {Suzuki}, {Suzuki}, {Tajima}, {Takada}, {Takakura},
  {Takei}, {Tomaru}, {Uzawa}, {Wada}, {Watanabe}, {Yoshida}, {Yamasaki},
  {Yoshida}, \& {Yotsumoto}}]{2014LiteBIRD}
{Matsumura}, T., {Akiba}, Y., {Borrill}, J., {et~al.}, {Mission Design of
  LiteBIRD}. 2014, J. Low Temp. Phys., 176, 733, \eprint{1311.2847}

\bibitem[{{Namikawa} \& {Nagata}(2014)}]{Namikawa2014}
{Namikawa}, T. \& {Nagata}, R., {Lensing reconstruction from a patchwork of
  polarization maps}. 2014, \jcap, 9, 009, \eprint{1405.6568}

\bibitem[{{Okamoto} \& {Hu}(2003)}]{2003PhRvD..67h3002O}
{Okamoto}, T. \& {Hu}, W., {Cosmic microwave background lensing reconstruction
  on the full sky}. 2003, \prd, 67, 083002, \eprint{astro-ph/0301031}

\bibitem[{{Perotto} {et~al.}(2010){Perotto}, {Bobin}, {Plaszczynski}, {Starck},
  \& {Lavabre}}]{2010AA...519A...4P}
{Perotto}, L., {Bobin}, J., {Plaszczynski}, S., {Starck}, J.-L., \& {Lavabre},
  A., {Reconstruction of the cosmic microwave background lensing for Planck}.
  2010, \aap, 519, A4

\bibitem[{{\sorthelp{Planck Collaboration 2011G}}{Planck Collaboration
  VII}(2011)}]{planck2011-1.10}
{\sorthelp{Planck Collaboration 2011G}}{Planck Collaboration VII},
  {\textit{Planck} early results. VII. The Early Release Compact Source
  Catalogue}. 2011, \aap, 536, A7, \eprint{1101.2041}

\bibitem[{{\sorthelp{Planck Collaboration 2011R}}{Planck Collaboration
  XVIII}(2011)}]{planck2011-6.6}
{\sorthelp{Planck Collaboration 2011R}}{Planck Collaboration XVIII},
  {\textit{Planck} early results. XVIII. The power spectrum of cosmic infrared
  background anisotropies}. 2011, \aap, 536, A18, \eprint{1101.2028}

\bibitem[{{\sorthelp{Planck Collaboration 2014L}}{Planck Collaboration
  XII}(2014)}]{planck2013-p06}
{\sorthelp{Planck Collaboration 2014L}}{Planck Collaboration XII},
  {\textit{Planck} 2013 results. XII. Diffuse component separation}. 2014,
  \aap, 571, A12, \eprint{1303.5072}

\bibitem[{{\sorthelp{Planck Collaboration 2014M}}{Planck Collaboration
  XIII}(2014)}]{planck2013-p03a}
{\sorthelp{Planck Collaboration 2014M}}{Planck Collaboration XIII},
  {\textit{Planck} 2013 results. XIII. Galactic CO emission}. 2014, \aap, 571,
  A13, \eprint{1303.5073}

\bibitem[{{\sorthelp{Planck Collaboration 2014P}}{Planck Collaboration
  XVI}(2014)}]{planck2013-p11}
{\sorthelp{Planck Collaboration 2014P}}{Planck Collaboration XVI},
  {\textit{Planck} 2013 results. XVI. Cosmological parameters}. 2014, \aap,
  571, A16, \eprint{1303.5076}

\bibitem[{{\sorthelp{Planck Collaboration 2014Q}}{Planck Collaboration
  XVII}(2014)}]{planck2013-p12}
{\sorthelp{Planck Collaboration 2014Q}}{Planck Collaboration XVII},
  {\textit{Planck} 2013 results. XVII. Gravitational lensing by large-scale
  structure}. 2014, \aap, 571, A17, \eprint{1303.5077}

\bibitem[{{\sorthelp{Planck Collaboration 2014R}}{Planck Collaboration
  XVIII}(2014)}]{planck2013-p13}
{\sorthelp{Planck Collaboration 2014R}}{Planck Collaboration XVIII},
  {\textit{Planck} 2013 results. XVIII. The gravitational lensing-infrared
  background correlation}. 2014, \aap, 571, A18, \eprint{1303.5078}

\bibitem[{{\sorthelp{Planck Collaboration 2014ZC}}{Planck Collaboration
  XXVIII}(2014)}]{planck2013-p05}
{\sorthelp{Planck Collaboration 2014ZC}}{Planck Collaboration XXVIII},
  {\textit{Planck} 2013 results. XXVIII. The Planck Catalogue of Compact
  Sources}. 2014, \aap, 571, A28, \eprint{1303.5088}

\bibitem[{{\sorthelp{Planck Collaboration 2014ZD}}{Planck Collaboration
  XXIX}(2014)}]{planck2013-p05a}
{\sorthelp{Planck Collaboration 2014ZD}}{Planck Collaboration XXIX},
  {\textit{Planck} 2013 results. XXIX. The Planck catalogue of
  Sunyaev-Zeldovich sources}. 2014, \aap, 571, A29, \eprint{1303.5089}

\bibitem[{{\sorthelp{Planck Collaboration 2015A}}{Planck Collaboration
  I}(2016)}]{planck2014-a01}
{\sorthelp{Planck Collaboration 2015A}}{Planck Collaboration I},
  {\textit{Planck} 2015 results. I. Overview of products and results}. 2016,
  \aap, in press, \eprint{1502.01582}

\bibitem[{{\sorthelp{Planck Collaboration 2015B}}{Planck Collaboration
  II}(2016)}]{planck2014-a03}
{\sorthelp{Planck Collaboration 2015B}}{Planck Collaboration II},
  {\textit{Planck} 2015 results. II. Low Frequency Instrument data processing}.
  2016, \aap, in press, \eprint{1502.01583}

\bibitem[{{\sorthelp{Planck Collaboration 2015C}}{Planck Collaboration
  III}(2016)}]{planck2014-a04}
{\sorthelp{Planck Collaboration 2015C}}{Planck Collaboration III},
  {\textit{Planck} 2015 results. III. LFI systematic uncertainties}. 2016,
  \aap, in press, \eprint{1507.08853}

\bibitem[{{\sorthelp{Planck Collaboration 2015D}}{Planck Collaboration
  IV}(2016)}]{planck2014-a05}
{\sorthelp{Planck Collaboration 2015D}}{Planck Collaboration IV},
  {\textit{Planck} 2015 results. IV. LFI beams and window functions}. 2016,
  \aap, in press, \eprint{1502.01584}

\bibitem[{{\sorthelp{Planck Collaboration 2015E}}{Planck Collaboration
  V}(2016)}]{planck2014-a06}
{\sorthelp{Planck Collaboration 2015E}}{Planck Collaboration V},
  {\textit{Planck} 2015 results. V. LFI calibration}. 2016, \aap, in press,
  \eprint{1505.08022}

\bibitem[{{\sorthelp{Planck Collaboration 2015F}}{Planck Collaboration
  VI}(2016)}]{planck2014-a07}
{\sorthelp{Planck Collaboration 2015F}}{Planck Collaboration VI},
  {\textit{Planck} 2015 results. VI. LFI maps}. 2016, \aap, in press,
  \eprint{1502.01585}

\bibitem[{{\sorthelp{Planck Collaboration 2015G}}{Planck Collaboration
  VII}(2016)}]{planck2014-a08}
{\sorthelp{Planck Collaboration 2015G}}{Planck Collaboration VII},
  {\textit{Planck} 2015 results. VII. High Frequency Instrument data
  processing: Time-ordered information and beam processing}. 2016, \aap, in
  press, \eprint{1502.01586}

\bibitem[{{\sorthelp{Planck Collaboration 2015H}}{Planck Collaboration
  VIII}(2016)}]{planck2014-a09}
{\sorthelp{Planck Collaboration 2015H}}{Planck Collaboration VIII},
  {\textit{Planck} 2015 results. VIII. High Frequency Instrument data
  processing: Calibration and maps}. 2016, \aap, in press, \eprint{1502.01587}

\bibitem[{{\sorthelp{Planck Collaboration 2015I}}{Planck Collaboration
  IX}(2016)}]{planck2014-a11}
{\sorthelp{Planck Collaboration 2015I}}{Planck Collaboration IX},
  {\textit{Planck} 2015 results. IX. Diffuse component separation: CMB maps}.
  2016, \aap, in press, \eprint{1502.05956}

\bibitem[{{\sorthelp{Planck Collaboration 2015J}}{Planck Collaboration
  X}(2016)}]{planck2014-a12}
{\sorthelp{Planck Collaboration 2015J}}{Planck Collaboration X},
  {\textit{Planck} 2015 results. X. Diffuse component separation: Foreground
  maps}. 2016, \aap, in press, \eprint{1502.01588}

\bibitem[{{\sorthelp{Planck Collaboration 2015K}}{Planck Collaboration
  XI}(2016)}]{planck2014-a13}
{\sorthelp{Planck Collaboration 2015K}}{Planck Collaboration XI},
  {\textit{Planck} 2015 results. XI. CMB power spectra, likelihoods, and
  robustness of parameters}. 2016, \aap, in press, \eprint{1507.02704}

\bibitem[{{\sorthelp{Planck Collaboration 2015L}}{Planck Collaboration
  XII}(2016)}]{planck2014-a14}
{\sorthelp{Planck Collaboration 2015L}}{Planck Collaboration XII},
  {\textit{Planck} 2015 results. XII. Full Focal Plane simulations}. 2016,
  \aap, in press, \eprint{1509.06348}

\bibitem[{{\sorthelp{Planck Collaboration 2015M}}{Planck Collaboration
  XIII}(2016)}]{planck2014-a15}
{\sorthelp{Planck Collaboration 2015M}}{Planck Collaboration XIII},
  {\textit{Planck} 2015 results. XIII. Cosmological parameters}. 2016, \aap, in
  press, \eprint{1502.01589}

\bibitem[{{\sorthelp{Planck Collaboration 2015O}}{Planck Collaboration
  XV}(2016)}]{planck2014-a17}
{\sorthelp{Planck Collaboration 2015O}}{Planck Collaboration XV},
  {\textit{Planck} 2015 results. XV. Gravitational lensing}. 2016, \aap, in
  press, \eprint{1502.01591}

\bibitem[{{\sorthelp{Planck Collaboration 2015T}}{Planck Collaboration
  XX}(2016)}]{planck2014-a24}
{\sorthelp{Planck Collaboration 2015T}}{Planck Collaboration XX},
  {\textit{Planck} 2015 results. XX. Constraints on inflation}. 2016, \aap, in
  press, \eprint{1502.02114}

\bibitem[{{\sorthelp{Planck Collaboration 2015V}}{Planck Collaboration
  XXII}(2016)}]{planck2014-a28}
{\sorthelp{Planck Collaboration 2015V}}{Planck Collaboration XXII},
  {\textit{Planck} 2015 results. XXII. A map of the thermal Sunyaev-Zeldovich
  effect}. 2016, \aap, in press, \eprint{1502.01596}

\bibitem[{{\sorthelp{Planck Collaboration 2015ZA}}{Planck Collaboration
  XXVI}(2016)}]{planck2014-a35}
{\sorthelp{Planck Collaboration 2015ZA}}{Planck Collaboration XXVI},
  {\textit{Planck} 2015 results. XXVI. The Second Planck Catalogue of Compact
  Sources}. 2016, \aap, in press, \eprint{1507.02058}

\bibitem[{{\sorthelp{Planck Collaboration 2015ZB}}{Planck Collaboration
  XXVII}(2016)}]{planck2014-a36}
{\sorthelp{Planck Collaboration 2015ZB}}{Planck Collaboration XXVII},
  {\textit{Planck} 2015 results. XXVII. The Second Planck Catalogue of
  Sunyaev-Zeldovich Sources}. 2016, \aap, in press, \eprint{1502.01598}

\bibitem[{{\sorthelp{Planck Collaboration IntZE}}{Planck Collaboration Int.
  XXX}(2016)}]{planck2014-XXX}
{\sorthelp{Planck Collaboration IntZE}}{Planck Collaboration Int. XXX},
  {\textit{Planck} intermediate results. XXX. The angular power spectrum of
  polarized dust emission at intermediate and high Galactic latitudes}. 2016,
  \aap, 586, A133, \eprint{1409.5738}

\bibitem[{{Plaszczynski} {et~al.}(2012){Plaszczynski}, {Lavabre}, {Perotto}, \&
  {Starck}}]{2012AA...544A..27P}
{Plaszczynski}, S., {Lavabre}, A., {Perotto}, L., \& {Starck}, J.-L., {A hybrid
  approach to cosmic microwave background lensing reconstruction from all-sky
  intensity maps}. 2012, \aap, 544, A27, \eprint{1201.5779}

\bibitem[{{Polnarev}(1985)}]{1985SvA....29..607P}
{Polnarev}, A.~G., {Polarization and Anisotropy Induced in the Microwave
  Background by Cosmological Gravitational Waves}. 1985, \sovast, 29, 607

\bibitem[{{Seljak} \& {Zaldarriaga}(1997)}]{1997PhRvL..78.2054S}
{Seljak}, U. \& {Zaldarriaga}, M., {Signature of Gravity Waves in the
  Polarization of the Microwave Background}. 1997, \prl, 78, 2054,
  \eprint{astro-ph/9609169}

\bibitem[{{Sherwin} \& {Schmittfull}(2015)}]{Sherwin2015}
{Sherwin}, B.~D. \& {Schmittfull}, M., {Delensing the CMB with the cosmic
  infrared background}. 2015, \prd, 92, 043005, \eprint{1502.05356}

\bibitem[{{Simard} {et~al.}(2015){Simard}, {Hanson}, \& {Holder}}]{Simard2015}
{Simard}, G., {Hanson}, D., \& {Holder}, G., {Prospects for Delensing the
  Cosmic Microwave Background for Studying Inflation}. 2015, \apj, 807, 166,
  \eprint{1410.0691}

\bibitem[{{Smith} {et~al.}(2009{\natexlab{a}}){Smith}, {Cooray}, {Das},
  {Dor{\'e}}, {Hanson}, {Hirata}, {Kaplinghat}, {Keating}, {Loverde}, {Miller},
  {Rocha}, {Shimon}, \& {Zahn}}]{cmbpol}
{Smith}, K.~M., {Cooray}, A., {Das}, S., {et~al.} 2009{\natexlab{a}}, in
  American Institute of Physics Conference Series, Vol. 1141, American
  Institute of Physics Conference Series, ed. S.~{Dodelson}, D.~{Baumann},
  A.~{Cooray}, J.~{Dunkley}, A.~{Fraisse}, M.~G. {Jackson}, A.~{Kogut},
  L.~{Krauss}, M.~{Zaldarriaga}, \& K.~{Smith}, 121--178

\bibitem[{{Smith} {et~al.}(2009{\natexlab{b}}){Smith}, {Cooray}, {Das},
  {Dor{\'e}}, {Hanson}, {Hirata}, {Kaplinghat}, {Keating}, {Loverde}, {Miller},
  {Rocha}, {Shimon}, \& {Zahn}}]{Smith2009}
{Smith}, K.~M., {Cooray}, A., {Das}, S., {et~al.} 2009{\natexlab{b}}, in
  American Institute of Physics Conference Series, Vol. 1141, American
  Institute of Physics Conference Series, ed. S.~{Dodelson}, D.~{Baumann},
  A.~{Cooray}, J.~{Dunkley}, A.~{Fraisse}, M.~G. {Jackson}, A.~{Kogut},
  L.~{Krauss}, M.~{Zaldarriaga}, \& K.~{Smith}, 121--178

\bibitem[{{Song} {et~al.}(2003){Song}, {Cooray}, {Knox}, \&
  {Zaldarriaga}}]{Song2003}
{Song}, Y.-S., {Cooray}, A., {Knox}, L., \& {Zaldarriaga}, M., {The
  Far-Infrared Background Correlation with Cosmic Microwave Background
  Lensing}. 2003, \apj, 590, 664, \eprint{astro-ph/0209001}

\bibitem[{{Spergel} \& {Zaldarriaga}(1997)}]{1997PhRvL..79.2180S}
{Spergel}, D.~N. \& {Zaldarriaga}, M., {Cosmic Microwave Background
  Polarization as a Direct Test of Inflation}. 1997, \prl, 79, 2180,
  \eprint{astro-ph/9705182}

\bibitem[{{Starobinsky}(1979)}]{1979PZETF..30..719S}
{Starobinsky}, A.~A., {Spectrum of gravitational background radiation and
  initial state of the universe.} 1979, Pisma v Zh. Eksp. Teor. Fiz., 30, 719

\bibitem[{{Starobinsky}(1982)}]{1982PhLB..117..175S}
{Starobinsky}, A.~A., {Dynamics of phase transition in the new inflationary
  universe scenario and generation of perturbations}. 1982, Phys. Lett. B, 117,
  175

\bibitem[{{Tegmark} {et~al.}(1997){Tegmark}, {Taylor}, \&
  {Heavens}}]{1997Tegmark}
{Tegmark}, M., {Taylor}, A.~N., \& {Heavens}, A.~F., {Karhunen-Lo{\`e}ve
  Eigenvalue Problems in Cosmology: How Should We Tackle Large Data Sets?}
  1997, \apj, 480, 22, \eprint{astro-ph/9603021}

\bibitem[{{The Polarbear Collaboration: P.~A.~R.~Ade} {et~al.}(2014){The
  Polarbear Collaboration: P.~A.~R.~Ade}, {Akiba}, {Anthony}, {Arnold},
  {Atlas}, {Barron}, {Boettger}, {Borrill}, {Chapman}, {Chinone}, {Dobbs},
  {Elleflot}, {Errard}, {Fabbian}, {Feng}, {Flanigan}, {Gilbert}, {Grainger},
  {Halverson}, {Hasegawa}, {Hattori}, {Hazumi}, {Holzapfel}, {Hori}, {Howard},
  {Hyland}, {Inoue}, {Jaehnig}, {Jaffe}, {Keating}, {Kermish}, {Keskitalo},
  {Kisner}, {Le Jeune}, {Lee}, {Leitch}, {Linder}, {Lungu}, {Matsuda},
  {Matsumura}, {Meng}, {Miller}, {Morii}, {Moyerman}, {Myers}, {Navaroli},
  {Nishino}, {Orlando}, {Paar}, {Peloton}, {Poletti}, {Quealy}, {Rebeiz},
  {Reichardt}, {Richards}, {Ross}, {Schanning}, {Schenck}, {Sherwin},
  {Shimizu}, {Shimmin}, {Shimon}, {Siritanasak}, {Smecher}, {Spieler},
  {Stebor}, {Steinbach}, {Stompor}, {Suzuki}, {Takakura}, {Tomaru}, {Wilson},
  {Yadav}, \& {Zahn}}]{2014ApJ...794..171T}
{The Polarbear Collaboration: P.~A.~R.~Ade}, {Akiba}, Y., {Anthony}, A.~E.,
  {et~al.}, {A Measurement of the Cosmic Microwave Background B-mode
  Polarization Power Spectrum at Sub-degree Scales with POLARBEAR}. 2014, \apj,
  794, 171, \eprint{1403.2369}

\bibitem[{{van Engelen} {et~al.}(2015){van Engelen}, {Sherwin}, {Sehgal},
  {Addison}, {Allison}, {Battaglia}, {de Bernardis}, {Bond}, {Calabrese},
  {Coughlin}, {Crichton}, {Datta}, {Devlin}, {Dunkley}, {D{\"u}nner},
  {Gallardo}, {Grace}, {Gralla}, {Hajian}, {Hasselfield}, {Henderson}, {Hill},
  {Hilton}, {Hincks}, {Hlozek}, {Huffenberger}, {Hughes}, {Koopman},
  {Kosowsky}, {Louis}, {Lungu}, {Madhavacheril}, {Maurin}, {McMahon},
  {Moodley}, {Munson}, {Naess}, {Nati}, {Newburgh}, {Niemack}, {Nolta}, {Page},
  {Pappas}, {Partridge}, {Schmitt}, {Sievers}, {Simon}, {Spergel}, {Staggs},
  {Switzer}, {Ward}, \& {Wollack}}]{ACT2015}
{van Engelen}, A., {Sherwin}, B.~D., {Sehgal}, N., {et~al.}, {The Atacama
  Cosmology Telescope: Lensing of CMB Temperature and Polarization Derived from
  Cosmic Infrared Background Cross-correlation}. 2015, \apj, 808, 7,
  \eprint{1412.0626}

\bibitem[{{Zaldarriaga} \& {Seljak}(1998)}]{1998PhRvD..58b3003Z}
{Zaldarriaga}, M. \& {Seljak}, U., {Gravitational lensing effect on cosmic
  microwave background polarization}. 1998, \prd, 58, 023003,
  \eprint{astro-ph/9803150}

\bibitem[{{Zaldarriaga} \& {Seljak}(1999)}]{1999PhRvD..59l3507Z}
{Zaldarriaga}, M. \& {Seljak}, U., {Reconstructing projected matter density
  power spectrum from cosmic microwave background}. 1999, \prd, 59, 123507,
  \eprint{astro-ph/9810257}

\end{thebibliography}

\end{document}